\documentclass[a4paper,12pt,english]{article}

\pdfoutput=1

\usepackage[bindingoffset=0.3cm,textheight=22.5cm,hdivide={2.7cm,*,2.7cm}, 
vdivide={*,22cm,*}]{geometry}
\usepackage{cite}
\usepackage[bookmarksnumbered=true,breaklinks=true]{hyperref}
\usepackage{youngtab}
\usepackage{amsmath,amsfonts,amssymb,babel,slashed,graphicx,color,empheq}

\usepackage[toc,page]{appendix}
\newcommand{\stoptocwriting}{%
  \addtocontents{toc}{\protect\setcounter{tocdepth}{-5}}}
\newcommand{\resumetocwriting}{%
  \addtocontents{toc}{\protect\setcounter{tocdepth}{\arabic{tocdepth}}}}

\makeatletter

\newcommand{\bbm}{\left(\begin{matrix}}
\newcommand{\ebm}{\end{matrix}\right)}
\newcommand{\beq}{\begin{eqnarray}}
\newcommand{\eeq}{\end{eqnarray}}
\makeatother

\newcommand{\del}{\partial}

\newcommand{\diverg}{\rm div}

\newcommand{\ad}{\mathrm{ad}}

\newcommand{\be}{\begin{equation}}
\newcommand{\ee}{\end{equation}}

\newcommand{\beqa}{\begin{eqnarray}}
\newcommand{\eeqa}{\end{eqnarray}} \newcommand{\eq}[1]{(\ref{#1})}
\def\nn{\nonumber} \def \bea{\begin{eqnarray}} \def\eea{\end{eqnarray}}
\def\obar{\overline}
\newcommand{\barr}{\begin{array}}
\newcommand{\earr}{\end{array}}
\numberwithin{equation}{section}

\def\a{\alpha}  \def\b{\beta}
 \def\g{\gamma} 
 \def\d{\delta} 
    \def\k{\kappa}
\def\l{\lambda} \def\L{\Lambda}  
    \def\r{\rho}
\def\s{\sigma}  

\def\cA{{\cal A}} \def\cB{{\cal B}} \def\cC{{\cal C}} 
\def\cD{{\cal D}}  \def\cF{{\cal F}} 
\def\cG{{\cal G}} \def\cH{{\cal H}} \def\cI{{\cal I}} 
\def\cJ{{\cal J}} \def\cK{{\cal K}}  
\def\cM{{\cal M}} \def\cN{{\cal N}} \def\cO{{\cal O}} 
\def\cP{{\cal P}} \def\cQ{{\cal Q}} \def\cR{{\cal R}} 
\def\cS{{\cal S}}

\def\R{{\mathbb R}} \def\C{{\mathbb C}} 

 \def\one{\mbox{1 \kern-.59em {\rm l}}}

\def\mh{\mathfrak{h}}

\def\msu{\mathfrak{su}}
\def\mso{\mathfrak{so}}
\def\hs{\mathfrak{hs}}

\def\A{{\bf A}}

\newcommand{\vol}{\mathrm{vol}}

\def\bit{\begin{itemize}} \def\eit{\end{itemize}} \def\Tr{\mbox{Tr}}

\def\({\left(} \def\){\right)}    

\def\und{\underline}

 \allowdisplaybreaks[3]

\begin{document}

\makeatother


\parindent=0cm

\renewcommand{\title}[1]{\vspace{10mm}\noindent{\Large{\bf

#1}}\vspace{8mm}} \newcommand{\authors}[1]{\noindent{\large

#1}\vspace{5mm}} \newcommand{\address}[1]{{\itshape #1\vspace{2mm}}}


\begin{titlepage}
\begin{flushright}
 UWThPh-2017-17 
\end{flushright}
\begin{center}
\title{ {\Large Higher spin gauge theory on fuzzy $S^4_N$}  }

\vskip 3mm

\authors{Marcus Sperling{\footnote{marcus.sperling@univie.ac.at}} and Harold C. 
Steinacker{\footnote{harold.steinacker@univie.ac.at}}}

\vskip 3mm

 \address{ 
{\it Faculty of Physics, University of Vienna\\
Boltzmanngasse 5, A-1090 Vienna, Austria  }  
  }

\bigskip

\vskip 1.4cm

\textbf{Abstract}
\vskip 3mm

\begin{minipage}{14cm}%

We examine in detail the higher spin fields which arise on the basic fuzzy sphere $S^4_N$ 
in the semi-classical limit. The space of functions can be identified with 
functions on classical $S^4$ taking values in a higher spin algebra associated 
to $\mso(5)$.
We derive an explicit and complete 
classification of the scalars and one-forms on the semi-classical limit of 
$S_N^4$. The resulting kinematics is reminiscent of Vasiliev theory.
Yang-Mills matrix models naturally provide an action formulation for higher spin gauge theory on $S^4$,
with 4 irreducible modes for each spin $s\geq 1$.
We diagonalize the quadratic part of the effective action and 
exactly 
evaluate the quadratic part in the spin $2$ sector. By identifying the linear 
perturbation of the effective metric, we obtain the exact kinetic term for all
graviton candidates.
At the classical level, matter $T_{\mu\nu}$ leads to  three different 
contributions to the linearized metric:
one consistent with linearized GR, one more rapidly decreasing contribution, and one non-propagating
contribution localized at $T_{\mu\nu}$. The latter is too large to be physically acceptable,
unless there is a significant induced quantum action.
This issue should be resolved on generalized fuzzy spaces.

\end{minipage}

\end{center}

\end{titlepage}

\tableofcontents

\section{Introduction} 
The fuzzy 4-sphere $S^4_N$ \cite{Grosse:1996mz,Castelino:1997rv} is a 
noncommutative space which can be viewed as a quantization of the round 
4-sphere.
It is characterized  by the radius $R$ and by an integer $N$, which 
sets the UV scale $L_{\rm NC} \sim \frac{R}{\sqrt{N}}$ where noncommutativity becomes important. 
Functions on the sphere are replaced by finite-dimensional matrices, which act on a large
irreducible representation (irrep)
$\cH_N$ of $\mso(5)$. This provides a finite 4-dimensional quantum geometry which is 
fully covariant\footnote{This is in contrast to  e.g.\ the Moyal--Weyl plane 
$\R^4_\theta$, which is not compatible with rotations.} under $SO(5)$.
The fuzzy 4-sphere has been considered in several different contexts, including 
string theory \cite{Castelino:1997rv,Ho:2001as,Ramgoolam:2001zx,Constable:2001ag},
matrix models \cite{Kimura:2002nq,Medina:2012cs}, 
and condensed matter theory \cite{Zhang:2001xs,Hasebe:2003gx}.
Geometrical and structural aspects were studied e.g.\ in 
\cite{Ramgoolam:2002wb,Kimura:2003ab,Abe:2004sa,Karczmarek:2015gda,Hasebe:2010vp}.

Due to the presence of an intrinsic UV scale $L_{\rm NC}$ as well as an IR scale $R$,
the fuzzy 4-sphere is a very promising background for 
formulating fundamental physical models, and to realize  ideas on 
emergent gravity \cite{Steinacker:2010rh} in a covariant (Euclidean) setting. 
However,  the non-trivial internal structure of $S^4_N$ leads
to some unusual features. Most notably, its 
algebra of ``functions'' $End(\cH_N)$ is much richer than the classical counterpart. 
Besides the usual scalar functions, it  contains further modes, which can be interpreted 
as higher spin modes with $s=1,2,\ldots , N$. This suggests that $S^4_N$ should 
naturally lead to a higher spin theory, 
as observed by several authors \cite{Kimura:2002nq,Hasebe:2003gx,Heckman:2014xha}
and further examined in \cite{deMedeiros:2004wb}.

A systematic study of the higher spin theories arising on fuzzy $S^4$
was recently initiated in \cite{Steinacker:2016vgf}, with focus on the gravity sector.
The natural framework for realizing gauge theory on $S^4_N$ is provided by 
matrix models, 
in particular the maximally supersymmetric IKKT model \cite{Ishibashi:1996xs}. 
In \cite{Steinacker:2016vgf}, spin 2 modes on fuzzy $S^4_N$ were identified  
which have the required features for gravitons including the appropriate 
coupling to matter, and the transformation  under diffeomorphisms.
However, it was found that the linearized Einstein equations 
arise only on  certain \emph{generalized} 
fuzzy spheres $S^4_\L$, with some assumptions and caveats. 
The underlying issue is a constraint between the translational and rotational spin 2 modes on 
the basic $S^4_N$. 
The analysis in \cite{Steinacker:2016vgf} was, however, incomplete due to the 
mixing of several different modes, which were not fully disentangled.

In the present paper, we provide a complete and systematic classification of 
the higher spin fields which arise on the basic fuzzy sphere $S^4_N$ 
in the semi-classical limit, completing the analysis in \cite{Steinacker:2016vgf}.
First, we realize in Section \ref{sec:scalar-fields} the space of 
functions in terms of suitable Young diagrams, or equivalently in terms of 
traceless rank $s$ tensor field on $S^4$.
There is one such mode for each spin $s$. This can be captured succinctly in terms of function on $S^4$ taking values 
in an infinite-dimensional higher spin algebra $\hs$ associated to $\mso(5)$. Locally, $\hs$ coincides with the 
semi-classical limit of Vasiliev's higher spin algebra,
but the global structure is more intricate. 
The fuzzy case provides a finite truncation of 
$\hs$.

Second,  we 
provide in Section \ref{sec:vec-harmonics} a complete and explicit classification of all ``vector'' fluctuation modes on $S^4_N$ in the 
framework of Yang-Mills matrix models. It turns out that there are 
4 distinct irreducible (off-shell, gauge-fixed) modes for each spin $s\geq 1$, which are explicitly realized in terms of suitable Young 
diagrams or, equivalently, in terms of  rank $s$ tensor fields.
All these modes can be arranged in terms of a tangential 1-form on $S^4$ taking 
values in $\hs$, and a (radial) function on 
$S^4$ taking values in $\hs$. This provides a kinematical link with Vasiliev theory \cite{Vasiliev:1990en,Vasiliev:2004qz}.
The local representation of these modes involves a combinations of 
gauge fields and their field strength, which is explicitly worked out for spin 1 and 2.

The next step is the formulation of physically interesting higher spin theories.
Matrix models provides a natural action for a  interacting
gauge theory on fuzzy $S^4_N$, where all fields transform under $\hs$-valued local gauge transformations.
This is  a remarkable statement, given the notorious difficulty in 
finding an action formulation 
of higher spin theories.
We explicitly diagonalize the quadratic part of this action in Section 
\ref{sec:metric_and_graviton}.
Focusing on the spin 2 sector, we identify the effective metric fluctuation (graviton) $h_{\mu\nu}$, which is a  
linear combination of the basic spin 2 modes. We recover the appropriate transformation law under diffeomorphisms, which 
are part of the higher spin gauge invariance. 

Given the full classification of the modes,
we compute exactly the  quadratic part of the effective action for these spin 2 modes coupled to matter via $T_{\mu\nu}$. 
It turns out that the quadratic action for $h_{\mu\nu}$ arises primarily via its spin connection, so that 
its dominant role is that of a non-propagating auxiliary field, 
which is strongly localized at the matter source.
However, there is a (sub-leading) mode  which does mediate linearized Einstein gravity, and yet another mode 
which is more rapidly decaying, but nevertheless large. 
This means that the classical higher spin theory on the basic $S^4_N$ in Yang-Mills matrix models 
does \emph{not} lead to realistic gravity, consistent with 
\cite{Steinacker:2016vgf}.

Nevertheless, we point out two possibilities which might lead to (more) 
realistic gravity in this context: 
First, the inclusion of one-loop quantum effects leads, as usual, to induced 
gravity terms in the effective action. If these induced terms are  large, 
the above conclusion is reversed:
the would-be auxiliary graviton is transmuted into a proper graviton governed by the appropriate linearized Einstein equations, 
while the other modes lead to sub-leading long-distance modifications, somewhat 
reminiscent of conformal gravity \cite{Mannheim:2010ti}.
However, this scenario requires special parameter regimes.

The second, perhaps more interesting possibility to obtain (more) realistic gravity is to replace the basic fuzzy sphere $S^4_N$ 
by the generalized fuzzy sphere $S^4_\L$, as suggested in \cite{Steinacker:2016vgf}. 
The point is that $S^4_\L$ admits translational modes which are independent of the rotational modes,
unlike in the basic case.
Consistent with the preliminary results in \cite{Steinacker:2016vgf}, we identify an appropriate graviton mode 
on $S^4_\L$ which appears to avoid the undesired behavior found on $S^4_N$. 
This could be investigated along the same lines as in the 
present paper, but is postponed to future work.

Even if the model under consideration may not yet lead to the desired physics, 
the main message is, nonetheless, 
remarkable and very promising: Matrix models provide a natural and simple framework for  
actions for higher spin gauge theories on (fuzzy) $S^4$, which arise from the twisted bundle structure of a
higher-dimensional noncommutative space over $S^4$.
 This provides a covariant quantization of a 4-dimensional space in a rigorous framework, and
a simple geometric origin for higher spin theories.
For example, the higher spin gauge transformations on $S^4$ are recognized as symplectomorphisms on $\C P^3$.

The present paper is restricted to the 
Euclidean case, having the advantage that the 
rich structure is completely under control.
Of course one would like to move on to Lorentzian signature. 
There are also some candidates for analogous covariant fuzzy spaces with Minkowski signature 
\cite{Gazeau:2009mi,Heckman:2014xha,Buric:2015wta}.
An analogous study for such spaces is postponed to future work.

\section{The basic fuzzy 4-sphere \texorpdfstring{$S^4_N$}{S4N}}
We are interested in covariant fuzzy 4-spheres, which are  defined in terms of 5  hermitian matrices 
$X^a, \ a=1,\ldots , 5$ acting on some finite-dimensional Hilbert space $\cH$, 
and transforming as vectors under $SO(5)$, i.e.
\begin{align}
 [\cM_{ab},X_c] &= i(\d_{ac} X_b - \d_{bc} X_a), \nn\\
  [\cM_{ab},\cM_{cd}] &=i(\d_{ac}\cM_{bd} - \d_{ad}\cM_{bc} - \d_{bc}\cM_{ad} + \d_{bd}\cM_{ac}) \ .
 \label{M-M-relations}
\end{align}
Throughout this paper, indices are raised and lowered with  $g_{ab} = \d_{ab}$.
The $\cM_{ab}=- \cM_{ba}$ for $a,b \in\{1,\ldots, 5\}$ generate a suitable
representation of $\mso(5)$ on $\cH$, and $X^a \in End(\cH)$ are operators
 interpreted as quantized embedding functions 
$X^a \sim x^a: \ S^4 \hookrightarrow \R^5$. Then the radius 
\begin{align}
 X^a X_a &= \cR^2 
 \label{XX-R-fuzzy}
\end{align}
is a scalar operator of dimension $L^2$. 
The commutator of the $X^a$ will be denoted by 
\begin{align}
  [X^a,X^b]  & \eqqcolon i \Theta^{ab} \ .
\end{align}
Such relations constitute a \emph{covariant quantum 4-sphere}.
Particular realizations of 
such fuzzy 4-spheres are obtained from generators $\cM^{ab}, \ a,b = 1,\ldots 
,6$ of $\mso(6) \cong \msu(4)$ via
\begin{align}
 X^a &= r \cM^{a6}, \qquad a = 1,\ldots ,5 \ , \qquad \Theta^{ab} = r^2 \cM^{ab} 
\ .
 \label{X-def}
\end{align}
Here $r$ is a scale parameter of dimension $L$, and $\cH$ is some irreducible representation (irrep) of $\mso(6)$. 
This class of quantum spheres was considered in \cite{Steinacker:2016vgf} as a promising basis for 
a higher spin theory including gravity, and their geometry was studied further in \cite{Sperling:2017dts}.

These covariant quantum 4-spheres can be viewed as compact versions of Snyder space \cite{Snyder:1946qz,Yang:1947ud}.
The crucial feature is that the classical isometry group  $SO(5)$ is fully realized.
This is in marked contrast to the basic quantum spaces such as the Moyal-Weyl quantum plane $\R^4_\theta$,
where the Poisson tensor $\theta^{ab}$ breaks this symmetry.
The price to pay is that the algebra of ``coordinates'' $X^a$  does not close, 
because extra generators $\Theta^{ab}$ are involved.
Nevertheless, one can define physical theories on such spaces via 
matrix models, leading to fully covariant higher spin theories
with large gauge symmetry, including a gauged version of $SO(5)$.

In this paper we will focus on
the simplest example of the above construction: the ''basic`` fuzzy four-sphere 
$\cS^4_N$ \cite{Grosse:1996mz,Castelino:1997rv,Ramgoolam:2001zx}.
This is obtained for the highest weight irrep  $\cH = \cH_\L$ of $\mso(6)$ with $\L = (N,0,0)$.
Throughout this paper we  denote  highest weights by their Dynkin indices. 
Then the following relations hold:
\begin{align}
  X^a X_a &= \cR^2 = r^2 R_N^2 \one \, , \qquad  R_N^2 = \frac{1}{4} N(N+4)  \, 
, \nn\\
 \{X_a,\Theta^{ab}\}_+ &=  0 \, , \nn\\
 \frac 12\{\Theta^{ab},\Theta^{a'c}\}_+ g_{aa'} &=  r^2\cR^2 \big(g^{bc} -\frac 
1{2\cR^2} \{X^b, X^c\}_+\big) \, , \nn\\
 \epsilon_{abcde} \Theta^{ab} \Theta^{cd}   &= 4(N+2) r^3\; X_e \, , 
  \label{XX-R-const-fuzzy}
\end{align}
for indices $a,b,\ldots =1,\ldots ,5$.
Here $\{\cdot,\cdot\}_+$ denotes the anti-commutator.
The first relation expresses the fact that $\cH_\L$ remains irreducible as 
representation of $\mso(5) \subset \mso(6)$, which no longer holds for generic $\L$.

\paragraph{Oscillator construction.}
It is worthwhile recalling the following oscillator construction of fuzzy $S^4_N$ \cite{Grosse:1996mz,Castelino:1997rv}.
Consider four bosonic oscillators 
\begin{align}
 [a^\b,a^\dagger_\a] = \d_\a^\b, \qquad \a,\b=1,\ldots ,4 \, ,
\end{align}
which transform in the spinorial representation of $\mso(6)$ (and $\mso(5)$).
Then define (for $r=1$)
\begin{align}
 X^c = \frac 12 a^\dagger \g^c a, \qquad \cM^{ab} =  a^\dagger  {\Sigma^{ab}}  a = -i[X^a,X^b]
 \label{JS-S4}
\end{align}
(suppressing spinorial indices), 
where $\g^c, \ c=1,\ldots ,5$ are the gamma matrices associated to $SO(5)$ 
acting on $\C^4$.
It is then easy to check that \eqref{M-M-relations} is satisfied, and 
\eqref{XX-R-fuzzy}, \eqref{XX-R-const-fuzzy} hold on 
the $N$-particle Hilbert space $\cH_N = a^\dagger_{\a_1} \ldots  
a^\dagger_{\a_N}|0\rangle \cong (0,N)_{SO(5)}$.

\subsection{Semi-classical limit \texorpdfstring{$\cS^4$}{SS4}}
To  understand the geometrical meaning of $\Theta^{ab}$, 
it is best to view the fuzzy sphere as 
quantization of the 6-dimensional coadjoint orbit $\C P^3$ of $SO(6)$; this 
viewpoint naturally extends to the generalized spheres $\cS_\Lambda^4$ of 
\cite{Steinacker:2016vgf,Sperling:2017dts}.
The generic construction is as follows\footnote{See e.g.\ 
\cite{Hawkins:1997gj} 
for a nice introduction to  (quantized) coadjoint orbits.}: 
For any 
given (finite-dimensional) irrep $\cH_\L$ of $SO(6)$ with highest weight $\L$, the generators $\cM^{ab} \in End(\cH_\L)$ 
of its Lie algebra $\mso(6)$
are viewed as quantized embedding functions
\begin{align}
 \cM^{ab} \sim m^{ab}: \quad \cO_\L \hookrightarrow \R^{15} \cong \mso(6) 
\ 
\end{align}
of the homogeneous space (coadjoint\footnote{For simplicity we identify the Lie algebra with its dual.} orbit)
\begin{align}
 \cO_\L = \{g\cdot \L \cdot g^{-1}; \ g\in SO(6)\} \ \cong SO(6)/\cK  \subset 
\R^{15} \,,
\end{align}
with $\cK$ denoting the stabilizer of $\L$ in $SO(6)$.
As customary, one can identify $\L$  with a Cartan generator 
of $\mso(6)$ via  $\L \in \mh^* \leftrightarrow H_\L \in \mh$.
For $\L = (N,0,0)$, this gives  $ \cO_\L \cong\C P^3$,  which is naturally a 
$S^2$-bundle over $S^4$ via the Hopf map
\begin{align}
  x^a \ = r\ m^{a6}: \quad   \C P^3  \to  S^4\hookrightarrow \R^5 \,. 
 \end{align}
We denote this $SO(5)$-equivariant bundle with $\cS^4 \cong \C P^3$ in this paper. 
Define $\theta^{ab} = r^2 m^{ab}$ for $a,b=1,\ldots,5$, then one can show the 
following semi-classical analogs of \eqref{XX-R-const-fuzzy}:
\begin{align}
  x^a x_a &= R^2  \, ,\nn\\
  x_a \theta^{ab} &=  0 \, , \nn\\
  \theta^{ab} \theta^{a'c} g_{aa'} &=  \frac{L_{NC}^4}{4} \big(g^{bc} -\frac 
1{R^2} x^b x^c \big) =   \frac{L_{NC}^4}{4}\, P_T^{bc} \, , \nn\\
 \epsilon_{abcde} \theta^{ab}\theta^{cd}  &=  2 L_{NC}^4\, \frac{x_e}{R} \, ,
  \label{XX-R-const}
\end{align}
for $ a,b=1,\ldots ,5$, where 
\begin{align}
 \theta = r^2, \qquad L_{NC}^2 = 2 r R \label{eq:def_L_NC}
\end{align}
are parameters of dimension $L^2$. We refer to the tensor $P_T^{bc}$ as 
\emph{tangential projector} because it satisfies $P_T^{bc}x^c = x^b$
and $P_T^{ab}P_T^{bc} = P_T^{ac}$.
\paragraph{Poisson structure.}
Any such coadjoint orbit  carries a (Kirillow--Kostant) Poisson structure,
\begin{align}
 \{\theta^{ab},\theta^{cd}\} &= \theta \Big(g^{ac}\theta^{bd}  - 
g^{ad}\theta^{bc} - g^{bc}\theta^{ad} + g^{bd}\theta^{ac} \Big) , \qquad 
a,b,c,d=1,\ldots ,5  \, , \nn\\
  \{\theta^{ab},x^{c}\} &=\theta \Big(g^{ac}x^{b}  - g^{bc}x^{a}  \Big) , \qquad 
a,b,c=1,\ldots ,5 \, ,  \nn\\
  \{x^{a},x^{c}\} &=   \theta^{ac} , \qquad a,c=1,\ldots ,5 \, ,
 \label{Poisson-theta}
\end{align}
which is  $SO(5)$-invariant.
This can also be obtained from an oscillator  construction as in \eqref{JS-S4},
replacing the creation- and annihilation generators $a_\a, a^\dagger_\a$ by 
holomorphic
coordinate functions on $\C^4$ 
with Poisson structure $\{a^\b,a^\dagger_\a\} = -i\d_\a^\b$.
In particular, we note
\begin{align}
  \{\theta^{ab},x^{b}\} &= - 4 \theta x^a \ ,
  \label{eom-class}
\end{align}
which are the equations of motion of the ''Poisson matrix model`` 
\eqref{bosonic-action} introduced on Section \ref{sec:vector-Laplacian}. 
Consequently, the $\cS^4$ generated by $x^a$ is a solution thereof.

For an arbitrary point $p\in S^4$  (e.g.\ the ``north pole'' $p=R(0,0,0,0,1)$), 
we can decompose  $\mso(5)$ into rotation generators 
$\cM^{\mu\nu}$ ($\mu,\nu =1,\ldots,4)$ and translation generators $P^\mu$, 
given by
\begin{align}
 P^\mu = \frac 1{R\theta}\theta^{\mu 5} \; ,\qquad \mu =1,\ldots,4 \ .
 \label{P-def}
\end{align}
Although they vanish as classical functions due to $P^\mu \propto \{x^\mu,x^a\} 
x_a = 0$, see also \cite{Steinacker:2016vgf}, they do not vanish as generators, 
and cannot be dropped.

\paragraph{Coherent states.}
It is well-known that the quantized coadjoint orbits $\cO_\Lambda$ allow for 
the introduction of coherent states, which lie on the $SO(6)$ orbits of the 
highest weight state $|\Lambda\rangle \in \cH_\Lambda$, i.e.
\begin{align}
 |x\rangle \equiv |x;\xi\rangle \coloneqq g_x \cdot |\Lambda\rangle  \,, \qquad 
 g_x \in SO(6) \; .
\end{align}
Here, we labelled the points on $\cO_\Lambda \cong \C P^3 $ by $x\in S^4$ and 
the fiber coordinate $\xi$, where the ``north pole'' $p$ corresponds to $ 
|\Lambda\rangle$. Coherent states are optimally localized, i.e.\ they minimize 
the uncertainty in position space. Using the defining relations 
\eqref{XX-R-const-fuzzy}, one computes in the large $N$ approximation
\begin{equation}
\begin{aligned}
 \Delta^2 &\coloneqq 
 \sum_{a=1}^5 \langle \left(X^a - \langle X^a \rangle \right)^2 \rangle 
 =\sum_{a=1}^5 \langle \left(X^a\right)^2 \rangle  - \left(\langle X^a \rangle
\right)^2 \\
&=r^2 \left( R_N^2 -\frac{N^2}{4} \right) 
\approx \frac{4}{N} R^2  \approx 2r R \eqqcolon L_{NC}^2 \;.
\end{aligned} 
\end{equation}
This then defines the length scale $L_{NC}$, which appeared in 
\eqref{eq:def_L_NC}, and highlights its role as non-commutativity scale.
\paragraph{Scales.}
Before proceeding we emphasize that there are two scales involved: 
the IR or ``cosmological'' scale $R$ giving the size of the sphere, and the  UV
scale $L_{NC} = \frac{R}{\sqrt{N}}$. This is the scale where 
non-commutative corrections in the star product become relevant, since
\begin{align}
  f \star g = f \cdot g + O\left( (L_{\mathrm{NC}} \cdot \partial)^2(f,g) 
\right) \; . \label{eq:Star-product}
\end{align}
Strictly speaking there is also a third scale $\frac{R}{N}$, which is the UV 
cutoff on $S^4_N$.

\subsection{Calculus and forms}
We want to develop a differential calculus which allows to efficiently work with this semi-classical $\cS^4$.
We introduce  formal Grassmann variables  $\xi^a$, which transform as a vector 
of $SO(5)$, and satisfy
\begin{align}
 \xi^a \xi^b = -  \xi^b \xi^a  \qquad a,b=1,\ldots,5\, .
\end{align}
The space $\Omega^n \cS^4$ is defined as the $\cS^4$-module of forms  of 
degree $n$ in $\xi^a$, and the (exterior) algebra of forms on $\cS^4$ is
\begin{align}
 \Omega^* \cS^4 = \bigoplus\limits_{n=0}^5  \Omega^n \cS^4, \qquad \quad
 \text{with } \quad  \Omega^0 \cS^4 \equiv \cC \ .
\end{align}
Here $\cC$ is the space of functions on $\cS^4 \cong \C P^3$ with generators $x^a$ and $\theta^{ab}$.
There are three special $SO(5)$-invariant forms 
\begin{align}
 \xi &\coloneqq  x_a \xi^a \qquad \in \Omega^1 \cS^4 \, , \nn\\
 \omega &\coloneqq  \theta_{ab}\xi^a \xi^b  \qquad\in \Omega^2 \cS^4  \, , \nn\\
 \Omega &\coloneqq \epsilon_{abcde}\xi^a \ldots  \xi^e   \qquad \in \Omega^5 
\cS^4 \, ,
\end{align}
which play a special role.
$\Omega$ is the 5-dimensional ''volume`` form.
Using the invariant metric, we can define the 4-dimensional cotangent space as orthogonal complement to $\xi$,
\begin{align}
 T^* \cS^4 \coloneqq  (\xi)^\perp \ = \{A_a \xi^a \in \Omega^1\cS^4 \ 
\mbox{with} \ A_a x^a = 0 \} \, . 
\end{align}
Consider the following $SO(5)$ intertwiners:
\begin{alignat}{2}
 \cQ: \ \Omega^n \cS^4 &\to \Omega^{n+1} \cS^4 , & \qquad \cQ(\a_n) &= 
\{\xi,\a_n\}_\pm \, , \nn\\
 \cJ:  T^* \cS^4 &\to  T^* \cS^4, & \qquad \cJ(\xi^a A_a) &= \xi_a\theta^{ab} 
A_b  \, , \nn\\
 \cI: \Omega^1 \cS^4 &\to \Omega^1\cS^4, & \qquad  \cI(\xi^a A_a) &= 
\xi_a\{\theta^{ab},A_b\} \, , \nn\\
 \cN:  \Omega^{1}\cS^4 &\to \cC , & \qquad  \cN(\xi^a A_a) &= x_a A^a \, ,
 \label{intertwiners}
\end{alignat}
where $\{\cdot , \cdot \}_\pm $ denotes the appropriately graded Poisson 
bracket.
They satisfy 
\begin{align}
 \cQ^2 \a &= \frac 12 \{\omega,\a\} \, , \nn\\
 \cJ^2 &= - \theta R^2 P_T, \qquad \cJ (\xi) = 0 \, .
 \label{eq:J-structure}
\end{align}
$\cJ$ arises from  the complex  (K\"ahler) structure on 
the bundle space $\C P^3$.
We will see in Section \ref{sec:gauge_trafo_functions} that $\cQ(\L)$ is 
the infinitesimal gauge transformation of the $\cS^4$ background.
Note that $\cQ$ and $\cJ$ are tangential, which means they are annihilated by 
$\cN$:
\begin{align}
 \cN(\cJ(\a)) &= 0 , \qquad \a\in  \Omega^1 \cS^4 \, , \nn\\
 \cN(\cQ(f)) &= 0,  \qquad \mbox{i.e.}\ \ \cQ f = \xi_a \{x^a,f\} \in T^* \cS^4,
\end{align}
because $x_a \{x^a,f\} = \frac 12\{R^2,f\} = 0$ and $f\in \cC$. Moreover, we can 
write
\begin{align}
 \cQ^2(f) \ =  \  \theta R\xi^5\xi^\mu P_\mu f, \qquad  f\in \cC \, ,
\end{align}
for instance at the north pole, since $\{\cM^{\mu\nu},\cdot \}$ vanishes for 
functions.
This might seem  reminiscent of supersymmetry, however $\cQ^2$ involves a 
commutator rather than an anti-commutator. 
Furthermore, using \eqref{eom-class}
one can show the following identities (see Appendix 
\ref{sec:app-proofs} for more details):
\begin{align}
   \cI(\xi \phi) &= - 4 \theta\xi \phi - \cJ(\cQ(\phi))    \label{I-N-identity} 
\, , \\
  \cI \circ\cJ (\xi^a A_a) &= - 4 \theta\cJ(\cA) +\theta \xi \{ x_c,A^c\} 
+\theta \cQ(\cN(\cA)) -  \cJ\circ \cI (\cA) \, .
  \label{I-J-identity}
\end{align}
One can also define a Hodge star operator either on  $\R^5$ or on $S^4$  as 
follows:
\begin{align}
*: \Omega^1 \cS^4 &\to \Omega^4\cS^4 , \qquad *(\xi^a) = c\,\epsilon^{abcde} 
\xi_b\xi_c\xi_d \xi_e\quad \mbox{etc.}  \nn\\
  *_4: \Omega^1 \cS^4 &\to \Omega^3 \cS^4, \qquad  *_4(\a) = *(\a \xi)  
\end{align}
normalized such that $* * = 1$; however this will not be important the present paper.

Now consider the  functional
\begin{align}
\cG : \Omega^1\cS^4 \to \cC \, , \qquad 
  \cG(\cA)  \coloneqq  \{x^a,A_a\}, \qquad \cA \equiv \xi_a A^a \in 
\Omega^1\cS^4 \, ,
\label{gauge-fixing-function}
\end{align}
which will be used for gauge fixing in Section \ref{sec:vector-Laplacian}.
The kernel of $\cG$  contains $\cJ(\cQ(\phi))$, because 
\begin{align}
 \cG[\cJ(\cQ(\phi))] &=  \{x^a,\theta^{ab} \{x_b,\phi\}\} \nn\\
  &=  \{x^a,\theta^{ab}\} \{x_b,\phi\}  +  \theta^{ab} \{x^a,\{x_b,\phi\}\}  \nn\\
  &=  \theta^{ab} \{x^a,\{x_b,\phi\}\} =0 \, ,
  \label{F-JQ-id}
\end{align}
noting that
\begin{align}
 \theta^{ab} \{x^a,\{x_b,\phi\}\}  
  &= - \theta^{ab}( \{x^b,\{\phi,x^a\}\} + \{\phi,\{x^a,x^b\}\}) \nn\\
  &= -\theta^{ab}\{x^a,\{x^b,\phi\}\} -\theta^{ab} \{\phi,\{x^a,x^b\}\} \, .
\end{align}
Hence
\begin{align}
 2\theta^{ab} \{x^a,\{x_b,\phi\}\}  
  &= -\theta^{ab} \{\phi,\theta^{ab}\} = 0 \, ,
\end{align}
which holds for any function $\phi\in\cC$.
Finally there is a natural ''Poisson`` Laplacian, given by
\begin{align}
  \Box f  \coloneqq  \{x^a,\{x_a,f\}\} ,  \qquad f \in \cC  
  \label{Poisson-Laplacian}
\end{align}
or equivalently $\Box f= c' *\cQ * \cQ f$, for some number $c'$.
The vector Laplacian is given by
\begin{align}
 \cD^2: \quad \Omega^1 \cS^4 \to \Omega^1 \cS^4, 
 \qquad \cD^2 \cA_a = \big(-\Box - 2\cI\big) \cA_a \, ,
 \label{vector-Laplacian}
\end{align}
which  will be discussed in Section \ref{sec:vector-Laplacian}.

\subsection{Derivation and connection}
\paragraph{Derivation.}
We can define the following $SO(5)$-covariant derivation on $\cC$:
\begin{align}
  \del   \coloneqq
  -\frac{1}{\theta R^2}\cJ\circ \cQ: \quad \cC \to T^*\cS^4
 \label{nabla-def}
\end{align}
or more explicitly 
\begin{align}
  \del^a \phi \coloneqq -\frac{1}{\theta R^2}\theta^{ab}\{x_b,\phi\}, \qquad 
\phi\in\cC \; ,
\end{align}
which is indeed tangential and satisfies the Leibniz rule.
The definition \eqref{nabla-def} is equivalent to
\begin{align}\boxed{
 \{x^a,\cdot\} = \theta^{ab}\del_b \, . \ 
 }
 \label{x-nabla-id}
\end{align}
In particular, 
\begin{align}
 \del^a x^c = -\frac{1}{\theta R^2}\theta^{ab}\{x_b,x^c\} = P_T^{ac} \, .
\end{align}
 Therefore $\del$ reduces to the ordinary tangential derivative $\del^\mu f|_p$
for scalar functions $f(x)$ at any given point $p\in S^4$, e.g. the north pole. 
The derivation acts on the $\theta^{ab}$ generators as
\begin{align}
 \del^a \theta^{cd} &= -\frac{1}{\theta R^2}\theta^{ab}\{x_b,\theta^{cd}\}
   = \frac{1}{R^2}\big(-\theta^{ad} x^c +\theta^{ac} x^d \big) \, ,
\end{align}
which at the north pole reduces to
\begin{align}
 \del^\mu \theta^{\nu\eta}  &= 0 \, ,  \nn\\
  \del^\mu P^{\nu} &=\frac{1}{R^2} \theta^{\mu \nu} , \qquad  
  P^\nu = \frac{1}{\theta R} \theta^{\nu 5} \ .
  \label{nabla-theta}
\end{align}
Note that the second relation connects $\theta^{\mu\nu}$ and $P^\mu$.
In particular, although the $P^\mu$  vanish as functions on $\C P^3$, they
do not vanish as generators, and cannot be dropped. 
As a consistency check, we note that 
\begin{align}
 0 = \del^\mu (\theta^{\nu a} x_a) = \frac{1}{R} \theta^{\mu \nu} R + 
\theta^{\nu a} \d^{\mu}_a \; .
\end{align}
Moreover, the following intertwiner $\diverg: \Omega^1 \cS^4 \to \cC$ defined 
as:
\begin{align}
 {\diverg} \cA &\coloneqq \del\cdot\cA = \del^a A_a \nn\\
 &= -\frac{1}{\theta R^2}\theta^{ab}\{x^a,A_b\} =   \frac{1}{\theta R^2}x^a\{\theta^{ab},A_b\} 
 =\frac{1}{\theta R^2} \cN(\cI(\cA)) 
   \label{N-I-A-id}
\end{align}
reduces to the divergence for tangential vector fields.
It satisfies
\begin{align}
 {\diverg} \cJ(\cA) &=  \frac{1}{\theta R^2}\cN(\cI(\cJ(\cA)))
  = \{ x^a,A_a\}  = \cF(\cA)  \nn\\
  &= \theta^{a\mu}\del_\mu A_a 
  = \frac 12\theta^{\nu\mu}(\del_\mu A_\nu - \del_\nu A_\mu) 
\end{align}
using \eqref{I-J-identity}. In particular,  $\{x^a,A_a\}$ is some component of 
the field strength $\cF(\cA)$ of.
\paragraph{Connection.}
Now consider tensor fields  on $S^4$ such as $A_a$, $A_{ab}$, 
$\ldots$ which are tangential, i.e.\ $A_{ab} x^a = 0$ etc.
Then $\del$ does not respect these constraints: 
for example, if $A_a$ is a tangential vector field, i.e.\ $x^a A_a = 0$, then
$\del_a A_b$ is not tangential in the index $b$, since
\begin{align}
 x^b \del_a A_b = \del_a(x^b A_b) -  A_b \del_a x^b = -  A_b P_T^{ba} = -A_a \neq 0 \ .
\end{align}
To remedy this, we project on the tangential indices with $P_T$,
and define
\begin{align}
 \nabla \coloneqq P_T \circ \del \, ,
 \label{nabla-proper-def}
\end{align}
where $P_T$ acts  on all components. For example if $A_a$ is tangential, then 
\begin{align}
 \nabla_a A_b = \del_a A_b + \frac{1}{R^2}x_b A_a 
 \label{nabla-a-phi-b}
\end{align}
 is indeed tangential. 
$\nabla$ is an $SO(5)$-equivariant connection on $S^4$ which does 
not respect the sub-bundle corresponding to $P^\mu$, due to the second relation 
in \eqref{nabla-theta}.

We conclude this section with two remarks. First, this calculus can be naturally extended 
to act on forms  $\cA \in \Omega^* \cS^4$
by defining 
\begin{align}
 \del_a \xi_b = 0 \ . 
\end{align}
This amounts to  $\{\theta^{ab},\xi_c\} = 0 = \{x^a,\xi^b\}$.
Second, there is another connection  besides the above, which is the
canonical $SO(5)$-equivariant connection given at the north pole by
\begin{align}
 \del'_\mu = \{P_\mu, \cdot \} \ .   
\end{align}
This derivation differs from $\del$ because
$\del' \theta^{\a\b} \sim P \neq \del \theta^{\a\b}$,  while $\del'P = \del P$.

\section{Functions on \texorpdfstring{$\cS^4$}{SS4} and higher spin}
\label{sec:scalar-fields}
It is well-known that 
the algebra of functions on $\cS^4$ decomposes into $SO(5)$ harmonics as follows 
\cite{Ramgoolam:2001zx}:
\begin{align}
\begin{aligned}
 \cC \equiv \cC^\infty(\C P^3) &\cong \bigoplus\limits_{s\geq0} \cC^s \, , \\
 \phi \quad &\mapsto \sum_{s\geq 0} \phi^{(s)} \, .
 \end{aligned}
 \label{full-functions-decomp}
 \end{align}
 Here
 \begin{align}
 \cC^s &\cong \bigoplus\limits_{n \geq 0} (n,2s)_{\mso(5)}
\end{align}
is a module over the algebra of scalar functions on $S^4$, corresponding to certain  spin $s$ fields.
The component of  $\phi\in\cC$ in the module $\cC^s$ will be denoted by $\phi^{(s)}$.
We will provide a more explicit interpretation of these modules below.
However, the full algebra
 respects  this gradation in $s$ only modulo 2, because of the 
constraints \eqref{XX-R-const}.
More details about the bundles $\cC^s$ and the corresponding field strength 
etc.\ shall be discussed elsewhere\footnote{Useful discussions and 
collaboration with S.\ Ramgoolam are gratefully acknowledged here.}.

\paragraph{Averaging.}
Consider the map \cite{Ramgoolam:2001zx}
\begin{align}
 \cC =  \cC^\infty(\cS^4) &\to \cC^0  \nn\\
   \phi^{(s)} &\mapsto [\phi^{(s)}]_0 \coloneqq  \d^{s,0}   \phi^{(s)}
\end{align}
which projects to the spin 0 scalar functions $\cC^0$.
This amounts to averaging over the $S^2$ fiber at each point $x\in S^4$.
Explicitly, this is given by (cf.\ \cite{Steinacker:2016vgf})
\begin{subequations}
\label{eq:averaging}
\begin{align}
 [\theta^{ab}]_0 &=  0  \,, \\
 [\theta^{ab}\theta^{cd}]_0 &= \frac 13 \theta R^2\big(P_T^{ac}  P_T^{bd} -  
P_T^{ad}   P_T^{bc} + \frac 1R\,\epsilon^{abcde} x^e    \big) \,, \\
  [\theta^{ab}\theta^{cd}\theta^{ef}]_0 &= 0 \,, \\
 [\theta^{ab} \theta^{cd } \theta^{ef} \theta^{gh}]_0 &= \frac{3}{5} \left(
 [\theta^{ab} \theta^{cd}]_0 [\theta^{ef} \theta^{gh}]_0
 +[\theta^{ab} \theta^{ef}]_0 [\theta^{cd} \theta^{gh}]_0 
 +[\theta^{ab} \theta^{gh}]_0 [\theta^{cd} \theta^{ef}]_0
\right) \,,
\end{align}
\end{subequations}
etc. 
Similarly, there is a natural integral over $\cS^4$ defined by projecting $\cC$ to the unique trivial component.
The normalization is given by the semi-classical limit of the trace:
\begin{align}
\Tr_{End(\cH)} \, \sim \int\limits_{\cS^4} d\Omega \ = 
\frac{\dim\cH}{\vol(\cS^4)}\int\limits_{\cS^4} \ 
=\frac{\dim\cH}{ \vol( S^4)}\int\limits_{S^4} [\cdot ]_0 \ . 
\label{trace-int}
\end{align}
This is an integral over $\cS^4 = \C P^3$ with the canonical symplectic measure 
$d\Omega$, which can be written 
as an integral over the 4-sphere $S^4$ after averaging $[\cdot]_0$ over the 
$S^2$ fiber. The appropriate 
factor $ \vol( \cS^4)$ or $\vol( S^4)$ is understood in the following, and 
we will drop $d\Omega$ if no confusion can arise.
\paragraph{$\cC^0$ as scalar fields  on $S^4$.}
For $n\geq0$, $(n,0)$ can be realized as space of  symmetric traceless $SO(5)$ 
tensors 
$\phi^{(0)}_{a_1 \ldots a_n}$
corresponding to Young diagrams ${\tiny \Yvcentermath1 \young(aaa)}$
consisting of only one line of length $n$. These are in one-to-one 
correspondence to polynomial functions on $S^4$,
\begin{align}
 \phi^{(0)}  = \phi^{(0)}_{a_1 \ldots a_n} x^{a_1} \ldots  x^{a_n} 
\eqqcolon \phi^{(0)}(x) \quad \in \cC^0 \, .
\end{align}
\paragraph{$\cC^1$ as vector fields  on $S^4$.}
Consider the space of $(n,2)$ functions on $\cS^4$, for $n\geq0$.
These modes can be similarly characterized by Young diagrams 
${\tiny \Yvcentermath1 \young(baaa,c)}$
with one row of length $n+1$ and one column of length $2$.
This defines irreducible representations 
\begin{align}
 \phi^{(1)}_{a_1 \ldots a_nb; c} \coloneqq (P_S \circ P_A)  \phi^{(1)}_{a_1 
\ldots a_nb; c} \quad \subset (\C^5)^{\otimes (n+2)} \, ,
\end{align}
which are totally symmetric in $a_1 \ldots a_nb$.
Here, we have chosen a basis of tensors exhibiting the symmetry of Young 
diagrams by first antisymmetrizing in columns by $P_A$ and subsequently 
symmetrizing in rows by $P_S$. 
By contracting these tensors with generators of $\cS^4$, they define $\cC^1$  
modes via
\begin{align}
 \phi^{(1)} = \phi^{(1)}_{a_1 \ldots a_nb; c} x^{a_1} \ldots  x^{a_n} \theta^{b c} 
 \eqqcolon \phi^{(1)}_{b;c}(x) \theta^{bc}  
 \qquad \in (n,2) \subset \cC^1 \, .
\end{align}
There is a canonical vector field (or one-form) on $S^4$ associated to 
such a $\phi^{(1)}\in 
\cC^1$, with components given by
\begin{align} 
\phi^{(1)}_{c}(x) \coloneqq   \phi^{(1)}_{a_1 \ldots a_nb;c} x^{a_1} \ldots  x^{a_n} 
x^{b} \,. 
\label{A_c_divfree-Young}
\end{align}
We will denote this $\phi^{(1)}_c(x)$ as \emph{symbol} for $\phi^{(1)}\in 
\cC^1$.
It amounts to a one-form
\begin{align}
 \cA^{(0)} &\coloneqq   \phi^{(1)}_{c}(x) \xi^c \, ,
\end{align}
which is  tangential and divergence-free 
\begin{align}
 \phi^{(1)}_{c}  x^{c} &= 0 = \cN(\cA^{(0)})  \, , \nn\\
  \del^a \phi^{(1)}_a(x) &= 0 =  {\diverg} \cA^{(0)} \, .
\end{align}
Hence $\cC^1$ can be identified with divergence-free rank 
$1$ tensor fields on $S^4$, via 
\begin{alignat}{4}
\begin{aligned}
&\Psi: & \quad &\cC^1 \ &  &\to \ &  &T^* S^4  \\
 & &  \phi^{(1)} &= \phi^{(1)}_{b;c}(x)\, \theta^{b c} &
 \ &\mapsto  & \  &\phi^{(1)}_{c}(x)  \, .
 \end{aligned}
 \label{C1-VF-map}
\end{alignat}
We will see that $\phi^{(1)}_{c}(x) \propto [\{x_c,\phi^{(1)}\}]_0$ in 
\eqref{pure-gauge-1-decomp-0}, and 
the inverse of this map is given by
\begin{align}
 \{x^{c},\phi^{(1)}_c(x)\}
 &= -(n+1) \phi^{(1)} 
 \label{C1-VF-explicit}
\end{align}
restricted to  divergence-free  $\phi_a$. 
Hence $\cC^1$ can be identified with \emph{volume-preserving 
diffeomorphisms} on $S^4$.

\paragraph{$\cC^s$  as tensor fields on $S^4$.}
For $n\geq 0$ , $(n,2s)$ is the space of totally traceless  $SO(5)$ tensors
$\phi^{(s)}_{a_1 \ldots a_n b_1\ldots b_s;c_1\ldots  c_s}$ corresponding to
Young diagrams ${\tiny \Yvcentermath1 \young(bbaa,cc)}$
which are first  antisymmetrized in each pair $(b_ic_i)$, and then
symmetrized in $(a_1 \ldots  a_nb_1\ldots b_s)$ and $(c_1 \ldots  c_s)$.
Then define
\begin{align}
 \phi^{(s)} \coloneqq  \phi^{(s)}_{a_1 \ldots  a_n b_1\ldots b_s;c_1\ldots  c_s} 
x^{a_1} \ldots   x^{a_n} \theta^{b_1 c_1} \ldots    \theta^{b_s c_s}  
  \eqqcolon \phi^{(s)}_{b_1\ldots b_s;c_1 \ldots c_s}(x) \theta^{b_1 c_1} 
\ldots   \theta^{b_s c_s} 
  \quad \in \cC^s  .
  \label{phi-s-def}
\end{align}
 We associate\footnote{We will often drop the superscript ${}^{(s)}$ if no confusion can arise.} 
 to each such  $\phi^{(s)}$ a symmetric rank $s$ tensor field 
 on $S^4$ via
\begin{align}
 \phi^{(s)}_{c_1 \ldots  c_s}(x) \coloneqq   \phi^{(s)}_{a_1 \ldots  a_n 
b_1\ldots b_s;c_1\ldots  c_s} x^{a_1} \ldots   x^{a_n} x^{b_1} \ldots  x^{b_s}
 \label{A-tensor-general}
\end{align}
denoted as \emph{symbol} for $\phi^{(s)}\in \cC^s$.
These are traceless, tangential and divergence-free, 
\begin{align}
 \phi_{c_1 \ldots  c_s}(x)  x^{c_i} &= 0  \, ,\nn\\
 \phi_{c_1 \ldots  c_s}(x)  g^{c_1 c_2} &= 0 \, , \nn\\
 \del^{c_i} \phi_{c_1 \ldots  c_s}(x) &= 0 \, .
\end{align}
Hence, $\cC^s$ can be identified\footnote{It is important  that the 
identification \eqref{psi-iso}  only applies to those tensor fields which are 
obtained from irreducible Young diagrams as in \eqref{phi-s-def}.}  
with symmetric traceless divergence-free rank 
$s$ tensor fields on $S^4$,
\begin{align}
\boxed{ \
\begin{aligned}
\Psi:\quad
 \cC^s \  &\to \  T^{*\otimes_{\mathrm{Sym}} s} S^4  \\
  \phi^{(s)} = \phi^{(s)}_{b_1\ldots b_s;c_1 \ldots c_s}(x)\,  \theta^{b_1 c_1} \ldots    \theta^{b_s c_s} 
 \ &\mapsto  \  \phi^{(s)}_{c_1 \ldots  c_s}(x)  
 = \phi^{(s)}_{b_1\ldots b_s;c_1 \ldots c_s}(x) x^{b_1} \ldots x^{b_s} \, .
  \end{aligned} \
  }
 \label{psi-iso}
\end{align}
The $\phi^{(s)} \in \cC^s$ are potentials of the tensor fields, in the sense 
that
\begin{align}
 \phi_{a_1 \ldots  a_s}(x)  \propto [\{x_{a_1},\ldots \{x_{a_s},\phi^{(s)}\} 
\ldots \}]_{0} \,.
 \label{phi-s-tensor-map}
\end{align}
Note that the projection on $\cC^0$  entails symmetrization, as for instance
\begin{align}
 & \{x_{d_1},\{x_{d_2},\phi^{(2)}\} \} - \{x_{d_2},\{x_{d_1},\phi^{(2)}\} \} 
 = \{\theta_{d_1d_2},\phi^{(2)}\}  
\end{align}
is in $\cC^2$. Conversely, we have
\begin{align}
 \{x^{c_1},\ldots \{x^{c_s},\phi^{(s)}_{c_1 \ldots  c_s}(x)  \}\ldots .\} &= (-1)^s 
(n+s) \ldots  (n+1) \phi^{(s)} \, ,
\end{align}
because $\phi^{(s)}$ is defined in terms of traceless Young tensors. 
\eqref{psi-iso} constitutes the first important result of this paper.

\paragraph{Spin 2 identities.}
For $s=2$ and $\phi^{(2)} \in (n,4)$, we note the following identities:
\begin{align}
 \{x^c,\{x^{d},\phi^{(2)}_{dc}\}\} &= (n+2)(n+1) \phi^{(2)}   \, ,\nn\\
 \{x^{c},\phi^{(2)}_{c d}(x)  \} &= -(n+2) \phi^{(2)}_{a_1 \ldots  a_nb_1b_2;c 
d} x^{a_1} \ldots   x^{a_n}x^{b_1}\theta^{b_2c}  \, , \nn\\
 \{x^a,\{x_a,\phi^{(2)}\}\} &= -\theta ((n+2)(n+5)-6)\phi^{(2)} \, , \nn\\
 [\{x^a,\{x^b,\phi^{(2)}\}\}]_0 &= c_n \phi^{(2)}_{ab}  \, ,
 \label{potential-gaugefix-id-2} \\
c_n &= \frac{2}{15} \theta^2 \frac{(n+5)(n+4)(n+3)}{n+1} \ . \nn
\end{align}
The details of the computation of the constant $c_n$ are in Appendix 
\ref{sec:app-proofs}.
We further need the following integral identities\footnote{The integral here is 
simply the projector of $\cC$ to the unique trivial mode. More 
details will be given in Section \ref{sec:scalar-field}.}: 
\begin{align}
\int \{x^a,\phi^{(2)} _{ab}\} \{x^c, \phi^{(2)}_{cb}\}
  &=  \frac 13(n+3)(n+4)\theta \int \phi^{(2)}_{ab} \phi^{(2)}_{ab} 
 \ \approx  \frac 13\theta \int \phi^{(2)}_{ab} \Box \phi^{(2)}_{ab} \, , \nn\\
\int \phi^{(2)}  \phi^{(2)} &= 
   \frac{2}{15} \frac{(n+5)(n+4)(n+3)}{(n+1)^2(n+2)}  \theta^2 \int \phi^{(2)}_{ab}  \phi^{(2)}_{ab} \ .
  \label{spin2-deriv-contract-id}
\end{align}
Here, $\approx$ indicates statements valid for large $n$.
The second  line of \eqref{spin2-deriv-contract-id} is a consequence of 
\eqref{potential-gaugefix-id-2},  i.e.
\begin{align}
\begin{aligned}
 c_n \int \phi^{(2)}_{ab}  \phi^{(2)}_{ab} &=
 \int \phi^{(2)}_{ab} \{x^a,\{x^b,\phi^{(2)}\}\} =  \int 
\{x^a,\{x^b,\phi^{(2)}_{ab}\}\} \phi^{(2)} \\
  &= (n+2)(n+1) \int \phi^{(2)}  \phi^{(2)} \ .
  \end{aligned}
\end{align}
While the first line of \eqref{spin2-deriv-contract-id} is derived exactly in Appendix 
\ref{app:inner-prod}; it
can also be understood more intuitively using the following semi-classical 
leading-order computation:
\begin{align}
  \int \{x^a, \phi^{(2)}_{ab}\} \{x^c, \phi^{(2)}_{cb}\} 
   &= \int \theta^{a\mu}\theta^{c\nu}\del_\mu \phi^{(2)}_{ab}\del_\nu 
\phi^{(2)}_{cb} \nn\\
     &= \frac 13 \theta R^2\int (g^{ac} g^{\mu\nu} - g^{a\nu}g^{c\mu} + 
\varepsilon(x)) \del_\mu \phi^{(2)}_{ab}\del_\nu \phi^{(2)}_{cb} \nn\\
     &\approx \frac{1}{3} \theta R^2\int\del_\mu \phi^{(2)}_{ab}\del^\mu 
\phi^{ab} \  +  O\left(\frac{ 1}{R}\right)  \nn\\
     &\approx \frac 13 \theta R^2\int  \phi^{(2)}_{ab} \Box \phi^{(2)}_{ab} \; ,
\end{align}
because $\phi_{ab}$ is divergence-free and radius $R$ is assumed to be 
sufficiently large.
In the second line we used the averaging \eqref{eq:averaging}. 

\subsection{Relation to higher spin algebras}
\label{sec:relation_HS}
According to \eqref{phi-s-def}, the module $\cC$ is described by functions on 
$S^4$ taking values in the 
direct sum of all rectangular Young diagrams with 2 rows.
This provides the relation to the higher spin algebra of Vasiliev theory.
To see this, we resort  to the Lie algebra normalization
 $\cM^{ab} = \theta^{-1}\theta^{ab}$  ,
and consider the subspace  $\hs \subset \cC$ with basis 
\begin{align}
 \phi_{b_1\ldots b_s;c_1\ldots  c_s} \cM^{b_1 c_1} \ldots    \cM^{b_s c_s}  
 \qquad \in \ (0,2s) \subset  \cC^s, \ s=1,2,3,\ldots 
\end{align}
where the  $\phi_{b_1\ldots b_s;c_1\ldots  c_s}$ are constant, totally traceless and have
the symmetries of a rectangular
two-row Young diagram ${\tiny \Yvcentermath1 \young(bbb,ccc)}\ $.
Thus as a vector space,
\begin{align}
 \oplus\ {\tiny \Yvcentermath1 \yng(3,3)} \  \cong  \ \hs :=  
\oplus_{s=1}^\infty\, (0,2s) \, .
 \label{HS-def}
\end{align}
Moreover, $\hs$ inherits a bracket from the Poisson structure
\eqref{Poisson-theta} on $\cC$, given by
\begin{subequations}
\label{Poisson-cM}
\begin{align}
\{\cM^{ab},\cM^{cd}\} &= g^{ac}\cM^{bd}  - 
g^{ad}\cM^{bc} - g^{bc}\cM^{ad} + g^{bd}\cM^{ac}  \;, \\
  \{\cM^{ab}, \phi_{b_1\ldots b_s;c_1\ldots  c_s} \cM^{b_1 c_1} \ldots    
 \cM^{b_s c_s}\}
  & = s \  g^{ab_1} \phi_{b_1\ldots b_s;c_1\ldots  c_s} \cM^{b c_1} 
\ldots    \cM^{b_s c_s} \ \pm \ldots 
\end{align}
\end{subequations}
which has the same form as in Vasiliev theory \cite{Joung:2014qya}.
This can be truncated to $\mso(5)$ at $s=1$, but not at any other finite $s$.
Due to the tangential projector in \eqref{XX-R-const}, 
$\hs$ does not close as (Poisson) algebra, which comes to no surprise given the 
origin of an $SO(5)$ covariant bundle, also known as Penrose twistor fibration.
Nonetheless, as vector spaces $\hs$ and the higher spin algebra 
$\widetilde{\hs}$ in Vasiliev theory coincide due to \eqref{HS-def}.
By construction, the latter arises from relations 
imposed on the vector space of rectangular traceless Young diagrams, i.e.\  
$\widetilde{\hs} = 
U(\mso(5))\slash \mathfrak{J}$  were $\mathfrak{J}$ is the Joseph ideal (see 
\cite{Joung:2014qya} for a review).

To make the relation to the Joseph ideal manifest, recall that
the definition of the fuzzy 4-sphere  entails similar relations 
\eqref{XX-R-const}
\begin{align}
 \cM^{ab}\cM^{ac} = \frac{R^2}{\theta} P_T^{bc} , \qquad \varepsilon_{abcde} 
\cM^{ab}\cM^{cd} = 8\frac{R}{\theta}  x^e \; ,
 \label{theta-relation-HS}
\end{align}
but besides the $\mso(5)$ generators $\cM^{ab}$ also the coordinate functions 
$x^a$ on $S^4$ are involved.
Now consider the quotient algebra $\mathfrak{chs}\coloneqq \cC \slash \cC^0$,
where $\cC \equiv \cC^{\infty} (\C P^3)$, $\cC^0 \equiv 
\cC^{\infty} (S^4)$ from \eqref{full-functions-decomp}.
Then the relations \eqref{theta-relation-HS} on $\cC$ 
imply the following relations on $\mathfrak{chs}$
\begin{align}
 \cM^{ab}\cM^{ac} = \frac{R^2}{\theta} g^{bc} , \qquad \varepsilon_{abcde} 
\cM^{ab}\cM^{cd} = 0 \; 
 \label{theta-relation-HS-quotient}
\end{align}
recovering the commutative limit of $\mathfrak{J}$.
However, the quotient does not respect the Poisson brackets of $\hs$.
Equivalently, one can define $\mathfrak{chs} \cong
\C[ \{ \cM^{ab}|a,b=1,\ldots,5\}]\slash \langle 
\eqref{theta-relation-HS-quotient} \rangle $ as a commutative quotient algebra. 
Consequently, $\mathfrak{chs}$ can
be understood as the Euclidean commutative or semi-classical vector space 
analog of the 
(Euclidean) higher spin Lie algebra of Vasiliev theory \cite{Vasiliev:2004qz}.
Thus locally, i.e.\ ''forgetting`` the functions on $S^4$, $\hs$ 
coincides with the  conventional $\widetilde{\hs}$, 
but not globally.
The  difference is tied to the presence of a scale $L_{\rm NC}$. 
For finite $N$, the fuzzy case yields a truncation of our  $\hs$.

A somewhat related approach in (A)dS signature has been elaborated in 
\cite{Iazeolla:2008ix}, based on a Lorentz-covariant slicing, where the role of 
our $X$ is taken over by ''momenta`` $P$.

Most importantly, we  obtain a geometrical interpretation of  $\hs$ or $\mathfrak{chs}$
as space of functions on $\C P^3$ which are constant along $S^4$.
More generally, $\hs$-valued functions 
on $S^4$ are  identified with the space of \emph{all} functions
on $\C P^3$,
\begin{align}
\cC \ni \
\phi = \phi_{\und{\a}}(x)\, \Xi^{\und{\a}}, \qquad 
 \Xi^{\und{\a}} =  \cM \ldots \cM \quad \in  \hs \, .
\end{align}
This  leads to $\hs$-valued 
gauge fields on $S^4$, as elaborated below.
Hence, the semi-classical  $\cS^4$ provides a natural realization of this higher 
spin algebra
(and associated gauge theories as we will see).
The $\theta^{ab}$ generators arise as functions on the $S^2$ fiber over $S^4$, and
$\hs$ is an $SO(5)$-invariant truncation of $\cC$.
However in the formulation of gauge theory discussed below, these  $\theta^{ab}$  also act on the $x^a$, unlike in 
Vasiliev's approach where classical space-time is added by hand.
In the fuzzy case, these  $\hs$ generators are inseparably linked to the 
$x^a$, because the analog of $P_T^{ab}$ of \eqref{theta-relation-HS}  becomes 
non-commutative.

\subsection{Local representation and constraints}
\label{sec:local-rep-scalar}
We have seen that  functions on $\cS^4$ can be viewed as functions on ordinary $S^4$
taking values in $\hs$. To understand better the meaning of this statement,
we decompose $\phi^{(s)}$ into ordinary tensor fields on $S^4$:
Fixing an arbitrary point $p\in S^4$  (e.g.\ the ``north pole'' 
$p=R(0,0,0,0,1)$), we can decompose $\mso(5)$ into translation generators 
$P^\mu$, see \eqref{P-def}, and rotation generators $\cM^{\mu\nu}$ for 
$\mu,\nu=1,\ldots,4$. This leads to an expansion of a spin $s$ mode of the form
 \begin{align*}
  \phi_{b_1\ldots b_s;c_1\ldots  c_s}(x) \theta^{b_1 c_1} \ldots \theta^{b_s 
c_s}\big|_p \equiv 
  \sum_{k=0}^{s} f_{\rho_1 \ldots\rho_k; \mu_1 \nu_1\ldots \mu_{s-k} \nu_{s-k} 
}(x) P^{\rho_1} \ldots P^{\rho_k} \cM^{\mu_1 \nu_1} \ldots \cM^{\mu_{s-k} 
\nu_{s-k}}
 \end{align*}
similar to \cite{Steinacker:2016vgf}.
The main consequence of the above classification of modes is that the 
coefficients $f_{\rho_1\ldots; \mu_1 \nu_1\ldots} $ 
of these generators are not independent; for example for $s=1$, both are 
determined by a single underlying divergence-free vector field.
This has important consequences for the resulting physics and we therefore 
elaborate these constraints explicitly.

\subsubsection{\texorpdfstring{$s=1$}{s=1} and field strength} 
\label{s1-field-strength}
With the above conventions, we can write
\begin{align}
 \phi^{(1)} =  \phi_{a_1 \ldots  a_nb; c} x^{a_1} \ldots   x^{a_n} \theta^{b c} 
  \eqqcolon A_{\mu}(x) P^{\mu} + \omega_{\mu\nu}(x) \cM^{\mu\nu} \qquad \in 
(n,2)\subset \cC^1 .
  \label{phi-1-local}
\end{align}
Here $\omega_{\mu\nu}(x)$ is
naturally defined to be antisymmetric (unlike the underlying  $\phi_{a_1 \ldots  
a_nb; c}$).
Carefully comparing the coefficients of $P^\mu$ using 
\eqref{F-x-xi-id}, we obtain\footnote{Note that one cannot simply read off 
$\omega_{\mu\nu}$ by comparing coefficients 
in \eqref{phi-1-local}, since  $\phi_{\ldots .b;c}$  is not anti-symmetric in 
$(bc)$, and $\cM_{\mu\nu}$ is self-dual.
However, the coefficient of $P^\mu$ is uniquely specified. 
Although $P^{\mu}$ vanishes as
function at $p\in S^4$, it does not vanish as generator in the Poisson algebra.} 
\begin{align}
 A_\mu(x) = -\frac{n+2}{n+1}\theta\phi_{a_1 \ldots  a_nb;\mu} x^{a_1} \ldots   
x^{a_n} x^b 
 = -\frac{n+2}{n+1}\theta\, \phi^{(1)}_{\mu}(x) 
 \quad  \text{with}  \quad
 \del^\mu A_\mu = 0, \ 
\end{align}
is nothing but the canonical tangential 
divergence-free vector field \eqref{A_c_divfree-Young}
associated to $\phi^{(1)}$. 
Since this vector field uniquely determines $\phi^{(1)}$, it must also determine
the tangential components $\omega_{\mu\nu}(x)$.
Indeed, we find
\begin{align}
 \del_\mu A_\nu &= -(n+2) \theta\phi_{a_1 \ldots  a_{n}\mu; \nu} x^{a_1} \ldots  
 x^{a_{n}} \, .
\end{align}
By further contracting $\del_\mu A_\nu$ with $\cM^{\mu\nu}$ and comparing to 
\eqref{phi-1-local}, we conclude
\begin{align}
 \omega_{\mu\nu} = -\frac{1}{2(n+2)}(\del_\mu A_\nu - \del_\nu A_\mu) ,
\end{align}
i.e.\ $\omega_{\mu\nu}$ is the field strength associated to the one-form 
$A_\mu$.
Conversely, it follows that\footnote{Here $\del^\mu$ amounts to the Levi-Civita connection on $S^4$.}
\begin{align}
 \del^\mu \omega_{\mu\nu} =-\frac{1}{2(n+2)}\del^\mu  (\del_\mu A_\nu - \del_\nu A_\mu) 
  = - \frac{(n+1)(n+4)}{2(n+2)}\ A_\nu \, ,
\end{align}
because $A_\nu$ is divergence-free, and 
$\del\cdot\del A_\nu = (n+1)(n+4)  A_\nu$.
Thus, $\phi^{(1)}$ encodes a multiplet consisting of a divergence-free vector 
field and its field strength.

\subsubsection{\texorpdfstring{$s=2$}{s=2} and curvature}
\label{s2-connect-curvature}
Similarly, for a mode $\phi^{(2)}$ the decomposition around the 
north pole yields
\begin{align}
 \phi^{(2)} &=  \phi_{a_1 \ldots  a_n bc; de} x^{a_1} \ldots   x^{a_n} \theta^{b 
d} \theta^{c e}
  \eqqcolon h_{\mu\nu}(x) P^{\mu}P^\nu + \omega_{\mu:\a\b}(x) P^\mu \cM^{\a\b}
  + \Omega_{\a\b;\mu\nu}(x) \cM^{\a\b}\cM^{\mu\nu} \nn\\
  &\in (n,4)\subset \cC^4 .
  \label{phi-2-local}
\end{align}
Here $\Omega_{\a\b;\mu\nu}(x)$ is naturally antisymmetric in $(\a\b)$ and $(\mu\nu)$ separately
(in contrast to the underlying $\phi_{}$). 
Carefully comparing coefficients of $P^{\mu}P^\nu$ at the north pole using 
\eqref{marcus-id}, we obtain
\begin{align}
 h_{\a\b} = \frac{n+3}{n+1}\theta^2 \phi_{a_1 \ldots  a_n bc; \a\b} x^{a_1} 
\ldots   x^{a_n}x^b x^c
 =  \theta^2 \phi^{(2)}_{\a\b}(x) 
 \quad  \text{with}  \quad
 \del^\mu h_{\mu\nu} = 0 \, ,
\end{align}
which is proportional to the symmetric rank 2 tensor $ \phi^{(2)}_{\mu\nu}(x)$ 
of \eqref{A-tensor-general} associated to $\phi^{(2)}$.
Since this tensor field uniquely determines $\phi^{(2)}$, it  also determines
$\omega_{\mu;\a\b}$ and $\Omega_{\a\b;\mu\nu}$. 
Indeed, consider derivatives of the above
\begin{align}
 \del_{\mu} h_{\a\b} &= \frac{(n+2)(n+3)}{n+1} \theta^2\phi_{a_1 \ldots  a_n 
c\mu;\a \b} x^{a_1} \ldots  x^{a_n} x^c  \, ,\nn\\
 \del_{\mu}\del_\nu h_{\a\b} &= (n+2)(n+3)\theta^2 \phi_{a_1 \ldots  
a_n\mu\nu;\a\b} x^{a_1} \ldots  x^{a_n}  \,.
\end{align}
Contracting these expressions with $\cM^{\mu\nu}$ and/or $P^\mu$ and comparing 
with \eqref{phi-2-local}, we conclude
\begin{align}
 -\omega_{\mu:\a\b}(x) P^\mu \cM^{\a\b} 
    &= 2 \theta^2\phi_{a_1 \ldots  a_n c\mu;\a \b} x^{a_1} \ldots  x^{a_n} x^c  
P^\a \cM^{\mu\b} 
     =  2\frac{n+1}{(n+2)(n+3)}\del_{\mu} h_{\a\b} P^\a \cM^{\mu\b}  \nn\\
 \omega_{\mu;\a\b} &= -\frac{n+1}{(n+2)(n+3)}(\del_\a h_{\mu\b} - \del_\b h_{\mu\a}) \ .
 \end{align}
 Hence, up to a factor, $\omega_{\mu;\a\b}$ is the spin connection defined by 
$h_{\a\b}$.
 Similarly,
 \begin{align}
 (n+2)(n+3) \Omega_{\mu\nu;\a\b}(x) \cM^{\mu\nu}\cM^{\a\b} &= 
  (n+2)(n+3)\theta^2\phi_{a_1 \ldots  a_n \mu\nu;\a\b} x^{a_1} \ldots   x^{a_n} 
\cM^{\mu \a} \cM^{\nu \b}  \nn\\
    &= \del_{\mu}\del_\nu h_{\a\b}\cM^{\mu \a} \cM^{\nu \b}  .
  \label{Omega-delh}
\end{align}
To understand the RHS, consider the linearized Riemann tensor associated to 
$h_{\mu\nu}$
\begin{align}
 {\cal R}_{\a \b \mu \nu} 
 &= \frac{1}{2}(\del_{\a\nu} h_{\b\mu}+\del_{\b\mu} h_{\a\nu} -\del_{\a\mu} 
h_{\b\nu}-\del_{\b\nu} h_{\a\mu}) \, . 
\end{align}
Contracting with $\cM^{\a\b} \cM^{\mu \nu}$
gives 
\begin{align}
 {\cal R}_{\a \b \mu \nu} \cM^{\a\b} \cM^{\mu \nu} 
 &= \frac{1}{2}(\del_{\a\nu} h_{\b\mu}+\del_{\b\mu} h_{\a\nu} -\del_{\a\mu} h_{\b\nu}-\del_{\b\nu} h_{\a\mu}) 
  \cM^{\a\b} \cM^{\mu \nu} \nn\\
   &= \del_{\a\nu} h_{\b\mu} \cM^{\a\b} \cM^{\mu \nu}
   = -  \del_{\mu}\del_\nu h_{\a\b}\cM^{\mu \a} \cM^{\nu \b}  \ .
\end{align}
Comparing with \eqref{Omega-delh}, we conclude
\begin{align}
 \Omega_{\a\b;\mu\nu}(x) = - \frac{1}{(n+2)(n+3)}{\cal R}_{\a \b \mu \nu} 
\end{align}
Thus, $\phi^{(2)}$ encodes a multiplet consisting of a divergence-free symmetric 
traceless tensor field (graviton),
its spin connection, and the (linearized) curvature tensor.
Similar relations hold between the tangential and radial components 
of general $\phi^{(s)}$.

\section{Vector harmonics on \texorpdfstring{$\cS^4$}{SS4} and higher spin 
fields on \texorpdfstring{$S^4$}{S4}}
\label{sec:vec-harmonics}
In this section, we derive the complete classification of one-forms (i.e.\ 
vector modes) on $\cS^4$. These are the basic degrees of freedom 
which arise in the semi-classical limit of matrix models on $S^4_N$.
We will first obtain the abstract classification from $\mso(5)$ representation 
theory, which for generic $s$ leads to five modes for each spin $s$.
In a second step, we will provide an explicit realization of these modes in 
terms of five ``Ans\"atze'' involving $\mso(5)$ tensors and  Young diagrams.
By elaborating their properties and comparing with the group-theoretical 
results, 
we show that this provides the complete set of modes.
This explicit realization will be the basis for the further analysis.

The tensor product decomposition of $\Omega^1 \cS^4$ is given by \cite{Steinacker:2015dra}
\begin{align}
 \cA = \xi^a  \cA_a \ &\in \ (1,0) \otimes (n,2s) \nn\\
  &= (n+1,2s) \oplus  (n-1,2s+2)   \oplus  (n,2s) \oplus  (n+1,2s-2) \oplus  (n-1,2s)  
  \label{mode-decomp-2}
\end{align}
for generic $(n,2s)$. For $s=0$ the decomposition truncates as follows:
\begin{align}
(1,0)\otimes (n,0)  &= (n+1,0) \oplus (n-1,2) \oplus (n-1,0)  , \qquad n \geq 1 
\, ,\nn\\
(1,0)\otimes (0,0)  &=(1,0) \, ,
 \label{mode-decomp-m=0}
\end{align}
and for $n=0$
\begin{align}
(1,0)\otimes (0,2s)  &= (1,2s) \oplus(0,2s)    \oplus  (1,2s-2),   \qquad s \geq 1 \ .
\end{align}
The irreducible components are characterized by their eigenvalues of the 
intertwiner $\cI$ of \eqref{intertwiners}, which commutes\footnote{because 
$\{\theta^{ab},\cdot \}$ is the adjoint action of a $\mso(5)$ generator on 
$\cC$, which commutes with the Casimir $\Box$.} with
the Laplacian $\Box$ defined in \eqref{Poisson-Laplacian}. This fact can be 
seen by expressing $\cI$ as follows:
\begin{align}
 -\theta (M_{cd}^{(\ad)} \otimes M_{cd}^{(5)}\cA)_a =
 -\big(M^{(5)}_{cd}\big)^a_b \{\theta^{cd},.\}  \cA_b 
 = 2\{\theta_{ab},\cA_b\} = 2\, \cI(\cA)_a \, .
 \label{I-MM-id}
\end{align}
Here 
\begin{align}
(M_{ab}^{(5)})^c_d &= \d^c_b \d_{ad} - \d^c_a \d_{bd}\, 
\label{M-rep-5}
\end{align}
is the  vector generator of $\mso(5)$, 
and $M_{bc}^{(\ad)} = \{\cM_{bc},\cdot \}$ denotes the representation of 
$\mso(5)$ induced by the Poisson structure on $\cS^4$
(cf.\ \eqref{Poisson-theta}).
Therefore $\cI$ measures the product of internal and space-time (angular) 
momentum, analogous to the spin-orbit coupling of the modes.
Now, we find
\begin{align}
 -M_{bc}^{(\ad)} \otimes M_{bc}^{(5)}
  &= -C^2[\mso(5)]^{(5)\otimes(\ad)} + C^2[\mso(5)]^{(\ad)} + 
C^2[\mso(5)]^{(5)} 
  \label{intertwiner}
\end{align}
i.e.\ $\cI$ is the difference between the total Casimir and the orbital and spin 
Casimirs.
This yields \cite{Steinacker:2015dra}
\begin{align}
\cI(\cA^{(n+1,2s)})  &= \theta(n+s) \cA^{(n+1,2s)} \; ,\nn\\
\cI(\cA^{(n-1,2s+2)})  &= \theta(s -1) \cA^{(n-1,2s+2)} \; ,\nn\\
\cI(\cA^{(n,2s)}) &=  - 2 \theta\cA^{(n,2s)} \, , \nn\\
\cI(\cA^{(n+1,2s-2)})  &= -\theta(s+2)\cA^{(n+1,2s-2)} \, ,\nn\\
\cI(\cA^{(n-1,2s)})  &= -\theta((n+s)+3)\cA^{(n-1,2s)} \, ,
  \label{vector-term-eom}
\end{align}
for $\cI$ acting on $\cA\in(n,2s)\otimes (1,0)$. To identify these  modes 
explicitly, we will define five intertwiners
\begin{alignat}{2}
 \cA^{(i)}[\cdot]: \quad \cC^s  \ &\cong  \quad \oplus \,
 {\footnotesize \Yvcentermath1 \young(1{\ldots}{.}{\ldots}*,1{\ldots}s) } &
\qquad\quad &\to \quad 
(*,2s)_i \subset (1,0)\otimes \cC, \qquad i=0,1,2,3,R  \nn\\
      \phi^{(s)} &\mapsto \quad  \phi^{(s)}_{a_1\ldots  a_*;c_1\ldots c_s} & 
&\mapsto \ \ \cA^{(i)}[\phi^{(s)}]
 \label{A-i-funktor}
\end{alignat}
for each $s=0,1,2,\ldots $ (except for $s=0$ where only $i=2,3,R$ arise).
These five modes $\cA^{(i)}$ provide a one-to-one realization of all vector 
fluctuations \eqref{vector-term-eom} on $\cS^4$.
This will be established by  diagonalizing $\cI$,
which will recover precisely the above eigenvalues.
In the following paragraphs we discuss separately the cases $s=0,1,2$ and higher 
$s$.

\paragraph{Notation.}
As noted at the beginning of this section, we refer to one-forms on $\cS^4$ 
also as \emph{vector fields} because they are tangential fluctuations around a 
background and are the degrees of freedom in the matrix model. The 
complementary group theoretical content of the one-forms/vector fields is 
captured by their spin. Hence, there are tangential fluctuations of arbitrary 
spin $s$, which are referred to as spin $s$ vector fields.

\subsection{Spin 0 vector fields \texorpdfstring{$\cA$}{A}}
First consider the three spin 0 vector modes $\cA$. They correspond to a Young 
diagram with one line, or equivalently $(n,0)$.

There are three such vector modes for each $n$, two from $(n+1,0) \oplus (n-1,0) 
\subset (n,0) \otimes (1,0)$  and one from $(n+1,0) \oplus (n,0) \subset (n-1,2) 
\otimes (1,0)$. Explicitly, they are given as follows:
\begin{subequations}
\label{spin-0-A-modes}
\begin{align}
\cA^{(2)} &\coloneqq \xi_c \cA_{c}^{(2)} = \xi_c  \phi_{c a_2 \ldots a_n}^{(0)} 
x^{a_2} \ldots  x^{a_n}  \ \ \in \ (n+1,0) \subset (n,0)\otimes (1,0) \, ,
\\
\cA^{(3)} &\coloneqq \cJ(A^{(2)}) = \xi_d \theta^{dc}\phi^{(0)}_{c a_2 \ldots 
a_n} x^{a_2} \ldots  x^{a_n}  \qquad \in (n+1,0) \subset 
(n,2)\otimes (1,0) \,, \\
\cA^{(R)} &\coloneqq \xi \phi = \xi_a x^a \phi^{(0)}  \qquad \in (n,0) \subset 
(n+1,0)\otimes (1,0) \, .
\end{align}
\end{subequations}
The notation will become clear later when considering spin $s$ fields in 
\eqref{eq:modes_spin-s} and their transformation into $\cI$-eigenmodes 
\eqref{eq:EW_cI_spin-s_pt1} and \eqref{eq:EW_cI_spin-s_pt2}.
\subsubsection{Properties of spin 0 fields}
Consider first the  vector field $\cA^{(2)}$, which satisfies
\begin{align}
 \cN(\cA^{(2)}) &=x^c \phi^{(0)}_{c a_2 \ldots a_n}  x^{a_2} \ldots  x^{a_n}  
= \phi^{(0)} \, ,
\label{eq:cA_not_tangential} \\
\cG(\cA^{(2)})&=
\{x^a,\cA^{(2)}_a\} 
= 0 \, , \\
 \cI(\cA^{(2)}) &= \xi_a\{\theta^{ac},\cA_c^{(2)}\}
\nn\\ 
 &= (n-1)  \xi_a \phi^{(0)}_{c a_2 \ldots a_n }x^{a_3} \cdots  x^{a_n} 
\{\theta^{ac}, x^{a_2}\}  \nn\\ 
 &= (n-1) \theta \cA^{(2)} \, ,
 \label{I-VF-longitudinal}
\end{align}
using \eqref{F-JQ-id} and tracelessness of $\phi^{(0)}$. 
The last relation implies that $\cA^{(2)}$ is an example of the 
first line in \eqref{vector-term-eom} for the case $s=0$. 
We observe from \eqref{eq:cA_not_tangential} that $\cA^{(2)}$ is not 
tangential, but its tangential projection can be straight forwardly worked 
out to read 
\begin{align}
 \cP_T \cA^{(2)} = \frac 1n \del \phi^{(0)} \, .
 \label{eq:tang_proj_A0_spin0}
\end{align}
Hence $\cP_T \cA^{(2)}$ is essentially the differential of a function on $S^4$.
Next, $\cA^{(3)}$ satisfies
\begin{align}
\cN(\cA^{(3)}) &= 0 \, , \nn\\
\cG(\cA^{(3)})&=
\{x^a, \cA^{(3)}_a\} 
= 0 \, ,\nn\\
  \cI(\cA^{(3)}) &= \cI(\cJ(\cA^{(2)})) = - 4\theta \cJ(\cA^{(2)}) + \theta 
\cQ(\cN(\cA^{(2)}))  -  \cJ\circ \cI (\cA^{(2)})  \nn\\
   &= -3\theta \cA^{(3)} \, ,
\end{align}
using \eqref{I-J-identity} and \eqref{pure-gauge-longitudinal} for the last 
relation, where
we also need the following expression for the spin 0 pure gauge modes 
\begin{align}
 \cQ(\phi^{(0)}) &=  \cQ(\phi^{(0)}_{a_1 \ldots  a_n} x^{a_1} \ldots   x^{a_n} )
   = n \cA^{(3)} \; .
   \label{pure-gauge-longitudinal}
\end{align}
Hence, one recognizes $\cA^{(3)}$ as an example of the fourth 
line in \eqref{vector-term-eom} for the case $s=1$, and, most notably, 
$\cA^{(3)}$ equals a gauge transformation generated by $\tfrac{1}{n}\phi^{(0)}$.
Finally, for the radial mode $ \cA^{(R)}$ we find
\begin{align}
 \cI(\cA^{(R)}) &= - 4 \theta\xi \phi^{(0)} - \cJ(\cQ(\phi^{(0)})) \nn\\
   &= - (4 +n)\theta \cA^{(R)} + n \theta R^2 \cA^{(2)} \, ,
\end{align}
using \eqref{I-N-identity}. 
Thus, we have explicitly realized  all three $(n,0)$ vector modes  $\cA$
in terms of irreducible Young tableaux or tensors with one line. 
In particular, $\cA^{(3)}$ is recognized as pure gauge field 
in noncommutative $U(1)$ Yang-Mills gauge theory\footnote{In the semi-classical 
limit the $U(1)$ gauge field  is a Maxwell field, but it becomes 
necessarily Yang-Mills in the non-commutative setting.}, 
and $\cA^{(2)}$ is completely determined by a function and its differential. 
\subsubsection{Diagonalization of \texorpdfstring{$\cI$}{I}}
Collecting the results of the $\cI$ action on the modes $\cA^{(R)}$, $R^2 
\cA^{(2)}$, and $\cA^{(3)} $ we find\footnote{The clumsy-looking organization 
will become more transparent once we proceed to higher spin fields.}
\begin{subequations}
\begin{align}
 \cI \begin{pmatrix}
      \cA^{(R)} \\ R^2 \cA^{(2)} 
     \end{pmatrix}
&= \theta 
\begin{pmatrix}
  -(n+4) & n \\
  0 & (n-1)  
\end{pmatrix}
\begin{pmatrix}
       \cA^{(R)} \\ R^2 \cA^{(2)}
     \end{pmatrix}  \; ,\\
 \cI (\cA^{(3)})
&= -3 \theta \ \cA^{(3)}  \; .
\end{align}
\end{subequations}
The eigenvalues of $\cI$ are $(n-1)$, $-(n+4)$, and  $-3$ with corresponding 
eigenmodes given by
\begin{subequations}
\label{eq:cI_eigenmodes_s=0}
\begin{align}
 C[\phi^{(0)}] &\coloneqq \cA^{(R)} - \frac{n}{2n+3}\ (R^2 \cA^{(2)}) \; ,\\
D[\phi^{(0)}] &\coloneqq  R^2 \cA^{(2)} \; ,\\
F[\phi^{(0)}] &\coloneqq \cA^{(3)} \; .
\end{align}
\end{subequations}
We observe that $C$ exemplifies the fifth line in \eqref{vector-term-eom} for 
the case of $s=0$.
\subsection{Spin 1 vector fields \texorpdfstring{$\cA$}{A}}
Now we account for all five $s=1$ modes $\cA$ in $(n,2)$ in a systematic 
fashion. They are determined in different ways in terms of mixed Young diagrams 
or corresponding tensors $\phi_{a_1 \ldots  a_nb;c}$.
We have the following spin 1 vector modes:
\begin{align}
\label{spin-1-modes}
\begin{aligned}
  \cA^{(0)} &\coloneqq \xi_c A_{c}^{(0)} =  \phi^{(1)}_{a_1 \ldots a_nb;c}  
\xi_c x^{a_1} \ldots  x^{a_n} x^{b}
  \quad \in \ (n,2) \subset (n+1,0)\otimes (1,0)  \, ,\\
  \cA^{(1)} &\coloneqq \cJ(A^{(0)}) = \xi_d \phi^{(1)}_{a_1 \ldots a_nb;c} 
\theta^{dc} x^{a_1} \ldots  x^{a_n} x^{b} 
 \quad \in \ (n,2) \subset (*,2)\otimes (1,0) \,, \\
  \cA^{(2)} &\coloneqq \xi^a  \phi^{(1)}_{aa_2 \ldots a_nb;c} x^{a_2} \ldots  
x^{a_n} \cM^{bc}
 \quad \in (n,2) \subset (*,2)\otimes (1,0) \, , \\
 \cA^{(3)} &\coloneqq \cJ(\cA^{(2)}) = \xi^d \theta^{da}  \phi^{(1)}_{aa_2 
\ldots a_nb;c} x^{a_2} \ldots  x^{a_n} \cM^{bc} 
  \quad \in (n,2) \subset (*,4)\otimes (1,0) \, ,  \\
  \cA^{(R)} &\coloneqq \frac{\xi \phi^{(1)}}{\theta} =  \xi_d x^d  
\phi^{(1)}_{a_1 \ldots 
a_nb;c}  x^{a_1} \ldots  x^{a_n}\cM^{bc} \quad \in (n,2) \subset 
(*,2)\otimes (1,0) \, .
\end{aligned}
\end{align}
\subsubsection{Properties of spin 1 fields}
Start with the divergence-free tangential vector field $\cA^{(0)}$,  see also
\eqref{A_c_divfree-Young}.
We can identify this with the second line in \eqref{vector-term-eom} for $s=0$, 
because
\begin{align}
\cN(\cA^{(0)}) &= 0 \, ,\\
\cG(\cA^{(0)})
&= -  (n+1) \phi^{(1)} \, , \\
 {\diverg} \cA^{(0)}  &= 0 \, ,
 \label{A0-div-free} \\
 \cI(\cA^{(0)}) &=  \phi^{(1)}_{a_1 \ldots a_nb;c} \xi_h\{\theta^{hc}, x^{a_1} 
\ldots x^{a_n} x^{b}\} \nn\\
   &= \theta (n+1) \xi_h g^{ha_1}  \phi^{(1)}_{a_1 \ldots a_nb;c}x^{a_2} \ldots 
x^{a_n} x^{b}  x^c \nn\\ 
     &=  - \theta  \cA^{(0)} \, ,
\end{align}
using \eqref{C1-VF-explicit} and \eqref{F-x-xi-id}.
Furthermore, we find that $\cA^{(1)}$ satisfies
\begin{align}
 \cN(\cA^{(1)}) &=0 \, ,\\
 \cG(\cA^{(1)}) 
 &= \phi^{(1)}_{a_1 \ldots a_nb;c} \{x^d,  
\theta^{dc} 
x^{a_1} \ldots  x^{a_n} x^{b} \} \nn\\
 &= \phi^{(1)}_{a_1 \ldots  a_nb;c}\Big((n+1) \theta R^2 x^{a_1} \ldots   
x^{a_n} 
\cP_T^{bc} \Big) 
  = 0 \, , \\
 \cI(\cA^{(1)}) &=  \cI(\cJ(\cA^{(0)})) 
  = - 4 \theta\cJ(\cA^{(0)}) +\theta \xi \{ x^c,A_c^{(0)}\} +\theta \cQ(\cN(\cA^{(0)})) 
  -  \cJ\circ \cI (\cA^{(0)})  \nn\\
  &=  - 3 \theta \cA^{(1)} - (n+1)\theta \xi \phi^{(1)} \,,
\end{align}
using \eqref{I-J-identity}. Thus,  $\cA^{(1)}$ is not an eigenvector of $\cI$.
For the mode $ \cA^{(2)}$, we obtain 
\begin{align}
\cN(\cA^{(2)}) &= \frac{\phi^{(1)}}{\theta}  \, ,\\
\cG(\cA^{(2)})
&=  \phi^{(1)}_{aa_2 \ldots  a_nb;c}\{x^a,  x^{a_2} \ldots   
x^{a_n} \cM^{bc}\} 
    = 0 \, ,\\
  \cI(\cA^{(2)}) &= \xi^d  \phi^{(1)}_{aa_2 \ldots a_nb;c} \{\theta^{da}, 
x^{a_2} 
\ldots  x^{a_n} \cM^{bc} \} \nn\\
   &= (n-1)\theta\cA^{(2)} + \theta\xi^a \phi^{(1)}_{aa_2 \ldots a_nb;c} 
x^{a_2} \ldots  x^{a_n} \cM^{bc} \nn\\
   &= n \theta \cA^{(2)} \, .
   \label{I-A2}
\end{align}
This is consistent with the first line in \eqref{vector-term-eom}.
Next, the vector mode $\cA^{(3)}$ satisfies 
\begin{align}
\cG(\cA^{(3)})
 &=  \{x_d,\theta^{da}  \phi^{(1)}_{aa_2 \ldots a_nb;c} 
x^{a_2} 
\ldots  x^{a_n} \cM^{bc} \}  \nn\\
  &= -(n+4) \phi^{(1)} \, ,\\
\cI(\cA^{(3)})= \cI(\cJ(\cA^{(2)}))  &= - 4 \theta\cJ(\cA^{(2)}) +\theta \xi 
\{x^c, 
A^{(2)}_c\} 
 +\theta \cQ(\cN(\cA^{(2)})) 
 -  \cJ\circ \cI (\cA^{(2)}) \nn\\
  &=  - 4 \theta\cJ(\cA^{(2)})  + \cQ( \phi^{(1)}) -  
n\theta\cJ(\cA^{(2)}) \nn\\
  &=  - (4+n) \theta\cJ(\cA^{(2)})  + \cQ(\phi^{(1)}) \nn\\
   &=  - 4 \theta\cA^{(3)}  + \frac{n+2}{n+1}\theta\cA^{(0)} \, ,
\end{align}
using \eqref{I-J-identity}.
Finally, for the radial mode $\cA^{(R)}$ we compute
\begin{align}
\cG(\cA^{(R)})
 &=  x_a  \{x^a,\phi^{(1)}\} = 0 \, ,\\
 \cI(\cA^{(R)} ) &=  - 4 \xi \phi^{(1)} 
-\frac{1}{\theta} \cJ(\cQ(\phi^{(1)}))  \nn\\
   &= -(4 +n)\xi \phi^{(1)} + n\theta R^2\cA^{(2)} - \frac{n+2}{n+1} 
\cA^{(1)} \, ,
\end{align}
using \eqref{I-N-identity}, \eqref{I-A2}, and $\cA^{(3)} = \cJ \cA^{(2)}$.
Again $\cI$ does not diagonalize on the radial mode, but decomposes into radial 
and tangential components.

\subsubsection{Diagonalization of \texorpdfstring{$\cI$}{I}}
Collecting the results, $\cI$ acts on the modes $\cA^{(1)}\slash \theta$, 
$\cA^{(R)}$, and $R^2 \cA^{(2)}$ as follows:
\begin{align}
 \cI \begin{pmatrix} \cA^{(1)}\slash \theta \\ \cA^{(R)} \\  R^2 \cA^{(2)}   
\end{pmatrix}
 = \theta\begin{pmatrix}
            -3 & -(n+1) & 0 \\
             -\frac{n+2}{n+1}& -(4+n) & n  \\
             0 & 0 & n
               \end{pmatrix}
\begin{pmatrix} \cA^{(1)}\slash \theta \\ \cA^{(R)} \\ R^2 \cA^{(2)}   
\end{pmatrix}
\end{align}
This matrix has indeed eigenvalues $-2,-(n+5), n$, in complete agreement with 
lines three, five and one of \eqref{vector-term-eom}. The corresponding  
eigenvectors are 
\begin{subequations}
\label{eigenmodes-spin-1}
\begin{align}
 B[\phi^{(1)}] &\coloneqq \frac{\cA^{(1)}}{\theta} - 
\frac{n+1}{n+2} \cA^{(R)} + 
\frac{n(n+1)}{(n+2)^2} \ R^2 \cA^{(2)} \, ,\\
 C[\phi^{(1)}] &\coloneqq  \frac 1{n+1} \frac{\cA^{(1)}}{\theta} 
+ \cA^{(R)} - 
\frac{n}{2n+5} \ R^2 \cA^{(2)} \, , \\
 D[\phi^{(1)}] &\coloneqq  R^2 \cA^{(2)}  \, .
\end{align}
\end{subequations}
All of these are physical, i.e.\ they are annihilated by $\cG$ which 
means they are gauge fixed, see Section \ref{sec:vector-Laplacian}.
Similarly, the action of $\cI$ on $\cA^{(0)}$, $\cA^{(3)}$ can be diagonalized 
as follows:
\begin{align}
 \cI \begin{pmatrix} \cA^{(0)}  \\ \cA^{(3)}   
\end{pmatrix}
 = \theta\begin{pmatrix}
            -1 & 0 \\
            \frac{n+2}{n+1} & -4
               \end{pmatrix}
\begin{pmatrix} \cA^{(0)}  \\ \cA^{(3)}   
\end{pmatrix} \, ,
\end{align}
which gives rise to the eigenvectors
\begin{subequations}
 \label{eigenmodes-spin-1-rest}
\begin{align}
 E[\phi^{(1)}] &\coloneqq \cA^{(0)} \, , \\
 F[\phi^{(1)}] &\coloneqq \cA^{(3)} - \frac{1}{3} \frac{n+2}{n+1} \cA^{(0)} \, ,
\end{align}
\end{subequations}
with eigenvalues $-1$ and $-4$, respectively. This corresponds to line two and 
four in \eqref{vector-term-eom}. Hence, we have a complete description of all 
spin 1 modes. In particular, this means that the $F$ modes live in $(1,0)\otimes 
\cC^2$.

Of course these $B,C,D,E,F$ eigenmodes are mutually orthogonal.
This is one of the main points of going to the above basis, and 
it will be elaborated in detail in the spin 2 case.

\paragraph{Pure gauge spin 1 vector fields.}
Consider the pure gauge modes  $\cQ(\phi^{(1)})$
for $\phi^{(1)} \in (n,2)$ 
\begin{align}
 \cQ(\phi^{(1)}) &= \cQ(\phi^{(1)}_{a_1 \ldots a_nb;c}  x^{a_1} \ldots 
x^{a_n}\theta^{bc})  \nn\\
  &=  \theta n\cA^{(3)} + \theta \frac{n+2}{n+1} \cA^{(0)}  \qquad \in \cC^2 
\oplus \cC^0
 \label{pure-gauge-1-decomp-2}
\end{align}
using \eqref{F-x-xi-id}.
As a check, we compute 
\begin{align}
\{x^a,\cQ(\phi^{(1)})_a\} 
  = \Box \phi^{(1)}   = -\theta (n^2+5n+2) \phi^{(1)} \, ,
\end{align}
which has the correct eigenvalue for $(n,2)$.
Note that the $\cA^{(3)}$ term contains some $\cC^0$ components, and after
a projection $[\theta^{da}\theta^{bc} ]_0$, using 
\eqref{eigenmodes-spin-1-rest}, one obtains
\begin{align}
  [\cQ(\phi^{(1)})]_0 = \theta \frac{1}{3} \frac{(n+2)(n+3)}{n+1} \cA^{(0)} \,.
   \label{pure-gauge-1-decomp-0}
\end{align}
We will see in Section \ref{sec:symmetries} that this $\cC^0$ contribution 
corresponds to volume-preserving diffeomorphisms, while the contribution in  
$\cC^2$
leads to the corresponding gauge transformation of the graviton.

\paragraph{Gauge fixing.}
Imposing the gauge fixing condition $\cG(\cA)=\{x^a,A_a\}=0$ in the $E,F$ sector 
leaves one physical mode
\begin{align}
\cG \left( \cA^{(3)} + \a \cA^{(0)} \right) = 0  \quad \text{for} \quad  \a = - 
\frac{n+4}{n+1}  \, .
\end{align}
Note that  the $\cQ[\phi]$ are exact zero modes before gauge fixing, and 
imposing $\cG(\cA) = 0$ removes these modes.
In the Euclidean case, there is no need to further factor out  pure gauge modes, because 
 \begin{align}
  \{x^a,Q(\phi)\} = \{x^a,\{x_a,\phi\}\} = \Box \phi
 \end{align}
is positive definite and invertible on $\cS^4$. Therefore any $\cA^a$
can indeed be gauge fixed uniquely via $\cA^a \to \cA^a + \cQ(\phi)$,
and gauge fixing removes only one mode  in the Euclidean case. In the Minkowski case, this story would be 
somewhat different.

%
%
\subsection{Spin 2 vector fields \texorpdfstring{$\cA$}{A}}
We have the following spin 2 vector modes:
\begin{align} 
\label{eq:modes_spin-2}
\begin{aligned} 
  \cA^{(0)} &\coloneqq \xi_c  \phi^{(2)}_{a_1 \ldots a_nbd;ce} x^{a_1} \ldots 
x^{a_n} x^{b}\cM^{de} 
  \quad \in \ (n,4) \subset (n+1,2)\otimes (1,0) \, , \\
   \cA^{(1)} &\coloneqq  \cJ(A^{(0)}) = \xi_f \theta^{fc} \phi^{(2)}_{a_1 \ldots 
a_nbd;ce}  x^{a_1} \ldots x^{a_n} x^{b} \cM^{de}
 \quad \in \ (n,4) \subset (*,4)\otimes (1,0)   \,, \\
  \cA^{(2)} &\coloneqq \xi^a  \phi^{(2)}_{aa_2 \ldots a_nbd;ce} x^{a_2} \ldots  
x^{a_n} \cM^{bc}\cM^{de} 
 \quad \in (n,4) \subset (*,4)\otimes (1,0) \, , \\
  \cA^{(3)} &\coloneqq \cJ(\cA^{(2)}) = \xi^d \theta^{da}  \phi^{(2)}_{aa_2 
\ldots a_nbd;ce} x^{a_2} \ldots  x^{a_n} \cM^{bc} \cM^{de} 
  \quad \in (n,4) \subset (*,6)\otimes (1,0) \,, \\
 \cA^{(R)} &\coloneqq \frac{1}{\theta^2}\xi \phi^{(2)} =  \xi_d x^d  
\phi^{(2)}_{a_1 \ldots 
a_nbd;ce}  x^{a_1} \ldots  x^{a_n}\cM^{bc} \cM^{de}  \quad \in (n,4) 
\subset (*,4)\otimes (1,0) \, .
\end{aligned}
\end{align}
We will again compute $\cI$ on these modes and diagonalize it.
This will result in the eigenmodes of the vector Laplacian 
\eqref{vector-Laplacian}.

\subsubsection{Properties of spin 2 fields}
First we note the representation of $\cA^{(0)}$ which follows from 
\eqref{potential-gaugefix-id-2}
\begin{align}
 A^{(0)}_a = -\frac{1}{\theta} \frac{1}{n+2}\{x^b,\phi^{(2)}_{ba}\} \, .
 \label{A0-x-phiab}
\end{align}
We have 
\begin{align}
 \cN(\cA^{(0)}) &= x^c \phi^{(2)}_{a_1 \ldots a_nbd;ce} x^{a_1} \ldots x^{a_n} 
x^{b}\cM^{de} = 0 \,,\\
\cG(\cA^{(0)})
 &= \phi^{(2)}_{a_1 \ldots a_nbd;ae} \{x^a,  x^{a_1} 
\ldots x^{a_n} x^{b}\cM^{de}\} 
   = -\frac{1}{\theta} (n+1) \phi^{(2)} \, ,
 \label{A0-2-gaugecond} \\
 \cI(\cA^{(0)}) &=  \phi^{(2)}_{a_1 \ldots a_nbd;ce} \xi_h\{\theta^{hc}, 
x^{a_1} \ldots x^{a_n} x^{b}\cM^{de}\} 
     = 0 \,,
\end{align}
using \eqref{F-x-xi-id} and the symmetry in $(ce)$ in the last step.
We can identify this with the second line of \eqref{vector-term-eom} for $s=1$.
For $\cA^{(1)}$, we compute
\begin{align}
\cN( \cA^{(1)}) &= 0 \, ,\\
\cG(\cA^{(1)})
 &= \phi^{(2)}_{a_1 \ldots  a_nbd;ce} \{x^f,  \theta^{fc} 
x^{a_1} \ldots   x^{a_n} x^{b} \cM^{de}\} 
  = 0 \, , \\
 \cI( \cA^{(1)}) &=  \cI(\cJ(A^{(0)}))
  = - 4 \theta\cJ(\cA^{(0)})  -(n+1) \xi \phi^{(2)}  +\theta 
\cQ(\cN(\cA^{(0)})) -  \cJ\circ \cI (\cA^{(0)})  \nn\\
  &= - 4 \theta \cA^{(1)} -(n+1) \xi \phi^{(2)} \, ,
\end{align}
using \eqref{I-J-identity}.
This is not an eigenvector, but this is addressed below.
Furthermore,
\begin{align}
\cN(\cA^{(2)}) &=  \phi^{(2)}_{fa_2 \ldots  a_nbd;ce} x^f x^{a_2} \ldots   
x^{a_n} 
\cM^{bc}\cM^{de} = \frac{1}{\theta^2}\phi^{(2)} \, ,
  \label{N-A2} \\
\cG(A^{(2)})
 &=  \phi^{(2)}_{aa_2 \ldots  a_nbd;ce}\{x^a,  x^{a_2} \ldots 
x^{a_n} \cM^{bc}\cM^{de}\}  
    = 0 \, , \\
 \cI(\cA^{(2)}) &= \xi_h \phi^{(2)}_{fa_2 \ldots a_nbd;ce} \{\theta^{hf}, 
x^{a_2} 
\ldots  x^{a_n} \cM^{bc}\cM^{de} \} 
    = (n+1)\theta \cA^{(2)} \, ,
\end{align}
because $\phi^{(2)}$ is traceless. Similarly, we can identify this with line 
one of \eqref{vector-term-eom} for $s=2$. Next, for $\cA^{(3)}$ we obtain 
\begin{align}
\cN(\cA^{(3)}) &= 0 \, ,\\
\cG(A^{(3)})
 &=  \{x_f,\theta^{fa}  \phi^{(2)}_{aa_2 \ldots a_nbd;ce} 
x^{a_2} \ldots  x^{a_n} \cM^{bc}\cM^{de} \} 
  = -\frac{(n+5)}{\theta} \phi^{(2)} \, ,
 \label{A3-gaugefix} \\
 \cI( \cA^{(3)}) &=  \cI(\cJ(A^{(2)}))
  = - 4 \theta\cJ(\cA^{(2)}) +\theta \cQ(\cN(\cA^{(2)})) -  \cJ\circ \cI 
(\cA^{(2)})  \nn\\
  &=  - 4 \theta\cA^{(3)} +\frac{1}{\theta} \cQ(\phi^{(2)}) - 
(n+1)\theta\cA^{(3)} \nn\\
  &=  - 5\theta\cA^{(3)} +2\theta \frac{n+2}{n+1}\cA^{(0)} \, ,
\end{align}
using \eqref{I-J-identity} (for the basic $S^4_N$) and \eqref{Q-F-2}.
Finally, for the radial mode $\cI$ does not diagonalize, but decomposes into 
radial and tangential components. In detail, we obtain
\begin{align}
\cG(A^{(R)})
&=
\frac{1}{\theta^2} x^a \{x_a, \phi^{(2)}\} = 0 \, ,\\
  \cI(\cA^{(R)}) &= 
  - 4 \frac{1}{\theta} \xi \phi^{(2)} 
  - \frac{1}{\theta^2}  \cJ(\cQ(\phi^{(2)})) \nn\\
   &= -\theta(n +4) \cA^{(R)} + n\theta R^2\cA^{(2)} - 2 
\frac{n+2}{n+1}\cA^{(1)} \, ,
\end{align}
using \eqref{I-N-identity}, \eqref{I-A2}, and $\cA^{(3)} = \cJ \cA^{(2)}$.
\subsubsection{Diagonalization of \texorpdfstring{$\cI$}{I}}
Now we can diagonalize $\cI$ as in the spin 1 case.
Collecting the above results, we obtain for 
the modes $\cA^{(1)} \slash \theta$, $\cA^{(R)}$, and $R^2 \cA^{(2)}$ the 
following:
\begin{align}
 \cI \begin{pmatrix} \cA^{(1)} \slash \theta \\ \cA^{(R)} \\ R^2 \cA^{(2)} 
\end{pmatrix}
 = \theta\begin{pmatrix}
            -4 & -(n+1) & 0 \\
             -2\frac{n+2}{n+1}& -(n+4) & n \\
             0 & 0 & (n+1)
               \end{pmatrix}
\begin{pmatrix} \cA^{(1)} \slash \theta \\ \cA^{(R)} \\ 
R^2 \cA^{(2)} 
\end{pmatrix} \, .
\end{align}
This matrix has eigenvalues $-2,-(n + 6), 1 + n$, in complete agreement with 
lines three, five and one of \eqref{vector-term-eom}. The corresponding 
eigenvectors are 
\begin{subequations}
\label{eigenmodes-spin-2}
\begin{align}
 B[\phi^{(2)}] &\coloneqq \frac{\cA^{(1)}}{\theta} - \frac{n+1}{n+2} 
\cA^{(R)} + \frac{n(n+1)}{(n+2)(n+3)} \ R^2 \cA^{(2)} \ , \\
 C[\phi^{(2)}] &\coloneqq  \frac{2}{n+1} \frac{\cA^{(1)}}{\theta} + \cA^{(R)} 
- \frac{n}{2n+7}\ R^2 \cA^{(2)}  \ , \\
 D[\phi^{(2)}] &\coloneqq R^2 \cA^{(2)} \ .
\end{align}
\end{subequations}
They all satisfy the gauge fixing condition $\cG(\cdot) = 0$.
We have therefore obtained a complete basis of spin 2 eigenmodes of $\cI$.
Similarly, we can compute the $\cI$ action on $\cA^{(0)}$ and $\cA^{(3)}$
\begin{align}
  \cI \begin{pmatrix} \cA^{(0)} \\ \cA^{(3)} 
\end{pmatrix}
 = \theta\begin{pmatrix}
            0 & 0 \\
            2 \frac{n+2}{n+1} & -5
               \end{pmatrix}
\begin{pmatrix} \cA^{(0)}   \\ \cA^{(3)} 
\end{pmatrix} \, ,
\end{align}
which has eigenvalues $0$ and $-5$. The corresponding eigenvectors are
\begin{subequations}
 \label{eigenmodes-spin-2-rest}
\begin{align}
 E[\phi^{(2)}] &\coloneqq \cA^{(0)} \, , \\
 F[\phi^{(2)}] &\coloneqq \cA^{(3)} - \frac{2}{5} \frac{n+2}{n+1} \cA^{(1)} \, .
\end{align}
\end{subequations}
We can identify this  with  line 2 and 4 of \eqref{vector-term-eom}.
In particular, this means that the $F$ modes live in $\cC^2 \otimes (1,0)$.

\paragraph{Pure gauge spin 2 vector modes.}
The gauge transformations generated by $\phi^{(2)}$ read
\begin{align}
  \cQ(\phi^{(2)}) &= \cQ(\phi^{(2)}_{a_1 \ldots a_nbd;ce}  x^{a_1} \ldots  
x^{a_n}\theta^{bc} \theta^{de} ) \nn\\
  &= \theta^2 n \cA^{(3)} +  \xi_f \phi^{(2)}_{a_1 \ldots  a_nbd;ce}  x^{a_1} 
\ldots   
x^{a_n}\{x^f,\theta^{bc} \theta^{de}\}  \nn\\
  &=\theta^2  n \cA^{(3)} +2 \theta^2 \frac{n+2}{n+1}\cA^{(0)}
  \label{Q-F-2}
\end{align}
using  \eqref{F-x-xi-id}, cf.\ \eqref{pure-gauge-1-decomp-2}. Hence, these are 
not new modes.

\subsubsection{Inner product matrix}
Later, when computing the kinetic terms in the action \eqref{fluct-action-nogf}, 
we will need the following expressions:
\begin{align}
 \int \cA^{(i)}[\phi^{(2)}]  \cA^{(j)}[\phi^{(2)}]  = K^{ij} \int \phi^{(2)}_{ab} \phi^{(2)}_{ab}, \qquad i,j\in\{0,1,2,3,R\}.
\end{align}
We only present the results here, and delegate the derivation to Appendix 
\ref{app:inner-prod}:
\begin{subequations}
 \label{inner-product-h-matrix}
\begin{align}
 \int \cA^{(0)}[\phi]  \cA^{(0)}[\phi] &=
  \frac{(n+3)(n+4)}{3(n+2)^2}\frac{1}{\theta} \int \phi _{ab}\phi _{ab}  \;,
\label{inner-product-h-matrix-0}\\
  \int \cA^{(0)}[\phi ]  \cA^{(3)}[\phi ] &= \frac{2}{15} \frac 
{(n+4)(n+3)}{(n+1)(n+2)} \frac{1}{\theta} \int \phi _{ab}\phi _{ab} \;,\\
  \int \cA^{(3)}[\phi ]  \cA^{(3)}[\phi ] &=  \frac 2{15} 
\frac{(n+3)(n+4)(n^2+8n+21)}{n(n+1)^2(n+2)} \frac{1}{\theta}  \int \phi_{ab}
\phi_{ab}  \;,\\
 \int \cA^{(1)}[\phi ]  \cA^{(1)}[\phi ] &=
    \frac{(n+3)(n+4)}{3(n+2)^2} R^2 \int \phi _{ab}\phi _{ab} \;,\\
 \int \cA^{(1)}[\phi ]  \cA^{(2)} [\phi ] &= -\frac{2}{15} \frac 
{(n+4)(n+3)}{(n+1)(n+2)} \frac{1}{\theta} \int \phi_{ab}\phi_{ab}  \;,\\
 \int \cA^{(1)}[\phi ]  \cA^{(R)}[\phi ] &=  0 \;,\\
 \int \cA^{(2)}[\phi ]  \cA^{(2)}[\phi ] &=
 \frac{1}{\theta^2} \frac{2  (n+3)^2 (n+4) (2 n+7)}{15 n (n+1)^2 
(n+2) R^2}\int \phi_{ab}\phi_{ab} \;,\\
 \int \cA^{(2)}[\phi ]  \cA^{(R)}[\phi ] &=  \frac{2}{15} 
\frac{(n+5)(n+4)(n+3)}{(n+1)^2(n+2)}  \frac{1}{\theta^2} \int \phi^{(2)}_{ab}  
\phi^{(2)}_{ab}     \;,\\
 \int \cA^{(R)}[\phi ]  \cA^{(R)}[\phi ] &= \frac{2}{15} 
\frac{(n+5)(n+4)(n+3)}{(n+1)^2(n+2)} R^2 \frac{1}{\theta^2}  \int 
\phi^{(2)}_{ab}  \phi^{(2)}_{ab}\;,
 \end{align}
 \end{subequations}
where $\phi \equiv \phi^{(2)}$. All other inner products vanish.
To gain some insights, we will give a 
more transparent derivation of e.g.\ \eqref{inner-product-h-matrix-0} in 
\eqref{inner-prod-A0-spin1-asympt}.
 In the basis of the $\cI$-eigenmodes \eqref{eigenmodes-spin-2}, 
\eqref{eigenmodes-spin-2-rest} we find 
the inner product matrix $K^{IJ}$ for $I,J \in \{B,C,D,E,F\}$, defined 
via
\begin{align}
 \int \cB_I[\phi^{(2)}]  \cB_J[\phi^{(2)}]  = K^{IJ} \int 
\phi^{I}_{ab} \phi^{J}_{ab}, \qquad I,J\in\{B,C,D,E,F\} \; ,
\label{eq:inner_prod_new_basis}
\end{align}
to be diagonal $K^{IJ} = \delta^{IJ} K^{I}$ with coefficients
\begin{subequations}
\label{eq:inner_prod_I-eigenmodes-spin2}
\begin{align}
 K^{B} &= \frac{ (n+4)^2 (n+5) }{5 (n+2)^3}\  \frac{R^2}{\theta^2} 
  \ \approx  \frac{R^2}{\theta^2} \frac 15 \;,   \\
K^{C} &= \frac{2  (n+3) (n+4)^2 (n+5)(n+7)}{15 (n+1)^2 
(n+2)^2 (2 n+7)}\ \frac{R^2}{\theta^2}
  \ \approx  \frac{R^2}{\theta^2} \frac 1{15}   \;,  \\
K^{D} &= \frac{2 (n+3)^2 (n+4) (2 n+7)}{15 n (n+1)^2 (n+2) }\ 
\frac{R^2}{ \theta^2} 
 \ \approx  \frac{R^2}{\theta^2} \frac 4{15} \;,   \\
K^{E} &= \frac{ (n+3) (n+4)}{3 (n+2)^2} \frac{1}{\theta} 
 \ \approx  \frac{1}{\theta} \frac 13  \;, \\
K^{F} &= \frac{2  (n+3) (n+4) \left(n^2+12 n+35\right)}{25 n 
(n+1)^2 (n+2)}\ \frac{1}{\theta} 
 \ \approx  \frac{1}{\theta} \frac 2{25} \; .
\end{align}
\end{subequations}
Here, $\approx$ indicates the leading contributions for large $n$.
This calculation provides a non-trivial consistency check.
Note that $\phi^{I}_{ab}$ is defined in terms of the same linear combination of 
$\phi_{ab}^i$ as the mode $\cB_I$ in terms of $\cA_i$.

\subsection{Higher spin vector fields \texorpdfstring{$\cA$}{A}}
\label{sec:recombination}
Now we briefly discuss the general structure of the fluctuations with generic 
spin $s$. From the examples above, it is clear that there 
are five vector modes for each spin $s$, realized by
\begin{align}
  \cA^{(0)} &= \xi_c  \phi^{(s)}_{a_1 \ldots a_nb\ldots d;c\ldots e} x^{a_1} 
\ldots x^{a_n} 
x^{b}\cM \ldots \cM^{de} 
  \quad \in \ (n,2s) \subset (*,2s{-}2)\otimes (1,0) \,, \nn\\
   \cA^{(1)} &=  \cJ(A^{(0)}) = \xi_f \theta^{fc} \phi^{(s)}_{a_1 \ldots 
a_nb\ldots d;c\ldots e}  x^{a_1} \ldots x^{a_n} x^{b} 
   \cM \ldots \cM^{de}
 \quad \in \ (n,2s) \subset (*,2s)\otimes (1,0)  \,,  \nn \\
  \cA^{(2)} &= \xi^a  \phi^{(s)}_{aa_2 \ldots a_nb\ldots d;c\ldots e} x^{a_2} 
\ldots  x^{a_n} 
\cM^{bc}\ldots  \cM^{de} 
 \quad \in (n,2s) \subset (*,2s)\otimes (1,0) \,, 
 \label{eq:modes_spin-s} \\
  \cA^{(3)} &= \cJ(\cA^{(2)}) = \xi^f \theta^{fa}  \phi^{(s)}_{aa_2 \ldots 
a_nb\ldots d;c\ldots e} x^{a_2} \ldots  x^{a_n} \cM^{bc} \ldots \cM^{de} 
  \quad \in (n,2s) \subset (*,2s{+}2)\otimes (1,0) \,, \nn \\
 \cA^{(R)} &= \frac{1}{\theta^s}\xi \phi^{(s)} =  \xi_f x^f  \phi^{(s)}_{a_1 
\ldots a_nb\ldots d;c\ldots e}  
x^{a_1} \ldots x^{a_n}\cM^{bc} \ldots \cM^{de} 
 \quad \in (n,2s) \subset (*,2s)\otimes (1,0) \,.\nn
\end{align}

\paragraph{Properties spin s fields}
Analogous to the previous calculations, one can evaluate $\cI$ on the five 
generic spin $s$ modes (some details are in Appendix 
\ref{app:eigenvalues_modes}). One obtains
\begin{align}
\label{eq:EW_cI_spin-s_pt1}
\cI
 \begin{pmatrix}
  \cA^{(1)} \slash \theta \\ \cA^{(R)} \\ R^2 \cA^{(2)}
 \end{pmatrix}
 = \theta 
\begin{pmatrix}
 -(s+2) & -(n+1) & 0 \\
 -s\frac{n+2 }{n+1} & -(n+4) & n  \\
 0 & 0 & n+s-1 \\
\end{pmatrix}
  \begin{pmatrix}
 \cA^{(1)} \slash \theta \\ \cA^{(R)} \\ R^2 \cA^{(2)}
 \end{pmatrix} \, ,
\end{align}
which has eigenvalues $-2$, $-(n+s+4)$, $n+s-1$. This agrees with Table 
\ref{tab:eigenvalues_I-modes} of Appendix \ref{app:eigenvalues_modes}.
The corresponding eigenvectors are defined as 
\begin{subequations}
\label{eq:trafo_spin_s_pt1}
\begin{align}
 B[\phi^{(s)}] &\coloneqq \frac{\cA^{(1)}}{\theta} - \frac{n+1}{n+2} \cA^{(R)}  
+ \frac{n(n+1) }{(n+2)(n+s+1)} \  (R^2  \cA^{(2)}) \ ,\\
 C[\phi^{(s)}] &\coloneqq \frac{2}{n+1} \frac{\cA^{(1)}}{\theta} + \cA^{(R)} - 
\frac{n }{2(n+s)+3} \ ( R^2 \cA^{(2)}) \ , \\
 D[\phi^{(s)}] &\coloneqq R^2 \cA^{(2)} \ .
\end{align}
\end{subequations}
Likewise we find
\begin{align}
\label{eq:EW_cI_spin-s_pt2}
\cI
 \begin{pmatrix}
  \cA^{(0)}  \\ \cA^{(3)}
 \end{pmatrix}
 = \theta 
\begin{pmatrix}
 s-2 &  0 \\
 s\frac{n+2 }{n+1} & -(s+3)
\end{pmatrix}
  \begin{pmatrix}
  \cA^{(0)} \\ \cA^{(3)}
 \end{pmatrix} \, ,
\end{align}
having eigenvalues $s-2$ and $-(s+3)$, see again Table 
\ref{tab:eigenvalues_I-modes}. The corresponding eigenvectors are given by
\begin{subequations}
\label{eq:trafo_spin_s_pt2}
\begin{align}
 E[\phi^{(s)}] &\coloneqq \cA^{(0)} \, , \\
 F[\phi^{(s)}] &\coloneqq \cA^{(3)} - \frac{s(n+2)}{(2s+1)(n+1)} \cA^{(0)} \, .
\end{align}
\end{subequations}
\subsection{Recombination and relation with Vasiliev theory}
To make contact with the standard formalism of Vasiliev theory and with the 
previous work \cite{Steinacker:2016vgf}, we 
observe that $\cA^{(0)}$ can be absorbed in a trace part of $\cA^{(1)}$, and 
the tangential part of $\cA^{(2)}$ can be absorbed in a trace part of 
$\cA^{(3)}$. To start with we rewrite
\begin{align}
 \cA^{(0)}_a = P_T^{ab} \cA^{(0)}_b &= -\frac{1}{\theta 
R^2}\theta^{ac}\big(\theta^{cd}\cA^{(0)}_d\big)  \, , \nn\\
 P_T^{ab} \cA^{(2)}_b &= -\frac{1}{\theta 
R^2}\theta^{ac}\big(\theta^{cd}\cA^{(2)}_d\big) \, .
\end{align}
For example,  the spin 1 mode of type $ \cA^{(0)}$ can be written as
\begin{align}
 \cA^{(0)}[\phi^{(1)}] &= -\frac{1}{\theta R^2} \xi_g \theta^{gh} 
 \big( \phi^{(1)}_{a_1 \ldots  a_nb;c} x^{a_1} \ldots  x^{a_n} x^{b}\theta^{hc} 
\big) \nn\\
  &= -\frac{1}{\theta R^2} \xi_g \theta^{gh} 
 \big( g_{hh'} \phi^{(1)}_{a_1 \ldots  a_nb;c}\big) x^{a_1} \ldots  x^{a_n} 
x^{b} \theta^{h'c} \nn\\
 &\equiv   \xi_g \theta^{gh} 
 \big(\tilde\phi^{(1)}_{a_1 \ldots  a_nbh';hc} x^{a_1} \ldots  x^{a_n} x^{b} 
\theta^{h'c} \big) \nn\\[1ex]
 &= \cA^{(1)}[\tilde\phi^{(1)}]  \, ,  
\end{align}
where 
\begin{align}
\tilde\phi^{(1)}_{a_1 \ldots  a_nbh';hc} &= - \frac{1}{\theta R^2} \, 
P^S_{a_1\ldots a_nbh'}
  \big( g_{hh'} \phi^{(1)}_{a_1 \ldots  a_nb;c} + (h\leftrightarrow c) \big) 
\, ,  \nn\\
   g^{hh'}\tilde\phi^{(1)}_{a_1 \ldots  a_nbh';hc} &= \phi^{(1)}_{a_1 \ldots  
a_nb;c} \, .
   \label{phi-1-tilde}
\end{align}
This is associated via $\cA^{(1)}[{\tiny \Yvcentermath1 \yng(4,2)}]$ of
\eqref{A-i-funktor} to
 a Young diagram which is no longer traceless, but double traceless
(but traceless within the horizontal lines).

Similarly, the tangential part of $\cA^{(2)}$ can be absorbed in a trace contribution 
to $\cA^{(3)}$, which we illustrate 
again for the spin 1 case
\begin{align}
 \cP_T\cA^{(2)}[\phi^{(1)}] &= -\frac{1}{\theta R^2} \xi_g \theta^{gh}   
 \big( \phi^{(1)}_{aa_2 \ldots  a_nb;c} x^{a_2} \ldots   x^{a_n} 
\theta^{bc}\theta^{ha}  \big) \nn\\
 &= \frac{1}{\theta R^2} \xi_g \theta^{gh}   
 \big( g_{hh'}\phi^{(1)}_{a_2 \ldots  a_nab;c} x^{a_2} \ldots   
x^{a_n}\theta^{ah'}  \theta^{bc} \big) \nn\\
 &=  \xi_g \theta^{gh}   
 \big(\tilde\phi^{(1)}_{ha_2 \ldots  a_nab;h'c} x^{a_2} \ldots   
x^{a_n}\theta^{ah'}  \theta^{bc} \big) \nn\\
 &= \cA^{(3)}[\tilde\phi^{(1)}] \, ,
\end{align}
where $\tilde\phi^{(1)}_{ha_2 \ldots  a_nab;h'c}$ coincides with 
\eqref{phi-1-tilde}. Likewise, the radial part of $\cA^{(2)}$ can be absorbed 
in the radial mode $\cA^{(R)}$.

In general, $\cA^{(0)}[\cdot]$ is absorbed in the trace part of the  
$\cA^{(1)}[\cdot]$, and  
$\cP_T \cA^{(2)}[\cdot]$ is absorbed in the trace part of $\cA^{(3)}[\cdot]$.
However, as exemplified in \eqref{phi-1-tilde} the underlying tensor of the 
vector fields is different. In the classification of fluctuation modes 
\eqref{vector-term-eom}, \eqref{eq:modes_spin-s}
we used irreducible Young diagrams, whereas the class of tensors for the 
recombinations needs to be generalized. These tensors (or Young diagrams) do 
not 
 correspond to irreducible representations any more, but have to advantage to 
repackage the tangential modes into only two objects.

Thus, we can collect \textbf{all} vector modes into a single form 
\begin{align}
\boxed{
\cA^a =  P_T^{ab}\cA^b +  \frac 1{R^2} x^a (x_b\cA^b) \equiv \theta^{ab} \A_b + 
x^a \phi  \,,
}
  \label{A-tang-radial}
\end{align}
separated into tangential gauge fields and a transversal scalar field
\begin{align}
  \A_b &= \frac{1}{\theta R^2}\theta^{ab}\cA^b \  = A_{b,\und{\b}}(x)\, 
\Xi^{\und\b}\; , \qquad  
 \Xi^{\und{\a}}  \in  \hs \, ,\nn\\
  \phi &= \frac 1{R^2} \, x_b\cA^b \ = \phi_{\und{\b}}(x)\, \Xi^{\und\b} \; , 
\qquad  
 \Xi^{\und{\a}}  \in  \hs \, ,
\end{align}
using \eqref{HS-def}, where $\A_a$ is a $\hs$-valued one-form on $S^4$ 
corresponding to double-traceless rectangular Young diagrams. $\A_a$ encodes 
all $\cA^{(0)}$, $\cA^{(1)}$, $\cP_T\cA^{(2)}$, $\cA^{(3)}$ associated to 
irreducible Young diagrams via the recombination into $\cA^{(1)}$ and 
$\cA^{(3)}$ build from more general diagrams.
The only difference between the $\cA^{(1)a}$ and the $\cA^{(3)a}$ modes is that
the external vector index $a$
is linked via $\theta^{ab}$ either to the second line or the first line of the Young diagram,
leading to a different number of generators $\cM$  in $\Xi^{\und{\a}}$.
In other words, there are two traceless
contributions of the same form  $\cA^a = \theta^{ab} A_{a;\und{\a}}(x) \Xi^{\und{\a}}$,
one describing an irreducible spin $s$ gauge field, and one spin $s-1$ contribution which will be recognized as
pure gauge sector.

In addition, the transversal degrees of freedom $(1-\cP_T)\cA^{(2)}$ and 
$\cA^{(R)}$ are encoded in the $\hs$-valued scalar $\phi$ on $S^4$.

The 1-form $\A_{a}$ provides the kinematical link to Vasiliev theory, cf.\ 
\cite{Steinacker:2016vgf}.
\subsection{Local representation and constraints}
Now we will express the spin 1 and spin 2 modes \eqref{spin-1-modes}, 
\eqref{eq:modes_spin-2} in terms of ordinary tensor fields near the north pole 
as in Section \ref{sec:local-rep-scalar}.
As for the scalar fields,  we will find that the $P^\mu$  and the $\cM^{\mu\nu}$ components
of the $\mso(5)$-valued  fields are not independent.
This leads again to constraints, which are illustrated in some examples.

\paragraph{Spin 1 mode  $\cA^{(2)}$.}
The decomposition for $\cA^{(0)}$ and $\cA^{(1)}$ involves only a vector field 
and has already been discussed. Thus, consider
the spin 1 mode $\cA^{(2)}$. Decomposed near the north pole of $S^4$ 
into tangential and radial components, it reads
\begin{align}
  \cA^{(2)}_a &= \phi^{(1)}_{a a_2 \ldots  a_nb;c}  x^{a_2} \ldots  x^{a_n} 
\theta^{bc}
  =  A_{\nu}^a P^\nu + F_{\mu\nu}^a \cM^{\mu\nu} \,.
 \label{A2-local-1}
\end{align}
For any fixed $a$, $A^{(2)}_a$ can be viewed as an element in $\cC^1$, which 
using the results of Section \ref{s1-field-strength} can be written as
\begin{align}
 F_{\mu\nu}^a = -\frac{1}{2(n+1)}(\del_\mu A^a _\nu - \del_\nu A^a _\mu) \, .
\end{align}
On the other hand, $\cA^{(2)}_a$ is fully determined by its radial component 
\begin{align}
 \phi^{(1)}\coloneqq  x^a\cA^{(2)}_a 
 \eqqcolon A_{\mu}(x) P^{\mu} + F_{\mu\nu}(x) \cM^{\mu\nu} \, ,
\end{align}
which, by using \eqref{nabla-theta}, yields at the north pole
\begin{align}
 A_{\mu}^\a &= \frac 1n\ \del^\a A_{\mu} , \qquad  
 F_{\mu\nu}^\a  = -\frac 1{n(n+2)}  \ \del^\a (\del_\mu A_\nu - \del_\nu A_\mu) 
\, .
 \label{A2-local-1-explicit}
\end{align}
Hence, these spin 1 modes also reduce to a vector field and its field strength 
tensor.
\paragraph{Spin 1 mode  $\cA^{(3)}$.}
The above result implies immediately
\begin{align}
  \cA^{(3)a} &=  \cJ(\cA^{(2)}) 
 =  \theta^{a\r} \Big(A_{\r\mu} P^{\mu} +  F_{\r;\mu\nu} \cM^{\mu\nu}        \Big)
 \label{A3-local-1}
\end{align}
at the north pole, where $A_{\r\nu}$ and $F_{\r;\mu\nu}$ are as in 
\eqref{A2-local-1-explicit}.
Note that $A_{\r\mu}$ is not necessarily symmetric.

\paragraph{Spin 2 modes $\cA^{(0)}$.}

Now consider the spin 2 (graviton) modes $\cA^{(0)}$, decomposed into tangential and radial components 
at the north pole of $S^4$
\begin{align}
  \A^{(0)}_\mu &=  \phi^{(2)}_{a_1 \ldots  a_nd;\mu e}  x^{a_1} \ldots  x^{a_n} 
\cM^{de} 
  =  h_{\mu\nu} P^{\nu} + \omega_{\mu;\r\s}(x) \cM^{\r\s} \; ,
 \label{A0-local}
\end{align}
recall that $ \cA^{(0)}$ is tangential.
For fixed $\mu$, $A^{(0)}_\mu$ can be viewed as an element in $\cC^1$, and 
applying the results of Section \ref{sec:local-rep-scalar} we obtain 
\begin{align}
 h_{\mu\nu}  &= - \frac{n+2}{n+1} \phi^{(2)}_{\mu\nu} = h_{\nu\mu} \, ,\nn\\
 \omega_{\mu;\r\nu} &= -\frac{1}{2(n+2)}(\del_\r h_{\mu\nu} - \del_\nu 
h_{\mu\r}) \, . 
 \label{h-omega-relation-spin2}
\end{align}
Hence $\omega_{\mu;\a\b}$ is proportional to the spin connection defined by 
the linearized metric mode $ h_{\mu\nu}$.
As a check,
we asymptotically recover the relation \eqref{inner-product-h-matrix} for $\cA^{(0)}$
\begin{align}
  \int \cA^{(0)}_\mu \cA^{(0)}_\mu 
 &= \frac 23 \frac{R^2}{\theta} \int  \omega_{\mu;\r\nu} \omega_{\mu;\r\nu}
  \sim \ \frac 13  \frac{R^2}{\theta}\int  \phi^{(2)}_{\mu\b}\, 
\frac{1}{(n+1)^2}\del\cdot\del \, \phi^{(2)}_{\mu\b} \nn\\
  &\sim \ \frac 1{3\theta}\int   \phi^{(2)}_{\mu\b} \phi^{(2)}_{\mu\b} \, ,
  \label{inner-prod-A0-spin1-asympt}
\end{align} 
dropping  $O(\frac{1}{R})$ curvature terms,
using $[P^\mu P^\nu]_0=0$, see \eqref{eq:averaging},
and  $\del\cdot\del \sim -\frac{(n+1)^2}{R^2}$.
Note that only the $\omega_{\mu;\r\nu}$ contributes.
For the generalized $S^4_\L$, there are different modes where the $h_{\mu\nu}$ term
provides the dominant contribution. This leads to a very different behavior, which will be elaborated elsewhere.

\paragraph{Spin 2 modes $\cA^{(1)}$.}
Similarly, we can specialize $\cA^{(1)} = \cJ(\cA^{(0)})$ at the north pole to
\begin{align}
  \cA^{(1)^\mu} &= \theta^{\mu\nu}\Big( \phi^{(2)}_{a_1 \ldots  a_nd;\nu e}  
x^{a_1} \ldots  x^{a_n} \cM^{de} \Big)
  =  \theta^{\mu\nu}\Big(h_{\nu\r} P^{\r} + \omega_{\nu;\r\s}(x) 
\cM^{\r\s}\Big) \, ,
 \label{A1-local}
\end{align}
where $\omega_{\mu;\a\b}$ is proportional to the spin connection defined by 
the linearized metric  $h_{\mu\nu}$.
Hence $\cA^{(0)}$ and $\cA^{(1)}$ both encode some ''metric`` tensor  and the 
associated spin connection.

\paragraph{Spin 2 modes $\cA^{(2)}$.}
Now consider  the  $\cA^{(2)}$ spin 2 (graviton) modes, decomposed into tangential and radial components 
at the north pole of $S^4$ as in  \eqref{phi-2-local}
\begin{align}
  \cA^{(2)}_a &= \phi^{(2)}_{a a_2 \ldots  a_n bd;c e}  x^{a_2} \ldots  x^{a_n} 
\cM^{bc} \cM^{de} \nn\\
  &=: h_{\mu\nu}^a P^{\mu}P^\nu + \omega_{\mu;\a\b}^a P^\mu \cM^{\a\b} + 
\Omega_{\a\b;\mu\nu}^a \cM^{\a\b}\cM^{\mu\nu} \,.
 \label{A2-local}
\end{align}
This is  determined by its radial component 
\begin{align}
 \phi^{(2)}\coloneqq  x_a\cA^{(2)}_a 
 \eqqcolon h_{\mu\nu} P^{\mu}P^\nu + \omega_{\mu:\a\b} P^\mu \cM^{\a\b}
  + \Omega_{\a\b;\mu\nu} \cM^{\a\b}\cM^{\mu\nu}
\end{align}
via
\begin{align}
  h_{\mu\nu}^\r &= \del_\r h_{\mu\nu} .
\end{align}
Applying the results of Section \ref{s2-connect-curvature}, we obtain for 
example
\begin{align}
 \omega_{\mu;\a\b}^\r &= \del_\r \omega_{\mu:\a\b}  \, , \nn\\
 \Omega_{\a\b;\mu\nu}^\r&= \del_\r  \Omega_{\a\b;\mu\nu} \, ,
\end{align}
up to curvature contributions of order $\frac 1R$.
The analysis of $\cA^{(3)}$ is analogous.

\section{Action and equations of motion}
So far we have obtained the higher spin fluctuation modes on $\cS^4$. 
The next step is the formulation of physical higher spin theories. 
There is a simple and natural framework to establish such higher spin actions 
on $S^4_N$, given by matrix models. 
In the semi-classical limit, this leads to higher spin gauge theories on $S^4$.
\subsection{Scalar theory on \texorpdfstring{$\cS^4$}{SS4}}
\label{sec:scalar-field}
As a warm-up, consider first  a ''scalar field theory`` on $\cS^4$,
with an action given by the semi-classical limit of a matrix model
\begin{align}
 S = \Tr \big(-\Phi \Box \Phi + V(\Phi)\big) 
  \ \sim \ \frac{\dim\cH}{\rm Vol \cS^4}\int\limits_{\cS^4} \phi (-\Box) \phi + V(\phi) ,
   \quad \Phi \in End(\cH) \sim \phi \in \cC \, ,
   \label{scalar_action}
\end{align}
recalling the relation \eqref{trace-int}  between trace and integral.
The spin 0 sector $\phi^{(0)}\, \in \cC^0$ leads to a 
scalar field theory on $S^4$, deformed by the non-associativity of the commutative product \cite{Ramgoolam:2001zx}
which leads to slightly non-local interactions. This is supplemented by a tower 
of spin $s$ fields $\phi^{(s)}\, \in \cC^s$. 
Similar models have been considered for instance in 
\cite{Medina:2002pc,Medina:2012cs}.
However this is not a gauge theory, and we will not consider it any further here.

\subsection{Vector theory on \texorpdfstring{$\cS^4$}{SS4} and vector 
Laplacian}
\label{sec:vector-Laplacian}
Now consider a gauge theory for fields $\cA = \xi_a \cA^a\in \Omega^1\cS^4$.
Such a theory arises naturally 
as Poisson limit of Yang-Mills matrix models, such as the IKKT model.
The action of the "Poisson matrix model'' reads as follows:
\begin{align}
 S &= \frac 1{g^2}\int\limits_{\cS^4}d\Omega \Big(\{y_a,y_b\}\{y^a,y^b\}\, + 
\mu^2 y^a y_a \Big) \, , 
 \label{bosonic-action}
\end{align}
where\footnote{To stabilize $S^4_N$ with the classical action, one needs a 
negative mass; however taking into account quantum corrections, 
 a positive bare mass term suffices at one loop \cite{Steinacker:2015dra}.
Alternatively one may add other terms such as 
 $\int\epsilon^{abcde} \{y^a,y^b\}\{y^c,y^d\} y^e$, cf.\ \cite{Kimura:2002nq}. }
\begin{align}
 y^a = x^a + \cA^a
\end{align}
are functions on $\cS^4 \cong \C P^3$, and $x^a$ is the background which defines 
$\cS^4$. As above in \eqref{trace-int} and \eqref{scalar_action}, the integral 
is defined by the symplectic volume form on  $\C P^3$. Collecting  the variables 
in the formal one-forms $Y = \xi_a y^a,\  X = \xi_a x^a, \ \cA = \xi_a \cA^a$ 
and expanding the action up to second order in $\cA^a$, one obtains
\begin{equation}
\begin{aligned}
 S[Y]  = S[X]  + \frac{2}{g^2}\int\limits_{\cS^4} d\Omega \Big(2\cA^a (-\Box 
&+\frac 12\mu^2) x_a 
  +\cA_a (-\Box +\frac 12\mu^2) \cA_a \\
  &+ 2 \{\cA_a,\cA_b\}\{x^a,x^b\} + 
\cG(\cA)^2 + O(\cA^3) \Big) \ . 
\label{eff-S-expand}
\end{aligned} 
\end{equation}
Here $ \Box = \{x^a,\{x_a,\cdot \}\}$
is the Poisson-Laplacian defined in \eqref{Poisson-Laplacian}, and, recalling 
\eqref{gauge-fixing-function}, $\cG(\cA)=\{\cA^a,x_a\}$
can be viewed as gauge fixing function, which
transforms as 
 $\cG \to \cG + \Box \L$
under gauge transformations. 
Hence, the quadratic fluctuations $\cA^a$ are governed by the quadratic form
\begin{align}
\int_{\cS^4} d\Omega\, \cA_a \Big(\cD^2 + \frac 12 \mu^2\Big) \cA_a  \,,
\label{fluct-action-nogf}
\end{align}
where the contribution from  $\cG(\cA)^2$ has been canceled by adding a 
suitable Faddeev--Popov gauge-fixing term.
Therefore, we recover that the relevant operator for the 
vector fluctuations is the ``vector'' Laplacian \eqref{vector-Laplacian}, i.e.
\begin{align}
\boxed{
\ \cD^2 \cA =\big(-\Box  - 2 \cI \big)   \cA  \
}
\label{fluct-action}
\end{align}
where $\cI$ is  the intertwiner  defined in \eqref{intertwiners}.
Its eigenvalues can be determined 
by relating $\cD^2$ to basic group-theoretical operators  \eqref{intertwiner} and 
\eqref{I-MM-id},
 \begin{align}
 \Box  &= \{x_a,\{x^a,\cdot\}\} = \theta \sum_{a=1}^5  \cM_{a6}^{(\ad)} 
\cM_{a6}^{(\ad)}  
 =  \theta ( C^2[\mso(6)]^{(\ad)} - C^2[\mso(5)]^{(\ad)})  \,, \nn\\
 2\, \cI &= \theta (- C^2[\mso(5)]^{(5)\otimes(\ad)} + C^2[\mso(5)]^{(\ad)} + 
C^2[\mso(5)]^{(5)}) \, .
 \label{box-Casimirs}
\end{align}
Here $M_{ab}^{(5)}$ is the  vector generator of $\mso(5)$, 
and $M_{bc}^{(\ad)} = \{\cM_{bc},\cdot \}$ denotes the representation of 
$\mso(5)$ on $\cS^4$
induced by the Poisson structure on $\cS^4$, cf.\ \eqref{Poisson-theta}.
This gives 
\begin{align}
 - \cD^2 = \theta(C^2[\mso(6)]^{(\ad)} -  C^2[\mso(5)]^{(5)\otimes(\ad)} + 4) 
\end{align}
with $C^2[\mso(5)]^{(5)}= 4$.
Now $\cS^4 $ decomposes under $\mso(5) \subset \mso(6)$ as follows:
\begin{align}
 \cS^4 = \bigoplus\limits_{n=0}^\infty (n,0,n)_{\mso(6)}, \qquad (n,0,n) = 
\bigoplus\limits_{s=0}^n (n-s,2s)_{\mso(5)} \,.
\end{align}
Therefore the eigenvalues of  $C^2[\mso(6)]^{(\ad)}$   acting on $(n',2s) 
\subset (n,0,n)$ as
\begin{align}
 C^2[\mso(6)]^{(\ad)} \phi_{(n',2s)} = 2(n'+s)(n'+s+3)\phi_{(n',2s)} 
\end{align}
 depend only on the combination $n'+s$. 
 Thus, the three modes $\cA_{(n,2s)} \subset 
\big((n,2s)\oplus(n-1,2s+2)\oplus(n+1,2s-2)\big)\otimes (1,0)$  
 are degenerate under $\cD^2$, and their ``wavefunctions'' $\phi_{(n',2s)} $ are related by $SO(6)$;
these are the $B,E,F$-type modes discussed above.

The explicit eigenvalues are given in 
Appendix \ref{app:eigenvalues_modes} and in \cite{Steinacker:2015dra}.
$\cD^2$ turns out to be positive except for some zero modes, given by the $(1,2s)$ modes of type $D$.
These are the $\cA^{(2)}$ modes without any explicit $x$ factors. 

As a remark, the fluctuations or gauge fields $\cA$ take values in 
$\Omega^1 \cS^4$. One might be tempted to restrict $\cA$ to $T^* S^4$, i.e.\ 
purely tangential fluctuations, but that is inconsistent as one has to take all 
possible (matrix) fluctuations into account. In other words, one cannot 
eliminate the radial fluctuations in the (Poisson) matrix model 
\eqref{bosonic-action}.
\section{Metric and graviton}
\label{sec:metric_and_graviton}
Assuming an action of the above type \eqref{bosonic-action}, we can identify 
the effective metric and its linearized fluctuation along the lines of 
\cite{Steinacker:2010rh}, and decompose it into the above spin modes.
As always, the metric is encoded in the kinematics of the fluctuation modes on 
the background $y^a$ in the action. Their kinetic term arises from the bi-vector 
field
\begin{align}
 \g = \{y^c,\cdot\}\{y_c,\cdot\} = \g^{\mu\nu}\del_\mu\del_\nu \, ,
\end{align}
up to possibly a conformal factor.
We can accordingly identify a ''metric tensor`` (in $SO(5)$ notation) on 
$\cS^4$ via 
\begin{align}
 \g^{ab}(x) \coloneqq  \{y^c,x^a\}\{y_c,x^b\} \, .
\end{align}
This is indeed a tangential tensor field on $S^4$, since
\begin{align}
 \g^{ab} x_a = 0 \,,
\end{align}
due to the radial constraint \eqref{XX-R-const}.
For the unperturbed $\cS^4$, 
\begin{align}
 \obar \g \coloneqq \{x^c,\cdot\}\{x_c,\cdot\} = \frac{L_{NC}^4}{4} 
g^{\mu\nu}\del_\mu\del_\nu
 \label{obar-gamma}
\end{align}
is the round $S^4$ background metric. 
For a deformed background 
\begin{align}
 y^a = x^a + \cA^a(x)
 \label{eq:fluctuation_background}
\end{align}
the metric is perturbed, $\g = \obar \g + \d_\cA \g + O(\cA^2)$, with linearized metric perturbation
\begin{align}
 \d_\cA \g = \{x^a,\cdot\}\{\cA_a,\cdot\} \  +  \  ( \leftrightarrow ) 
\eqqcolon H[\cA] \ .
\end{align}
Explicitly, the corresponding metric fluctuation tensor  is
\begin{align}
 H^{ab}[\cA] &\coloneqq  \{x^c,x^a\}\{\cA_c,x^b\} + (a \leftrightarrow b)  \nn\\
  &= \theta^{ca}\{\cA_c,x^b\} + \theta^{cb}\{\cA_c,x^a\} \,,
  \label{gravitons-1}
\end{align}
which is tangential
\begin{align}
  H^{ab} x_b = 0 \ .
\end{align}
Hence, $H^{ab}[\cA]$ defines an $SO(5)$ intertwiner from $\Omega^1 \cS^4$ to 
tangential symmetric 2-tensor fields.
At low energies, only the $\cC^0$ modes of the energy-momentum tensor $T_{\mu\nu}$ are relevant, 
so that only the average metric is important, i.e.\ the projection 
\begin{align}
\boxed{
 h^{ab} \coloneqq \frac{4}{L_{NC}^4}  [H^{ab}]_0  \ \ \in (5)\otimes (5) \otimes \cC^0 
 }
\label{eq:def_graviton}
\end{align}
cf.\ \cite{Steinacker:2016vgf}. This will be called \textbf{graviton} in this 
paper.
The normalization factor is chosen consistent with \eqref{obar-gamma}, such 
that the full effective metric is 
\begin{align}
 \g^{\mu\nu} = g^{\mu\nu} + h^{\mu\nu} \; .
\end{align}
Then by construction, $h^{\mu\nu}$ couples to matter via its 
energy-momentum tensor\footnote{This defines the normalization of $T_{\mu\nu}$. 
The conformal factor is a bit tricky, for a discussion see  \cite{Steinacker:2016vgf}.},
\begin{align}
 \d_h S_{\rm matter} 
     = \frac 12 \int_{S^4} d^4 x\, h^{\mu\nu} \, T_{\mu\nu} \, .
 \label{def:matter-T-coupling}
\end{align}

\subsection{Gravitons and eigenmodes}
We can rewrite $H_{ab}[\cA]$ by means of
\begin{align}
 \theta^{ca}\{\cA_c,x_b\}  
  &= \{\theta^{ca}\cA_c,x_b\} - \theta (\cA_b x^a -  g^{ab} \cA_c x^c)  
\end{align}
as follows:
\begin{align}
 H_{ab}[\cA] &= \{\theta^{ca}\cA_c,x_b\} +  \{\theta^{cb}\cA_c,x_a\}
 - \theta (\cA_b x_a + \cA_a x_b )  + 2\theta g_{ab} (\cA_c x^c)   \; .
\label{gravitons-2}
\end{align}
In particular, the radial fluctuations $\cA_c^{(R)} = x_c\, \phi$ give rise to 
a conformal metric fluctuation
\begin{align}
 H_{ab}[\cA^{(R)}] &= 2\theta R^2\phi \Big(g_{ab}  - R^{-2} x_b x_a \Big)
  = 2\theta R^2\phi\, P_T^{ab} \, ,
\end{align}
while for tangential $\cA^a$, \eqref{gravitons-2} simplifies to
\begin{align}
 H_{ab}[\cA] &= \{\theta^{ca}\cA_c,x_b\} +  \{\theta^{cb}\cA_c,x_a\}
 - \theta (\cA_b x_a + \cA_a x_b ) \ . 
\label{gravitons-3}
\end{align}
We can now elaborate the graviton contributions of the 
different vector fluctuation modes considered in Section 
\ref{sec:vec-harmonics}:

\paragraph{Gravitons for spin 2 fields.}
To gain some intuition, consider first the graviton associated with the spin 2 
mode $\cA^{(1)}$
using the local representation \eqref{A1-local}, i.e.
\begin{align}
 H_{\mu\nu}[\cA^{(1)}] &= -\frac{L_{NC}^4}{4}\Big(\theta^{\nu \a}\del_\a \A_\mu + \ (\mu \leftrightarrow \nu)  \Big) \nn\\
  &= -R^2\theta\theta^{\nu \a}\del_\a \Big( h_{\mu\r} P^{\r} + \omega_{\mu;\r\s} 
\cM^{\r\s}   \Big)  \ 
  + \ (\mu \leftrightarrow \nu)  \, .
  \label{eq:grav_A1_spin_2}
\end{align}
Upon averaging \eqref{eq:averaging}, the $P^\r$ term drops out\footnote{This is 
precisely the problem with the basic $S^4_N$.
There are extra modes on generalized $S^4_\L$ which survive this step.}, and the leading term is
\begin{align}
 h_{\mu \nu}[\cA^{(1)}] &= \frac{2 R^2}{3} \del^\r\omega_{\mu;\r\nu}  \  + \ 
(\mu \leftrightarrow \nu)  \
     \sim - \frac{2}{3}  n\, \phi_{\mu\nu}  \, , 
  \label{h-A1-local}
\end{align}
using \eqref{h-omega-relation-spin2}, cf.\ \cite{Steinacker:2016vgf}.

We can derive exact expressions for all five modes $ H_{ab}[\cA^{(i)}]$ using 
the Young diagram representations,
which are given in  Appendix \ref{app:metric-fluct}.
The physical gravitons are then obtained by averaging these via \eqref{eq:averaging}. This leads to
\begin{align}
h_{ab}[\cA^{(0)}]&=0 \, ,\\
 h_{ab}[\cA^{(1)}]&=
- \frac{2}{3}  \frac{(n+3)(n+4)}{n+2}  
 \phi^{(2)}_{ab }(x) \, ,\\
 h_{ab}[\cA^{(2)}] &=\frac{4}{15} \frac{(n+3)(n+4)}{n+1}  \frac{1}{R^2}
 \phi^{(2)}_{ab}(x) \, ,\\
 h_{ab}[\cA^{(3)}] &=0 \, ,\\
 h_{ab}[\cA^{(R)}] &=0 \, ,
 \label{hab-modes}
\end{align}
which for large $n$ consistently reduces to the local derivation 
\eqref{h-A1-local}.
The main feature is the factor $n$ in $h_{ab} \sim n \phi_{ab}$, which arises from the 
derivative contributions  $\del^\a \omega_{\mu;\a\b}$ in \eqref{h-A1-local}.
This will imply that the quadratic action translates into  $h^{ab}h_{ab}$,
rather than $h^{ab}\Box h_{ab}$.
As explained in Section \ref{sec:outlook_generalised_S4}, we expect that this 
problem does not arise
for the generalized fuzzy sphere, due to extra momentum-type generators $t^a$ 
which are non-vanishing as functions.
This would lead to $h_{ab} \sim \phi_{ab}$
without factor $n$, hence to gravity at the classical level.

The gravitons of the $\cI$ eigenmodes \eqref{eigenmodes-spin-2} then read 
\begin{subequations}
\label{eq:phys_gravitons_spin-2}
\begin{align}
   h_{ab}^{(B)}[\phi^B]    
&= - \frac{2}{5}    \frac{(n+4)(n+5)}{n+2}   \frac{1}{\theta}
 \phi^B_{ab}(x) \ \approx \ 
 - \frac{2}{5}  \frac{1}{\theta} n \phi^B_{ab}(x) \;, \\
%
%
 h_{ab}^{(C)}[\phi^C] &=
-  \frac{4}{5}  \frac{(n+3)(n+4)(n+5)(n+7)}{(n+1)(n+2)(2n+7)} 
\frac{1}{\theta}
 \phi^C_{ab}(x)  \ \approx \
 - \frac{2}{5}  \frac{1}{\theta} n \phi^C_{ab}(x)
\;, \\
%
%
h_{ab}^{(D)}[\phi^D] &= \frac{4}{15} 
\frac{(n+3)(n+4)}{n+1} \frac{1}{\theta} \phi^D_{ab}(x)  \ \approx \  
\frac{4}{15}  \frac{1}{\theta} n \phi^D_{ab}(x)  \;,\\
h_{ab}^{(E)}[\phi^E] &=0 \;,\\
h_{ab}^{(F)}[\phi^F] &=0 \; ,
\end{align}
\end{subequations}
and the approximations are valid for large $n$.
Also, observe that all terms have a similar structure, including an explicit 
factor $n$.

\paragraph{Gravitons for spin $1$ fields.}
The gravitons for the spin $1$ modes \eqref{spin-1-modes} read as follows:
\begin{align}
 h_{ab}[A^{(0)}] &= -\frac{(n+1) }{3}  
 \left(\phi_{a_1 \ldots a_n a;b } + \phi_{a_1 \ldots a_n b;a } \right)
 x^{a_1} \ldots x^{a_n}
 +\frac{n}{3 R^2}  \left( x^a A_b^{(0)} +x^b A_a^{(0)} \right)  \; , \nn\\
h_{ab}[A^{(1)}] &=0 \; ,\nn\\
h_{ab}[A^{(2)}] &= 0 \; , \nn\\
h_{ab}[A^{(3)}] &= - \frac{(n+2)^2 }{3n}
\left( \phi_{a_1 \ldots a_n a;b } + \phi_{a_1 \ldots a_n b;a } \right)
x^{a_1} \ldots x^{a_n} +  \frac{(n+2)^2}{3(n+1) R^2}  \left(x^a A_b^{(0)} +x^b 
A_a^{(0)} 
\right) \; , \nn \\
h_{ab}[A^{(R)}] &=0 \; .
\end{align}
We observe that all spin $1$ gravitons are traceless and tangential.
One can rewrite the non-trivial modes as follows:
\begin{subequations}
\label{spin-1-graviton}
\begin{align}
 h_{ab}[A^{(0)}] &= -\frac{1 }{3} (\nabla_a \phi^{(1)}_b + \nabla_b 
\phi^{(1)}_a)   \; ,\\
 h_{ab}[A^{(3)}] &= -\frac{1}{3} \frac{(n+2)^2}{n(n+1)}(\nabla_a  \phi^{(1)}_b + 
\nabla_b \phi^{(1)}_a)   \; ,
\end{align}
\end{subequations}
which is recognized as pure gauge contribution to the graviton.
Here $\phi^{(1)}_a$ is the canonical vector field associated to the Young 
diagram,
and $\nabla$ is defined in \eqref{nabla-proper-def}.
In particular, the graviton contribution of the spin 1 pure gauge modes 
$\cQ(\phi^{(1)})$, cf.\ \eqref{pure-gauge-1-decomp-2}, is 
\begin{align}
 h_{ab}[\cQ(\phi^{(1)}] &=  
 -\frac{\theta  }{3} \frac{(n+2)(n +3)}{n+1}
 \big(\nabla_a  \phi^{(1)}_b + \nabla_b \phi^{(1)}_a\big)  \; .
\end{align}
\paragraph{Gravitons for spin $0$ fields.}
Finally for the spin $0$ modes \eqref{spin-0-A-modes},
we obtain the graviton contributions
\begin{subequations}
\begin{align}
 h_{ab}[A^{(2)}] 
 &=  - \frac{2}{3} (n-1)  \Big( \phi_{a b a_3 \ldots a_n} x^{a_3} \ldots 
x^{a_n} 
 - \frac{1}{R^2}\left( x^a A_b^{(0)} + x^b A_a^{(0)} \right) +\frac{g^{ab} 
\phi^{(0)}}{R^2} 
\Big)  \, ,\\
 h_{ab}[\cJ A^{(2)}] &= 0 \; , \\
  h_{ab}[A^{(R)}] &= 2  P_T^{ab} \phi^{(0)} \; .
\end{align}
\end{subequations}
The first mode can be written as 
\begin{align}
 h_{ab} [A^{(2)}] &= - \frac{2}{3n}  \big(\nabla_a\nabla_b \phi^{(0)} 
   +  \frac{n^2}{R^2}P^{ab}\phi^{(0)}\big) \; ,
\end{align}
which encodes the pure gauge graviton associated to the vector field $\del_a \phi^{(0)}$.
Similarly, $h_{ab}[A^{(R)}]$ is the conformal metric contribution.
Note that in the Einstein-Hilbert action, the conformal mode suffers from an instability,
see for instance \cite{Wetterich:1997bz}. There is no such instability in the 
present action.

\subsection{Spin \texorpdfstring{$2$}{2} action and equations of motion}
We want to understand the physics of the spin $2$ modes in the presence of 
matter. Before embarking on the detailed computation, we should have some idea 
of what to expect. For example, the quadratic action for 
$h^{\mu\nu}[\cA^{(1)}]$ is obtained from 
\eqref{h-A1-local} and \eqref{inner-prod-A0-spin1-asympt}
approximately as
\begin{subequations}
\label{eq:eff_action_graviton}
\begin{align}
 \frac{1}{g^2} \int d\Omega\, \cA^{(1)}_\mu \cD^2 \cA^{(1)}_\mu  
 &\approx  \frac 1{3g^2  \theta} \frac{L_{NC}^4}{4}\int d\Omega\,  
\phi^{(2)}_{\mu\nu} \, \theta n^2 \phi^{(2)}_{\mu\nu} 
   \approx   \frac {3L_{NC}^4}{4 g^2}\frac{\dim\cH}{\vol(\cS^4)}\int_{S^4} 
h^{\mu\nu} h_{\mu\nu} .
\end{align}
Combined with the coupling to matter \eqref{def:matter-T-coupling} 
\begin{align}
 \d_h S_{\rm matter} 
     = \frac 12 \int_{S^4} d^4 x\, h^{\mu\nu} \, T_{\mu\nu} \,,
\end{align}
\end{subequations}
we arrive at an equation of motion of the form 
\begin{align}
  h_{\mu\nu}[\cA^{(1)}] \ \approx \  - \frac {3 g^2}{4}  \frac{\vol(\cS^4)}{L_{NC}^4\dim\cH} \, T_{\mu\nu} \ .
  \label{h-A1-approx}
\end{align}
This means that $h_{\mu\nu}$ behaves like a non-propagating 
auxiliary field, rather than a graviton.
However, we need to take the mixing between the different modes $\cA^{(i)}$ into 
account; this 
is taken care of by using the eigenbasis $\widetilde{\cB}_I^{(s)}$. We can then solve exactly the quadratic action governing the spin 2 
sector. This will exhibit an interesting sub-leading behavior, and by, taking 
account of possible induced gravity terms from quantum corrections, it might 
even acquire the appropriate behavior of gravity.

Now we derive the precise equations of motion.
Consider the action of the vector fluctuations $\cA$ in the original matrix 
model in the semi-classical limit,
\begin{align}
 S= \frac{1}{g^2} \int\limits_{\cS^4} d\Omega \left( \cA \cD^2 \cA \right)
 \label{eq:action_vector-modes}
\end{align}
where $\cD^2$ has been defined in \eqref{fluct-action}.
The fluctuation $\cA$ can be expanded in the five (or three) independent spin 
$s$ fields for $s\geq1$ (or $s=0$) as follows:
\begin{align}
 \cA= \sum_{s\geq 0} \left( \cA^{(0,s)} +\frac{1}{\theta}\cA^{(1,s)} +R^2 
\cA^{(2,s)} + \cA^{(3,s)}+ \cA^{(R,s)}   \right)  
\equiv
\sum_{s\geq 0}
\sum_i \widetilde{\cA}^{(i,s)}
\; ,
\end{align}
where $\{\widetilde{\cA}^{(i,s)}\}_i= \{ \frac{1}{\theta}\cA^{(1,s)}, 
\cA^{(R,s)},
R^2 \cA^{(2,s)} ,  \cA^{(0,s)},  \cA^{(3,s)} \} $. Recall that for $s=0$ the 
modes $\cA^{(0,s=0)}$ and $\cA^{(1,s=0)}$ are absent, see 
\eqref{eq:cI_eigenmodes_s=0}. The modes $\widetilde{\cA}^{(i,s)}$ 
have a uniform dimension, unlike the modes $\cA^{(i,s)}$ of \eqref{eq:modes_spin-s}.
At each spin $s$, the transformation matrix 
\eqref{eq:trafo_spin_s_pt1}, \eqref{eq:trafo_spin_s_pt2} for the basis change 
into the 
$\cI$-eigenmodes $\{\cB_I^{(s)} \}_I \equiv \{ B^{(s)}, C^{(s)}, D^{(s)}, 
E^{(s)}, F^{(s)}\}$ can be cast into the form
\begin{align}
 \cB_I^{(s)}  = \sum_i M_{I i}^{(s)} \widetilde{\cA}^{(i,s)}
 \quad  \text{with} \quad
 \cD^2 \cB_I^{(s)} 
 = \theta \lambda^{(s)}_{I} \cB_I^{(s)} \ .
\end{align}
Again, $s=0$ has only a rank 3 transformations 
matrix, and the modes $B^{(0)}$, $E^{(0)}$ are 
absent.
Inserting this into the action \eqref{eq:action_vector-modes}, we obtain 
\begin{align}
 S&= 
\frac{1}{g^2} \sum_{s,s'} \sum_{i,j}\int\limits_{\cS^4} d\Omega  \left( \widetilde{\cA}^{(i,s)} \cD^2 
\widetilde{\cA}^{(j,s')} \right) \nn \\
&=\frac{1}{g^2} \sum_{s,s'} \sum_{I,J} 
\underbrace{\int\limits_{\cS^4} d\Omega \left( 
 \cB_I^{(s)} \cD^2  \cB_J^{(s')}
 \right)}_{\propto \ \theta \lambda^{(s)}_I \delta_{IJ} \delta_{s s'}}
 \sum_{i,j} (M^{-1})_{iI}^{(s)} (M^{-1})_{jJ}^{(s')}  \nn\\
&=\frac{1}{g^2} \sum_{s,I}
 \bigg( \underbrace{\sum_{i} (M^{-1})_{iI}^{(s)}}_{\equiv N_I^{(s)}} \bigg)^2
 \int\limits_{\cS^4} d\Omega \left( 
 \cB_I^{(s)} \cD^2  \cB_I^{(s)}
 \right) \; .
\end{align}
The $\cI$-eigenmodes $\cB_I^{(s)}$ can be
canonically normalized  by
absorbing the normalizations $N_I^{(s)}$ into the fluctuations via
\begin{align}
 \cB_I^{(s)} \mapsto \widetilde{\cB}_I^{(s)} \coloneqq N_I^{(s)}\cB_I^{(s)} \; .
\end{align}
Then the action reads
\begin{align}
  S= \frac{1}{g^2}  \sum_{s,I} \int\limits_{\cS^4} d\Omega \left( \widetilde{\cB}_I^{(s)} \cD^2 
\widetilde{\cB}_I^{(s)}\right) \; .
\label{eq:action_quadratic_part}
\end{align}
Focusing on the spin $s=2$ sector, we can  evaluate the action in the 
semi-classical limit as follows:
 \begin{align}
 S_{|s=2}&= \frac{1}{g^2} \sum_I \int\limits_{\cS^4} d\Omega \left( \widetilde{\cB}_I^{(2)} \cD^2 
\widetilde{\cB}_I^{(2)}\right)
= \frac{1}{g^2} \sum_I \theta \lambda^{(2)}_{I} \  \int\limits_{\cS^4} d\Omega \left( 
\widetilde{\cB}_I^{(2)} \widetilde{\cB}_I^{(2)}\right) \nn\\
 &\sim \frac{\dim(\cH)}{g^2 \ \vol(\cS^4)} \sum_{I}  
\theta \lambda^{(2)}_{I} \ K^{I} 
\int \widetilde{\phi}^{I}_{ab} \widetilde{\phi}^{I}_{ab}  \; .
\end{align}
We used the inner product \eqref{eq:inner_prod_I-eigenmodes-spin2} 
and absorbed the normalizations $N_I^{(2)}$ of the $\widetilde{\cB}_I^{(2)}$ 
into $\widetilde{\phi}^{I}_{ab}\coloneqq N_I^{(2)} \phi^{I}_{ab}$, 
so that \eqref{eq:inner_prod_new_basis} turns into
\begin{align}
 \int \widetilde{\cB}_I^{(s)} \widetilde{\cB}_J^{(s)} = K^{I}  
\delta_{I J} \int \widetilde{\phi}^{I}_{ab} \widetilde{\phi}^{I}_{ab}  \; ,
\end{align}
where all $K^I$ are order one. The eigenvalues 
$\lambda_I^{(2)}$ are given in
Appendix \ref{app:eigenvalues_modes} and read
\begin{align}
  \lambda^{(2)}_{B_n} &=  \lambda^{(2)}_{E_n} =  
\lambda^{(2)}_{F_n} = n(n+3) + 4(n+2) \, , \nn\\
  \lambda^{(2)}_{C_n} &=  \lambda^{(2)}_{D_{n+4}} =  (n+3)(n+8) \,. 
\end{align}
The appearing degeneracy has been explained in Section 
\ref{sec:vector-Laplacian}.

Now consider the coupling \eqref{def:matter-T-coupling} of the spin 2 modes to 
matter
\begin{subequations}
\label{matter-T-coupling}
\begin{align}
 \d_h S_{\rm matter}  &= \frac{1}{2} \int_{S^4} \, 
h^{\mu\nu}[\cA] \, T_{\mu\nu}  \; .
\end{align}
Since the gravitons depend linearly on the modes \eqref{eq:def_graviton}, we can write
\begin{align}
 \d_h S_{\rm matter}  &=  \sum_i \frac{1}{2} \int_{S^4} \, 
h^{\mu\nu}[\widetilde{\cA}^{(s,i)}] \, T_{\mu\nu} 
=  \sum_I \underbrace{\sum_i (M^{-1})_{i,I}}_{=N_I^{(s)}} \frac{1}{2} \int_{S^4} \, 
h^{\mu\nu}[\cB^{(s)}_I] \, T_{\mu\nu} \nn \\
&= \sum_I \frac{1}{2}  \int_{S^4} \, 
h^{\mu\nu}[\widetilde{\cB}^{(s)}_I] \, T_{\mu\nu} \ .
\end{align}
\end{subequations}
Restricting to $s=2$
we observe that the form of \eqref{eq:phys_gravitons_spin-2} remains untouched, 
i.e.
\begin{subequations}
\begin{align}
   h_{ab}^{(B)}[\widetilde{\phi}^B]    
&= - \frac{2}{5}  \frac{(n+4)(n+5)}{\theta (n+2)} \widetilde{\phi}^B_{ab}(x)
\; ,\\
%
%
 h_{ab}^{(C)}[\widetilde{\phi}^C] &=
 -  \frac{4}{5}    \frac{(n+3)(n+4)(n+5)(n+7)}{\theta (n+1)(n+2)(2n+7)} 
\widetilde{\phi}^C_{ab}(x) 
\; ,\\
%
%
h_{ab}^{(D)}[\widetilde{\phi}^D_{ab}] &= \frac{4}{15}  \frac{(n+3)(n+4)}{\theta 
(n+1)}  
\widetilde{\phi}^D_{ab}(x)  
\; .
\end{align}
\end{subequations}
Using the short hand notation $h_{ab}[\widetilde{\cB}_I^{(2)}] = \Xi^I 
\widetilde{\phi}^I_{ab}$, the equations of motion can be compactly written as
\begin{align}
 h_{ab}^{(I)}[\widetilde{\phi}^{I}] \ = \ - \frac{(\Xi^I)^2}{4\theta 
\lambda^{(2)}_I K^I} \  \frac{g^2\ \vol(\cS^4)}{\dim(\cH)} \ T_{ab} \qquad \text{for 
} I \in\{B,C,D\} \; .
\end{align}
The  coefficient in front of the energy momentum tensor is non-negative.
Since $\l \approx n^2$
this leads to $ \phi_{\mu\nu} \sim \frac 1n T$, which is somewhat strange and non-local.
The induced metric fluctuations are explicitly
\begin{align}
  h^{(B)}_{\mu\nu}  [\widetilde{\phi}^B]
    &= - \frac{4}{L_{NC}^4} \frac{g^2\ \vol(\cS^4)}{\dim(\cH)} \ \frac{1}{5}  
\left( 1+ \frac{2}{ n^2+7 n+8}   \right) \, T_{\mu\nu} \,, \nn \\
 h^{(C)}_{\mu\nu}[\widetilde{\phi}^C]
  &= - \frac{4}{L_{NC}^4} \frac{g^2\ \vol(\cS^4)}{\dim(\cH)} \ \frac{3}{5} 
\left( 1
  + \frac{7}{3 (2 n+7)}
  -\frac{2}{3 (n+8)} \right) 
\, T_{\mu\nu} \,, \nn \\
  h^{(D)}_{\mu\nu}  [\widetilde{\phi}^D]
    &= - \frac{4}{L_{NC}^4}  \frac{g^2\ \vol(\cS^4)}{\dim(\cH)} \ \frac{1}{15}
 \left( 1 +\frac{2}{3 (n-1)} -\frac{7}{3 (2 n+7)}  \right) \, T_{\mu\nu} \,,
\label{metric-T-eom}
\end{align}
which results in the total metric fluctuation
\begin{align}
 h_{\mu\nu} &= \sum_{I=B,C,D} h^{(I)}_{\mu\nu}[\widetilde{\phi}^I]
\\
 &= - \frac{4}{L_{NC}^4}   \frac{g^2\ \vol(\cS^4)}{\dim(\cH)} \ \frac{1}{45}
\left(39 -\frac{18}{n+8}+\frac{56}{2 n+7}+\frac{2}{n-1} 
 +\frac{18}{n(n+3)+4(n+2)}\right)T_{\mu\nu} \ . \nn
\end{align}
The leading contribution to this equation of motion is\footnote{Since we focus on the spin 2 sector, only the
traceless part of $T^{\mu\nu}$ enters here. }
\begin{align}
 h^{\mu\nu}_0 = - \frac{13}{15}  \frac{4}{L_{NC}^4}   \frac{g^2\ 
\vol(\cS^4)}{\dim(\cH)} \  \ T_{\mu\nu}  \ ,
\end{align}
which agrees with \eqref{h-A1-approx}.
This is a  non-propagating metric perturbation localized at the matter source,
consistent with \cite{Steinacker:2016vgf}.

\subsection{Flat limit}
To understand the meaning of the above results \eqref{metric-T-eom}, we focus on 
a region near the north pole $p$ of $S^4$,
and assume that the radius $R$ is much larger than any other relevant length 
scale.
We can then relate the kinetic parameters on $S^4$ to ordinary momenta on the tangential $\R^4$,
using the tangential coordinates $x^\mu, \ \mu = 1,\ldots,4$.
It is easy to see e.g.\ from \eqref{obar-gamma} that the Laplace operator 
becomes 
\cite{Steinacker:2016vgf}
\begin{align}
 \Box \sim \frac{L_{NC}^4}{4} g^{\mu\nu}\del_\mu \del_\nu \eqqcolon 
\frac{L_{NC}^4}{4} \Box_g      
\end{align}
neglecting curvature contributions $\sim \frac{1}{R}$. Recalling $ \frac{L_{NC}^4}{4} = R^2 r^2$
and
$\Box \approx - \theta\, n^2$ on $(n,0)$ modes, we can identify 
\begin{align}
 \frac{L_{NC}^4}{4} \Box_g  \ \cong \ - \theta \ n^2, \qquad - n^2 = R^2\,  \Box_g 
\end{align}
for modes with $n \gg 1$.
Now consider first the $\boldsymbol{B}$ \textbf{mode} of \eqref{metric-T-eom}, 
which satisfies 
\begin{align}
h^{(B)}_{\mu\nu}[\widetilde{\phi}^B] = \k \left(-1+\frac{1}{R^2 \Box_g }\right) 
T_{\mu\nu}, 
\end{align}
for some constant $\k = \frac{4}{5 L_{NC}^4} 
\frac{g^2 \ \vol(\cS^4)}{\dim(\cH)}$. Clearly the two source terms lead to two 
contributions
\begin{align}
h^{(B)}_{\mu\nu}[\widetilde{\phi}^B] 
=  h_{\mu\nu}^{(B,\mathrm{loc})} +  h_{\mu\nu}^{(B,\mathrm{grav})}\, ,
\end{align}
where 
\begin{align}
\label{eq:eom_modeB_flat}
\begin{aligned}
 \Box_g   h_{\mu\nu}^{(B,\mathrm{grav})} &= \frac{\k}{R^2} T_{\mu\nu} \ \equiv\ 
G_N T_{\mu\nu}  \; ,  \\
   h_{\mu\nu}^{(B,\mathrm{loc})} &= -\k T_{\mu\nu} \ \equiv \  - G_N R^2  
T_{\mu\nu} 
\; .
\end{aligned}
\end{align}
The first (sub-leading) term $h_{\mu\nu}^{(B,\mathrm{grav})}$ has indeed the structure of linearized gravity, and 
we have tentatively identified $\k = 16\pi G_N R^2$, such that\footnote{note 
that $g^2$ has dimension $L^4$}
$G_N = \frac{4}{5 R^2 L_{NC}^4} \frac{g^2 \ \vol(\cS^4)}{\dim(\cH)} \ \eqqcolon 
\ L_{pl}^2 $.
However, the local contribution $h_{\mu\nu}^{(B,\mathrm{loc})}$ is then too large: assuming $R \approx 10^{27} m$ 
($= 10^{11}$ ly) we would have $L_{pl}^2 R^2 \approx 10^{-16}\ m^4$,
which results in $h_{\mu\nu}^{(B,\mathrm{loc})} \gg 1$ even in the presence of 
modest energy densities, which is clearly unacceptable.
Hence, although the present mechanism does lead to a propagating graviton 
contribution, it is not realistic in the current form.
The most promising way to obtain more realistic gravity is by replacing $\cS^4$ 
with the generalized fuzzy sphere $S^4_\L$ 
\cite{Steinacker:2016vgf,Sperling:2017dts}.
Then extra spin 2 contributions arise, which avoid the 
dominant derivative contributions. This will be discussed briefly in Section \ref{sec:outlook}.
In the presence of an induced gravity term, this problem is also ameliorated, as we will see.

Now consider the $\boldsymbol{C}$ \textbf{and} $\boldsymbol{D}$ \textbf{modes} 
in \eqref{metric-T-eom}, which lead to 
\begin{align}
 h^{(I)}_{\mu\nu}[\widetilde{\phi}^I] = 
 - \k \left(1+\frac{1}{R \sqrt{|\Box_g| }}\right) T_{\mu\nu}, \quad \text{for } 
I=C,D 
\; . 
\end{align}
This leads again to two contributions
\begin{align}
h^{(I)}_{\mu\nu}[\widetilde{\phi}^I] =  h_{\mu\nu}^{(I,\mathrm{loc})} +  
h_{\mu\nu}^{(I,\mathrm{nonloc})} \; ,
\end{align}
where 
\begin{align}
\label{eq:eom_modeCD_flat}
\begin{aligned}
 h_{\mu\nu}^{(I,\mathrm{loc})} &= -\k T_{\mu\nu} = -G_N R^2  T_{\mu\nu} \,, \\
 \sqrt{|\Box_g| }\  h_{\mu\nu}^{(I,\mathrm{nonloc})} &= -\frac{\k}{R}  
T_{\mu\nu} \ = \ - G_N \, R \, T_{\mu\nu}  \, .
\end{aligned}
\end{align}
The relevant Green's functions are computed in Appendix \ref{App:propagators} in 
the flat limit $R\to\infty$.
To have a meaningful comparison between the various contribution, we assume  a 
Gaussian mass distribution with variance $a^2$. The results are summarized as follows:

\begin{enumerate}  
 \item The Green's function for the $B$ mode $h_{\mu\nu}^{(B,\mathrm{grav})}$ 
\eqref{eq:eom_modeB_flat}
 is $\sim 1\slash |k^2|$ in momentum space, which leads to  the standard 
$\frac{1}{r^2}$  behavior in $4$-dimensional space. In the presence of the 
Gaussian source term, this gives
\begin{align}
h_{\mu\nu}^{(B,\mathrm{grav})} = G_N \frac{1-e^{-\frac{r^2}{2 a^2}}}{4 \pi 
^2 r^2} \; ,
\end{align}
for any components $\mu\nu$. In the limit $a\to0$, the source reduces to a
$4$-dimensional Delta-distribution, and we recover the usual $\frac{1}{4\pi^2 
r^2}$ Green's function.
  \item The Green's function for the $C,D$ modes 
$h_{\mu\nu}^{(I,\mathrm{nonloc})}$, in \eqref{eq:eom_modeCD_flat},
  is $\sim 1\slash |\vec{k}|$ in momentum space,
 which for the same Gaussian source gives
 \begin{align}
h_{\mu\nu}^{(I,\mathrm{nonloc})}
&= \frac{G_N \,R}{8 \sqrt{2} \pi ^{3/2} a^3}e^{-\frac{r^2}{4 a^2}} 
\left(I_0\left(\frac{r^2}{4 a^2}\right)-I_1\left(\frac{r^2}{4 
a^2}\right)\right) \; ,
\end{align}
 for any $\mu\nu$, and $I_n(z)$ denotes the modified Bessel 
functions of first kind.
The  asymptotic behavior for small and large radial distances $r$ is
\begin{align}
h_{\mu\nu}^{(I,\mathrm{nonloc})} &=
\begin{cases}
 \frac{ G_N R}{8 \sqrt{2} \pi^{3/2} a^3}   e^{-\frac{r^2}{4 a^2}} \ \left(1 
-\frac{r^2}{8a^2}\right) &, 0 < r \ll 2a \; ,\\
 \frac{G_N R}{4 \pi ^2 r^3} \left( 1
 +\frac{3 a^2}{  r^2} 
 +\frac{95 a^4}{16  r^4} + \ldots \right)
&, \frac{r}{2a}\to \infty \; .
 \end{cases}
 \label{propagator-n-longdist}
\end{align}
\item All the fluctuations exhibit one mode $h_{\mu\nu}^{(I,\mathrm{loc})}$, 
for $I=B,C,D$, which has an algebraic equation of motion.
By the same analysis as in the previous cases, one finds
\begin{align}
 h_{\mu\nu}^{(I,\mathrm{loc})}= \frac{G_N \, R^2}{4 \pi ^2 a^4} 
e^{-\frac{r^2}{2 a^2}} \;,
 \label{eq:propagator_loc_flat}
\end{align}
for the Gaussian source, for any $\mu\nu$.
\end{enumerate}
Equipped with these results, we can consider two regimes:

\paragraph{Small $\boldsymbol{a}$.}
This is the regime where the observer is far from the source.
Then the mode $h_{\mu\nu}^{(B,\mathrm{grav})}$ behaves indeed like a
graviton. 
However, $h_{\mu\nu}^{(I,\mathrm{nonloc})}$ has a leading contribution $\sim 
\frac{1}{4 \pi^2 r^2} \frac{R}{r}$, which by far dominates over the graviton 
$h_{\mu\nu}^{(B,\mathrm{grav})}$. 
Sufficiently far from the source, the ''local`` term $h_{\mu\nu}^{(I,\mathrm{loc})}$ is exponentially 
suppressed and therefore irrelevant.

In summary, the basic fuzzy 4-sphere does not 
lead to a physical gravity for localized matter distributions, but it does lead 
to a long-range   $\sim \frac{R}{4 \pi^2 r^3}$  ''gravity``.
\paragraph{Large $\boldsymbol{a}$.}
This regime applies to observers within  mass distributions that are spread 
out over large scales $a$ (such as dust), but  still smaller than the extend of 
the entire space,  $a\ll R$.
Now the local contribution $h_{\mu\nu}^{(I,\mathrm{loc})}$ for $I=B,C,D$ is 
enhanced by $R^2 \slash a^2$ compared to $h_{\mu\nu}^{(B,\mathrm{grav})}$. 
Similarly,  $h_{\mu\nu}^{(I,\mathrm{nonloc})}$, 
for $I=C,D$ is enhanced by $R \slash a$. On the other hand, 
$h_{\mu\nu}^{(B,\mathrm{grav})}$ is of order 
$ \frac{G_N}{8\pi^2a^2}$, but otherwise still exhibits a graviton-like 
behavior.
Nevertheless, the ''local`` and ''non-local`` modes dominate the graviton-like 
mode, which renders the scenario unphysical.
\subsubsection{Induced gravity  effects}
\label{subsec:induced_gravity}
Now we want to take the leading quantum effects into account.
Integrating out any fields that couple to a background metric, one generically 
obtains induced gravity terms 
in the effective action, such as 
$\int d^4 x \L^2 R[\g]$. Here $\L^2$ is the effective cutoff, presumably determined by the SUSY breaking scale.
This would lead to a quadratic term $\int h_{ab} \Box h_{ab}$ in the action
(there cannot be any linear contribution for the traceless modes). There may be a 
linear contribution to the conformal factor, which we ignore here; 
this is basically the question whether the background is stable, which relies on quantum effects anyway \cite{Steinacker:2015dra}.
Hence the cosmological constant problem is tantamount to the issue of stability of the background.

Let us therefore add such an induced kinetic term to the action. 
For the $\boldsymbol{B}$ \textbf{mode}, the equation of motion becomes
\begin{align}
\label{eq:eom_modeB_induced}
 \left(1+ \sigma\bar R^2 \Box_g \right) (h^{(B)}[\widetilde{\phi}^{B}])_{\mu\nu} 
 = -\k \left(1+\frac{1}{-R^2 \Box_g }\right) T_{\mu\nu},
\end{align}
where $\bar{R}$
is the effective cutoff scale for induced gravity\footnote{$\bar R$ is some combination of the scale $\L$, which 
could e.g. be the scale of SUSY breaking, and the other scales in the model such as $R$ and $L_{\rm NC}$.}, 
and $\sigma = \pm 1$ depending on the precise field content of the model.
Now the two source terms lead to two contributions
\begin{align}
\label{eq:split_modeB}
h^{(B)}_{\mu\nu}[\widetilde{\phi}^{B}] =  h_{\mu\nu}^{(B,\mathrm{loc})} +  
h_{\mu\nu}^{(B,\mathrm{grav})} \, ,
\end{align}
where 
\begin{align}
\label{eq:eom_modeB_split}
\begin{aligned}
 \left(\sigma\Box_g  + \frac{1}{\bar{R}^2} \right) h_{\mu\nu}^{(B,\mathrm{loc})} &= 
 - G_N \frac{R^2}{\bar R^2} T_{\mu\nu} \,,  \\
  \left(\sigma\Box_g  + \frac{1}{\bar{R}^2} \right) \Box_g  
h_{\mu\nu}^{(B,\mathrm{grav})} &= G_N  \frac{1}{\bar{R}^2} T_{\mu\nu} \, .
\end{aligned}
\end{align}
For $\sigma=-1$,  $h_{\mu\nu}^{(B,\mathrm{loc})}$ behaves like a massive spin 2 graviton with mass 
$1\slash \bar{R}^2$, 
with appropriate coupling to matter by $\frac{R^2}{\bar R^2} G_N T_{\mu\nu}$. For $\s=1$, it would be a tachyonic mode, which we discard.

For the $\boldsymbol{C,D}$ \textbf{modes}, the equation of motion becomes
\begin{align}
\label{eq:eom_modeCD_induced}
 \left(1+ \s\bar R^2 \Box_g \right) (h^{(I)}[\widetilde{\phi}^{I}])_{\mu\nu} = 
-\k 
\left(1+\frac{1}{R \sqrt{|\Box_g| }} \right) T_{\mu\nu}, 
\qquad \text{for } I=C,D \; .
\end{align}
This leads to 
\begin{align}
\label{eq:split_modeCD}
h^{(I)}_{\mu\nu}[\widetilde{\phi}^{I}]=  h_{\mu\nu}^{(I,\mathrm{loc})} +  
h_{\mu\nu}^{(I,\mathrm{nonloc})} \, ,
\end{align}
where 
\begin{align}
\label{eq:eom_modeCD_split}
\begin{aligned}
 \left(\s\Box_g  + \frac{1}{\bar{R}^2} \right) h_{\mu\nu}^{(I,\mathrm{loc})} &= 
-\frac{1}{\bar{R}^2}\k T_{\mu\nu} = - G_N \frac{R^2}{\bar{R}^2}  T_{\mu\nu} \, 
,\\
 \left(\s\Box_g  + \frac{1}{\bar{R}^2} \right) \sqrt{\Box_g } 
h_{\mu\nu}^{(I,\mathrm{nonloc})} &= -G_N \frac{R}{\bar{R}^2}  T_{\mu\nu} \, .
\end{aligned}
\end{align}
Again for $\s=-1$, the first term $h_{\mu\nu}^{(I,\mathrm{loc})}$ behaves like a massive graviton with 
mass $1\slash \bar{R}^2$. We therefore assume $\s=-1$ from now on.

The induced gravity terms can have quite different implications depending on 
the 
scale of $\bar R$:
For small $\bar{R}$, one recovers effectively the same physics as described by 
\eqref{eq:eom_modeB_flat}, \eqref{eq:eom_modeCD_flat}.
However for large $\bar{R}$, the quantum corrections changes the nature of 
$h_{\mu \nu}^{\mathrm{loc}}$, $h_{\mu \nu}^{\mathrm{nonloc}}, h_{\mu 
\nu}^{\mathrm{grav}}$, such that $h_{\mu \nu}^{\mathrm{loc}}$ behaves like a 
graviton. In any case,
we note that the UV contribution  is suppressed, as the mass term acts 
as a UV regulator. Hence  we no longer need to assume a Gaussian source as above.
In Appendix \ref{App:induced_gravity} we compute the Green's functions for the 
PDEs in \eqref{eq:eom_modeB_split} and \eqref{eq:eom_modeCD_split}. With these 
results we find  the following behavior for the two scaling regimes:
\paragraph{$\bar R$ small.}
In this regime, $\bar R$ plays a similar role as $a$ in the previous section.
 $h_{\mu\nu}^{(B,\mathrm{grav})}$  behaves like gravity for large 
distances $r \gg \bar R$, while
$h_{\mu\nu}^{(I,\mathrm{loc})}$ (for $I=B,C,D$) is exponentially 
decaying at a scale $\bar R^{-1}$ away from matter, see 
\eqref{eq:Green_local_part} for $r\slash \bar{R} \to \infty$. 
However  within  uniformly distributed matter, 
it behaves the same as before, i.e.\ $h_{\mu\nu}^{(I,\mathrm{loc})}\sim - G_N R^2 
T_{\mu\nu}$ for $I=B,C,D$.
This is consistent with the behavior \eqref{eq:propagator_loc_flat}, but it is again too large and
the situation is not improved.

The remaining mode $h_{\mu\nu}^{(I,\mathrm{nonloc})}$ exhibits essentially the 
same 
long-distance behavior  as in \eqref{propagator-n-longdist} (see \eqref{eq:Green_nonloc_part} for $r \slash \bar{R} 
\to \infty$), with 
leading long-distance term $h_{\mu\nu}^{(I,\mathrm{nonloc})} \sim -\frac 1{r^3}$ and higher-order terms suppressed by the scale factor $\bar R$ (replacing $a$).
Again due to the explicit $R$ in the source, this is too large to be physically 
acceptable for $\k=G_N R$.

Hence the small $\bar R$ case
confirms the result that the basic fuzzy $4$-sphere does not lead to physical 
gravity,
and the spin 2 fluctuation modes of type $B,C,D$ should be viewed 
as auxiliary (or very massive) fields.
Somewhat remarkably, we have found that there are nevertheless sub-leading 
long-distance  interactions, with a large $-\frac{R}{r^3}$  contribution and a 
subdominant contribution which behaves like usual gravity $\frac{1}{r^2}$.

\paragraph{$\bar R$ large.}
Now consider the opposite limit where the induced gravity term $\bar R$ is so large that
the bare ''mass`` term in \eqref{eq:eom_modeB_split} can be neglected. Then the 
modes $h_{\mu\nu}^{(I,\mathrm{loc})}$ for any 
$I=B,C,D$, 
can indeed play the role of a physical graviton, see \eqref{eq:Green_local_part} for 
$0<r\slash \bar{R}  \ll 1$.
In this case, $h_{\mu\nu}^{(I,\mathrm{nonloc})}$ for $I=C,D$ and 
$h_{\mu\nu}^{(B,\mathrm{grav})}$ (being now misnomers) would lead to a sub-leading
very-long-range interactions somewhat reminiscent of conformal gravity, as apparent from \eqref{eq:Green_nonloc_part} 
and \eqref{eq:Green_grav_part} for $0<r\slash \bar{R} \ll 1$. 
Hence this scaling regime might indeed lead to interesting 
gravitational physics, as long as the graviton mass is sufficiently small and $\s=-1$.

\section{Symmetries and gauge transformations}
\label{sec:symmetries}
\subsection{Global symmetries}
The 5-dimensional matrix model \eqref{bosonic-action} has a global $ISO(5)$ 
symmetry.
Consider first the  $SO(5)$ symmetry. 
Since the $X^a$ and $\Theta^{ab}$ are tensor operators, the $SO(5)$ action on 
them  
\begin{align} 
 X^a &\to \L_a^b \, X^b =   U_\L X_a U_\L^{-1}
\end{align} 
(and similarly for $\Theta^{ab}$)
is equivalent to a gauge transformation.
In this sense, the background is ``covariant''. 
This implies that the vector 
modes $\A_a$ \eqref{A-tang-radial} transform indeed as vectors,
\begin{align}
 \A_a &\to  \L_a^b \, U_\L \A_b U_\L^{-1} .
\end{align}
It also implies that the corresponding rotational zero modes (Goldstone bosons) 
are unphysical.
However there is a tower of non-trivial zero modes, which arises from the 
$(1,2s) \subset (1,0)\otimes (0,2s)$ modes in \eqref{mode-decomp-2}, 
cf.\ \cite{Steinacker:2015dra}.
For $s=0$ these are
the translations $X^a \to X^a + c^a$ corresponding to the 5 zero modes in $(1,0) \otimes (0,0)$
in \eqref{mode-decomp-m=0}. For  $s\geq 1$ they are associated with higher spin 
symmetries.

\subsection{Gauge transformations of functions on 
\texorpdfstring{$\cS^4$}{SS4}}
\label{sec:gauge_trafo_functions}
Now suppose $\phi\in \cC$. Gauge transformations act on $\phi$ as
\begin{align}
 \phi \to U^{-1} \phi U, \qquad \d_\L\phi = \{\phi,\L\} \, .
\end{align}
This is simply the action of symplectomorphisms on $\C P^3$.
We can make this more transparent by viewing $\L$ as $\hs$-valued function on $S^4$,
\begin{align}
 \L = \L_{\und\a}(x)\, \Xi^{\und\a} \ .
 \label{gauge-HS}
\end{align}
Hence, gauge transformations are local (i.e.\ $x$-dependent) versions of the 
$\hs$ algebra
acting on fields on $S^4$. 
We discuss a few aspects of these transformations.

\paragraph{$\cC^0$ gauge transformations.}
Gauge transformations generated by $\L(x) \in \cC^0$ act on functions on $S^4$ via
\begin{align}
 \d_\L x^a  &=\{x^a,\L\} = \cQ(\L^{(1)}) = \theta^{a\mu}\del_\mu \L(x)  \qquad \in \cC^1 \ .
\end{align}
This is no longer a function, but a spin 1 field, and it
has the typical form of gauge transformation in noncommutative Yang-Mills gauge fields. 
Hence these transformations
are naturally interpreted as local $U(1)$ gauge transformations\footnote{Since 
$[\theta^{\mu\nu}]_0 = 0$ here, these are \emph{not} related to diffeomorphisms 
on $S^4$, in contrast to e.g.\ gauge theory on the Moyal-Weyl quantum plane 
$\R^4_\theta$ \cite{Steinacker:2010rh}.}
of noncommutative gauge theory on $S^4_N$.
There is no constant counterpart of that symmetry.

\paragraph{$\cC^1$ gauge transformations.}
Now consider 
gauge transformations generated by $\L = \L^{(1)}_{ab}(x) \theta^{ab} \in \cC^1$:
\begin{align}
 \d_{\L} x^a &= \{x^a,\L^{(1)}\} =  \L^{(1)}_{bc}(x)\{x^a, \theta^{bc}\} + \{x^a,\L^{(1)}_{bc}(x)\} \theta^{bc}  \nn\\
 &= \ \cQ(\L^{(1)}) \ \ = \ \  \theta \frac{n+2}{n+1} \cA^{(0)} \ \  +  \ \ \theta n\cA^{(3)}  
   \qquad \in \cC^0 \oplus \cC^2
   \label{spin-1-gaugetrafo-explicit}
\end{align}
using \eqref{pure-gauge-1-decomp-2}.
This contains in particular the global $SO(5)$ symmetry. 
We recall from Section \ref{sec:scalar-fields} 
that $\L \in \cC^1$ is characterized by a divergence-free vector field, which is  explicit
in the local representation \eqref{phi-1-local} 
\begin{align}
 \L = \L^{(1)}_{ab}(x) \theta^{ab} = v_\mu(x) P^\mu + \omega_{\mu\nu}(x)\cM^{\mu\nu}  .
\end{align}
Here $\omega_{\mu\nu}$ is the field strength of the  divergence-free vector field $v_\mu$. Then
\eqref{nabla-theta} gives\footnote{This is strictly speaking 
valid only at the north pole $p$, since \eqref{nabla-theta} holds only at $p$. 
However since $p$ is arbitrary, there is no loss of generality.}
\begin{align}
 \d_{\L} x^\mu &=  - v^\mu  \ + 
 \big(\del_\nu v_\r P^\r + \del_\nu\omega_{\r\s}\cM^{\r\s}  \big)\theta^{\mu\nu} 
  \quad \in \ \ \cC^0 \ + \ \cC^2
\end{align}
noting that $\{P_\mu,x^\nu\} = \d_\mu^\nu$ and $\{\cM^{\mu\nu},x^a\} = 0$ at the north pole.
Hence the non-derivative contribution in $\cC^0$ is a vector field $v$, which
can be interpreted as  diffeomorphism on $S^4$. This in turn determines  the 
 $\cC^2$ contribution, which
 involves derivatives of $v$, and will be recognized as gauge transformation of 
spin 2 gauge fields (gravitons) under diffeomorphisms.
Thus the $\cC^1$ gauge transformations amount to  (volume-preserving) 
diffeomorphisms, which act both on the $S^4$ space and the gauge fields on it.

Note that the $\cC^1$ gauge transformations are only a subset of the
local $SO(5)$ gauge transformations, given by the  group of area-preserving diffeomorphisms. 
The full local $SO(5)$ arises only on the generalized $\cS^4_\L$, which was 
one reason for introducing it in \cite{Steinacker:2016vgf}.

\paragraph{$\cC^s$ gauge transformations.}
Gauge transformations generated by $\cC^s$ act similarly as
\begin{align}
 \d_{\L} x^a &= \{x^a,\L^{(s)}\} =  \cQ(\L^{(s)})  \qquad \in \cC^{s-1} \oplus \cC^{s+1} \ .
\end{align}
This includes global transformations with  generators in the higher spin algebra
$\hs$ \eqref{HS-def}. 
As before, we expect that the non-derivative $\cC^{s-1}$ contribution can be given a geometrical meaning
on $S^4$ corresponding to some global symmetry, while the 
$\cC^{s+1}$ terms provide the associated (derivative) gauge transformations 
of spin $s+1$ gauge fields.

\subsection{Gauge transformations of gauge fields on 
\texorpdfstring{$\cS^4$}{SS4}}
Now consider an $\cS^4$  background with a generic perturbation $
 y^a = x^a + \cA^a $, as in \eqref{eq:fluctuation_background}.
Gauge transformations act on the background as
\begin{align}
 \d_\L y^a &\coloneqq   \{y^a,\L\}  =  \{x^a + \cA^a,\L\} \eqqcolon  \cD^a \L  
, \qquad 
\L \in \cC \, ,
\end{align}
which can be absorbed by a gauge transformation of the fluctuation by defining
\begin{align}
  \d_\L\cA^a =  \cD^a \L = \cQ(\L) + \{\cA^a,\L\} \ .
\end{align}
The inhomogeneous contribution 
\begin{align}
  \cQ(\L)^a &= \theta^{ab}\del_b \L =: \theta^{ab}\d_\L \A_a  \nn\\
   \d_\L \A_a &=  \del_a \L =  \del_a \big(\L_{\und\a}(x)\Xi^{\und\a}) \ ,
\end{align}
is a gauge transformation of the 
embedding function $x^a \in \cC^0$ as discussed above. 
We have seen that this has  the general form 
$\cQ(\L) \sim \cA^{(0)} + \cA^{(3)}$, where the ``non-derivative'' contribution $\cA^{(0)}$ 
can be interpreted as geometric transformation of $S^4$, and the ``derivative'' contribution  $\cA^{(3)}$
can be interpreted as pure gauge contribution to  $\A$.
E.g.\ for the  spin 1 gauge transformations \eqref{spin-1-gaugetrafo-explicit}, 
the $\cA^{(0)}$ contribution  is  the vector field $v_\mu$ corresponding to a (volume-preserving) 
diffeomorphism, while the $\cA^{(3)}$ contribution is combined with the spin 2 
gauge field $\cA^{(1)}$, see \eqref{A1-local},
and leads to the pure gauge part of a spin 2 gravitons  $h_{\mu\nu} = \del_\mu 
v_\nu + \del_\nu v_\mu$, see \eqref{spin-1-graviton}. 
The $\cA^{(0)}$ contribution is absorbed in the trace part of $\A$. 
A similar discussion should apply for the higher spin case, leading to
the Fronsdal form of gauge transformations of rank $s$ 
tensor fields  via the appropriate  
identifications\footnote{One can  associate a symmetric rank $s$ tensor field on $S^4$ to $\cA$ generalizing 
$h_{\mu\nu}$, which then transforms 
as in the Fronsdal form. An explicit elaboration is postponed to future work.}.

Now separate  $\cA^a=\theta^{ab} \A_b + x^a \phi$ into tangential  gauge fields and 
a transversal scalar field as in \eqref{A-tang-radial},
and consider the homogeneous contribution to its gauge transformation
\begin{align}
 \{\cA^a,\L\} &= \theta^{ab}\{\A_b, \L\} + \A_b \{ \theta^{ab}, \L\} 
  + \{x^a,\L\} \phi + x^a \{\phi,\L\} \ .
\end{align}
Re-grouping and combining with $\cQ(\L)$,  we can write the non-linear gauge transformation as
\begin{align}
 \d_\L\cA^a =  \cD^a \L 
  = \theta^{ab} D_b \L + x^a \d_\L \phi + A_b\d_\L \theta^{ab} 
  \label{gaugetrafo-A}
\end{align}
where
\begin{align} 
  \d_\L (\cdot) = \{\cdot, \L\}, \qquad  D_a = (1+\phi) \del_a  + \{ 
A_a,\cdot \} \ .
\end{align} 
We can organize the perturbations and the gauge parameter  in terms of $\hs$ 
generators \eqref{gauge-HS}:
\begin{align}
 \L = \L_{\und\a}(x)\, \Xi^{\und\a}\, , \qquad  
  \A_b =  A_{b,\und{\b}}(x)\, \Xi^{\und\b} \, ,\qquad  
  \phi = \phi_{\und{\b}}(x)\, \Xi^{\und\b} \, .
\end{align}
This suggests to view $\A_\mu$ as $\hs$-valued gauge field on $\cS^4$, with 
Yang-Mills-type gauge transformations $\d \A_\mu = D_\mu \L$,
\begin{align}
 \{A_\mu, \L\} =   A_{\mu,\und{\a}}(x) \L_{\und\b} \{\Xi^{\und\a},\Xi^{\und\b}\} 
 + \L_{\und\b}\{A_{\mu,\und{\a}}(x),\Xi^{\und\b}\}\Xi^{\und\a} + \ldots  .
 \label{A-L-trafo}
\end{align}
The first term has the usual form of a nonabelian gauge theory;
however the remaining terms are non-standard.
Similar non-standard terms arise\footnote{Another approach to gravity with similar 
non-standard actions of the generators on fields 
was recently discussed in \cite{Ho:2015bia}.} 
in \eqref{gaugetrafo-A}. These originate from  the action $\d_\L (.)$ of 
a local $\hs$ transformation on 
 $x^a$ and  on $\theta^{ab}$, which is precisely what allowed us to understand the inhomogeneous contributions
from $\cQ(\L)$ discussed above. 
Hence the present model does not coincide with a standard Yang-Mills 
formulation of $\hs$ gravity
(and with Vasiliev theory, to our understanding). However, the above features are 
crucial in the matrix model realization of $\hs$ gauge theory.

\paragraph{Field strength.}
As in noncommutative gauge theory, 
the gauge-covariant field strength arise from commutators or Poisson brackets
\begin{align}
  \{y^a,y^b\} = \{x^a+\cA^a,x^b+\cA^b\} = \theta^{ab} + \cF^{ab} \,,
 \end{align} 
which transforms in the adjoint of the gauge group.
Dropping the radial fluctuations for simplicity, i.e.\ $\cA^a x_a=0$, we obtain
 \begin{align} 
  \cF^{ab} &= \theta^{ac} \del_c \cA^b -  \theta^{bc} \del_c \cA^a 
+\{\cA^a,\cA^b\} 
\end{align}
and, by further employing that also $A_bx^b=0$ holds, we find   
\begin{subequations}
\begin{align}
 \cF^{ab} &= \theta^{ab}\ \theta A_c A^c + \theta^{ac}\theta^{bd} F_{cd}
+\theta \left( A_a (x^b + \theta^{bc} A_c) - A_b (x^a + \theta^{ac} A_c) 
\right) \\
&\qquad +\theta^{ac} \{A_c,\theta^{bd} \} A_d - \theta^{bc} \{A_c,\theta^{ad} 
\} A_d \nn \, ,\\
  F_{ab} &\coloneqq \del_{a} A_{b} -  \del_{b} A_{a} \ +\{A_a,A_b\}   \;.
  \label{flux-perturbed} 
\end{align}
\end{subequations}
By evaluating $\cF^{ab}$ at a point on the 4-sphere we can shed some light on 
its structure and make contact to conventional noncommutative Yang-Mills theory:
\begin{align}
 \cF^{\mu \nu} 
&=  \theta^{\mu \rho}\theta^{\nu \sigma} \left( F_{\rho \sigma}
+ \frac{4\theta }{L_{NC}^4}\left(
\theta^{\rho 
\sigma}  A_\gamma A^\gamma 
+\theta^{\gamma \rho} A_\gamma  A_\sigma 
-   \theta^{\gamma \sigma} A_\gamma A_\rho 
\right) \right)
\label{cF-full}
\end{align}
by means of $A_\mu = P_T^{\mu \rho} A_\rho  = \theta^{\mu 
\sigma} \  \frac{4}{L_{NC}^4}  \theta^{\rho \sigma} A_\rho $ and $\theta^{\mu 
\nu} = \theta^{\mu \rho} P_T^{\rho \nu} = \theta^{\mu \rho} \theta^{\nu 
\sigma} \ \frac{4}{L^4_{NC}} \theta^{\rho \sigma}$.
We observe that the corrections to the ``usual'' field strength 
$F_{\rho \sigma}$ are suppressed by the factor $4\theta \slash L_{NC}^2= r \slash R$,
hence the set-up resembles a Yang-Mills field strength.
However, despite the conventional appearance, some unusual terms arise  in $F_{\mu\nu}$,
because 
the $\mso(5)$ generators do act on the functions and derivatives. For example,
$\del^\mu P^\nu = \theta^{\mu\nu}$ arises from $P$-valued gauge fields.
These extra terms
\footnote{The fully non-linear case can
probably be made more transparent if some of the fluctuations are absorbed  in $\theta^{ab}$.
This will be discussed elsewhere.}  are responsible for the present mechanism for gravity,
cf.\ \cite{Steinacker:2016vgf}.

In any case, the quantities $\{y^a,y^b\}$ or $\cF^{ab}$ transform in the adjoint under 
 gauge transformations, and are therefore natural building blocks for 
a higher spin gauge theory. 
The action \eqref{bosonic-action} under consideration in the present paper is  
of that type. 
Of course other, more complicated terms are conceivable, such as 
$\int \{y^a,y^b\}\{y^c,y^d\} y^e \varepsilon_{abcde}$ cf.\ \cite{Kimura:2002nq}.

\section{Remarks and outlook}
\label{sec:outlook}
We briefly comment on some interesting aspects and open questions which we have to set aside for the 
moment, to keep the paper within bounds.
\subsection{The generalized fuzzy sphere}
\label{sec:outlook_generalised_S4}

As remarked earlier, the shortcomings of the gravitons 
\eqref{eq:phys_gravitons_spin-2} on the basic fuzzy 4-sphere may be overcome by 
considering the generalized fuzzy 4-sphere $S^4_\L$ as background instead of 
$S^4_N$.  
Following the geometric description presented in 
\cite{Steinacker:2016vgf,Sperling:2017dts}, the extended bundle structure leads to 
additional generators $t^b$ in the algebra functions such that new 
fluctuation modes arise. A particular promising mode is given by
\begin{align}
 \A_a = \phi_{c_1\ldots c_n;ab} x^{c_1} \ldots  x^{c_n} t^b = h_{ab}(x) t^b \; .
 \label{extra-modes-generalized}
\end{align}
The new feature is that derivative contributions to $h_{ab}$ are suppressed, 
in contrast to \eqref{eq:grav_A1_spin_2} -- \eqref{h-A1-local} for the basic 
4-sphere. More specifically, there is no accompanying 
``spin connection'' term $\omega_{a;\mu\nu} \cM^{\mu\nu}$ which would spoil its 
contribution. There is no such mode on $S^4_N$.

Moreover, there should be similar Goldstone bosons arising from the generalized background
\begin{align}
 Y^A = \begin{pmatrix}
        X^a \\ T^a
       \end{pmatrix}
\end{align}
found in \cite{Sperling:2017dts}, i.e.\  $X^a \to X^a + \L^{ab} T_b$.
These are physical (in contrast to the $SO(5)$ would-be Goldstone bosons), 
massless, and are closely related to the above gravitons.

\subsection{Decoupling and interactions of higher spin fields}
For the intents and purposes of this paper, we restricted the analysis of
fluctuation modes to the quadratic order. Nevertheless, the higher order terms in the action
will of course lead to interactions between the various higher spin modes, and should be 
considered in more detail. 

We focused on the spin 2 modes in this paper, which couple to other fields via the 
energy-momentum tensor \eqref{def:matter-T-coupling}. 
This involves two derivatives such as in $\{\cA,\phi(x)\}\{X,\phi(x)\}$, which
is suppressed by some dimensionful scale parameter (identified with the 
Newton constant). It is crucial that there is indeed a natural UV scale $L_{\rm NC}$
on the fuzzy sphere which can serve this purpose; that seems to be a major 
advantage over Vasiliev's higher spin theory.

Similarly, all spin $s$ fluctuation modes can interact with scalar fields only via $s$ derivatives,
or with spin $l$ fields via $s-l$ derivatives. Therefore the interactions of all higher spin fields 
should be suppressed via appropriate powers of that UV scale parameter,
see \eqref{eq:Star-product}. 
In contrast, the scale of higher derivative interactions in Vasiliev 
theory is given by the cosmological constant, which is the only available scale 
there. However, 
here we have two scales at our disposal: the UV and the cosmological scale.
It is therefore quite plausible that these higher spin fields decouple at low 
energies, as they should.

Nonetheless, we cannot make any definite statements about the higher spin 
fields at this stage. The framework presented has the intriguing feature of 
formulating a higher spin gauge theory in the presence of a UV scale 
$L_{\rm{NC}}$, which naturally defines energies below $L_{\rm{NC}}$ 
as ``low energies''. However, it is not at all clear 
whether the higher spin fields are massive or massless, i.e.\ whether they are 
part of the propagating low energy degrees of freedom or not. 
We have exemplified this in the spin 2 sector, whose physical behavior 
is more intricate than naively expected. 
To address this question more systematically one might need to refine the 
current framework.

There are also non-derivative interactions among the higher spin fields, which arise from their
structure as $\hs$-valued nonabelian Yang-Mills theory, cf. \eqref{cF-full}. It 
would be desirable to 
gain some insights into their significance and their relation to Vasiliev's formulation.

\section{Conclusion}
We have shown that the fuzzy 4-sphere $S^4_N$ and its semi-classical limit $\cS^4$
naturally carry an tower of higher spin fields, which is finite or infinite 
respectively.
We have provided an explicit classification of the scalar and vector 
fluctuation 
modes in Section \ref{sec:scalar-fields} and \ref{sec:vec-harmonics}, 
respectively.
The resulting kinematics is very similar to that of 
Vasiliev theory, including 
a structure which plays the role of a local higher spin algebra $\hs$, as discussed 
in Section \ref{sec:relation_HS}. 
However, there are some differences compared to Vasiliev theory.
The most distinctive new feature is the presence of an intrinsic UV scale $L_{\rm 
NC}$, which is the scale where the underlying noncommutativity becomes 
significant. This scale plays a crucial role throughout.

Moreover, it turns out that matrix models and their semi-classical limit as 
``Poisson models'' provide a natural formulation of higher spin 
gauge theory on $\cS^4$. In particular, the Yang-Mills matrix model action 
\eqref{bosonic-action} provides an action for interacting higher spin gauge fields,
which can be arranged in terms of a tangential 1-form and a scalar field taking values in $\hs$, 
as in Vasiliev theory. However, the resulting dynamics appears to be different. This is found by
explicitly diagonalizing the quadratic part \eqref{eq:action_quadratic_part} of 
the resulting action.

It is natural to expect that the spin 2 sector of a higher spin theory 
contains gravity.  We have checked this explicitly using the  classification of  fluctuation modes,
which allowed for the identification of the physical gravitions 
\eqref{eq:phys_gravitons_spin-2} together with their  effective 
action \eqref{eq:eff_action_graviton} including matter coupling. 
We found that although one of the three graviton modes mediates linearized Einstein gravity, the 
remaining modes are dominant, and behave as auxiliary non-propagating or short-range fields. 
This confirms the preliminary result in \cite{Steinacker:2016vgf} that  
Yang-Mills matrix models on the basic fuzzy 
4-sphere do not provide a relativistic version of gravity, at least at the classical level.
However,  induced 
gravity  due to one-loop effects might reverse this conclusion, by transmuting the
auxiliary spin 2 field into a realistic graviton,
as discussed in Section \ref{subsec:induced_gravity}. In that case, the remaining modes become, 
for a suitable parameter range, subdominant long-range phenomena.

In order to judge the physical relevance of the higher spin theory formulated 
by the Poisson matrix model, one would have to perform a full fledged analysis 
of all higher spin modes analogously to our considerations of the spin 2 modes. 
Only this would allow conclusive statements about the on-shell degrees of 
freedom and their relation to the low energy region set by $\L_{\mathrm{NC}}$. 
We have to leave this to future research.

In any case, the present paper provides the necessary tools to 
address the more promising generalized fuzzy sphere $S^4_\L$.
Following the preliminary analysis in \cite{Steinacker:2016vgf} and supported by the geometric 
results in \cite{Sperling:2017dts}, we expect that this background will  
lead to the linearized Einstein equations, due to extra modes \eq{extra-modes-generalized}. 
These should provide the required properties for physical gravity 
as sketched in Section \ref{sec:outlook_generalised_S4},  
along with an interesting non-abelian Yang-Mills gauge theory.
We hope to report on this  in the near future.

\paragraph{Acknowledgments.} Useful discussions with  S.\ Fredenhagen,  K.\ 
Mkrtchyan
and E.\ Skvortsov are gratefully acknowledged. 
HS would like to thank in particular S.\ Ramgoolam for an ongoing collaboration 
and discussions on
closely related topics.
This work was supported by the Austrian Science
Fund (FWF) grant P28590, and by the Action MP1405 QSPACE from the European
Cooperation in Science and Technology (COST).

\appendix
\section{Details}
\subsection{Explicit derivations}
\label{sec:app-proofs}
Equation \eqref{I-J-identity} is seen as follows:
\begin{align}
  \cI \circ\cJ (\xi^a A_a) &= \xi^a\{\theta^{ab},\theta_{bc} A^c\}   \nn\\
   &= \xi^a\big(-3\theta\theta^{ac} A^c + \{\theta^{ab}\theta_{bc},A^c\} - 
\theta^{ab}\{\theta_{bc},A^c\}  \big)\nn\\
    &= \xi^a\big(-3\theta\theta^{ac} A^c - \theta R^2 \{P^T_{ac},A^c\} \big) - 
\cJ\circ \cI (\xi^a A_a)  \nn\\  
  &= - 3 \theta\cJ(\xi^a A_a) + \theta\xi^a\{x_a x_c,A^c\}   -  \cJ\circ \cI 
(\cA) \nn\\  
  &= - 3\theta \cJ(\cA) +\theta \xi \{ x_c,A^c\} +\theta \xi^a\{x_a , x_c A^c\} 
- \theta \xi^a\theta_{ac} A^c   -  \cJ\circ \cI (\cA) \nn\\  
  &= - 4 \theta\cJ(\cA) +\theta \xi \{ x_c,A^c\} +\theta \cQ(\cN(\cA)) -  
\cJ\circ \cI (\cA) \, .
  \label{I-J-identity-proof}
\end{align}
\paragraph{Details spin $2$ fields.}
To derive the first identity in \eqref{spin2-deriv-contract-id}, consider 
\begin{align}
 [\{x^a,\{x^c, \phi^{(2)}_{cb}\}\}]_0 &= 
 - (n+2) \phi^{(2)}_{a_1 \ldots a_nb_1b_2;c b} x^{a_1} \ldots  x^{a_n} 
\{x^a,\theta^{b_1c}\}  x^{b_2} \nn\\
 & \quad - (n+2)(n+1) \phi^{(2)}_{a_1 \ldots a_nb_1b_2;c b} x^{a_1} \ldots  
x^{a_n}  [\theta^{ab_2} \theta^{b_1c}]_0 \nn\\ 
  &= (n+2)\theta  \phi^{(2)}_{a_1 \ldots a_n a b_2; cb} x^{a_1} \ldots  
x^{a_n}x^c x^{b_2} 
   - (n+2) \theta \phi^{(2)}_{a_1 \ldots a_n b_1b_2; ab} x^{a_1} \ldots  
x^{a_n}x^{b_1} x^{b_2} \nn\\
  & \quad - \frac 13 (n+2)(n+1) \theta \phi^{(2)}_{a_1 \ldots a_nb_1b_2; cb} 
x^{a_1} \ldots  x^{a_n} 
     (- P_T^{ab_1}x^{b_2}x^c + P_T^{ac}x^{b_2}x^{b_1}) \nn\\ 
 &= (n+2)\theta  \phi^{(2)}_{a_1 \ldots a_nab_2; cb} x^{a_1} \ldots  x^{a_n}x^c 
x^{b_2} 
   - (n+2)\theta  \phi^{(2)}_{ab} \\
  & \quad + \frac 13 (n+2)(n+1) \theta \phi^{(2)}_{a_1 \ldots a_nb_1b_2; cb} 
x^{a_1} \ldots  x^{a_n} 
     (g^{ab_1}x^{b_2}x^c - g^{ac}x^{b_2}x^{b_1}) \; . \nn 
\end{align}
Hence,
\begin{align}
 k^{ab} [\{x^a,\{x^c, \phi^{(2)}_{cb}\}\}]_0 
  &= \Big(-(n+3) - \frac 13 (n+3)(n+1)\Big)\theta  k^{ab} \phi^{(2)}_{ab}  \nn\\
  &= -\frac 13(n+3)(n+4)\theta  k^{ab} \phi^{(2)}_{ab} 
 \end{align}
 for any symmetric matrix $k^{ab}$,
 using \eqref{graviton-anomalous-relation}.
Thus,
\begin{align}
 \int \{x^a,\phi^{(2)} _{ab}\} \{x^c, \phi^{(2)}_{cb}\}
  &= - \int \phi^{(2)} _{ab} [\{x^a,\{x^c, \phi^{(2)}_{cb}\}\}]_0  
  =  \frac 13(n+3)(n+4)\theta \int \phi^{(2)}_{ab} \phi^{(2)}_{ab} \; .
\end{align}
In order to compute the coefficient $c_n$ in \eqref{potential-gaugefix-id-2}, we 
need to compute $\{x^a, \{x^b , \phi^{(2)}\}\}$. To start with
\begin{align}
\begin{aligned}
  \{x^b , \phi^{(2)}\} = \phi^{(2)}_{a_1 \ldots a_n c e ; d f }
  \Big(& 
  n x^{a_1} \cdots x^{a_{n-1}} \theta^{b a_n} \theta^{cd} \theta^{ef}
  -2 \theta x^{a_1} \cdots x^{a_{n}} x^f g^{eb} \theta^{cd} \\
  &+2 \theta x^{a_1} \cdots x^{a_{n}} x^e g^{fb} 
\theta^{cd}
  \Big) \; ,
\end{aligned}
\end{align}
then we find
\begin{align}
 \{x^a, \{x^b , \phi^{(2)}\}\} =& \
 n(n-1)\phi^{(2)}_{a_1 \ldots a_n c e ; d f } x^{a_1} \cdots x^{a_{n-2}}
\theta^{a a_{n-1}} \theta^{b a_n} \theta^{cd} \theta^{ef} \\
&- n g^{ab} \theta \phi^{(2)}
+n \theta x^b \phi^{(2)}_{a_1 \ldots a_{n-1} a  c e ; d f } x^{a_1} \cdots 
x^{a_{n-1}} \theta^{cd} \theta^{ef} \nn \\
&+2n\theta \big( 
\phi^{(2)}_{a_1 \ldots a_{n}   c e ; d a }
- \phi^{(2)}_{a_1 \ldots a_{n}   c a ; d e }
\big)
x^{a_1} \cdots x^{a_{n-1}} \theta^{b a_n} \theta^{cd} x^e \nn\\
&+2n\theta \big( 
\phi^{(2)}_{a_1 \ldots a_{n}   c e ; d b }
- \phi^{(2)}_{a_1 \ldots a_{n}   c b ; d e }
\big)
x^{a_1} \cdots x^{a_{n-1}} \theta^{a a_n} \theta^{cd} x^e \nn\\
&+2\theta \big( 
\phi^{(2)}_{a_1 \ldots a_{n}   c e ; d b }
- \phi^{(2)}_{a_1 \ldots a_{n}   c b ; d e }
\big)
x^{a_1} \cdots x^{a_{n}} \theta^{a e} \theta^{cd}  \nn\\
&+2 \theta^2 \big( 
\phi^{(2)}_{a_1 \ldots a_{n}   d a ; e b }
-\phi^{(2)}_{a_1 \ldots a_{n}   a d ; e b }
-\phi^{(2)}_{a_1 \ldots a_{n}   d a ; b e }
+\phi^{(2)}_{a_1 \ldots a_{n}   a d ; b e }
\big)
x^{a_1} \cdots x^{a_{n}} x^d x^e .\nn
\end{align}
As consistence check we perform
\begin{align}
 g_{ab}\{x^a, \{x^b , \phi^{(2)}\}\} = -\theta (n^2 +7n +4 ) \phi^{(2)} \; ,
\end{align}
which is the correct result.
Finally, we employ the averaging expressions \eqref{eq:averaging} and obtain
\begin{align}
 [\{x^a, \{x^b , \phi^{(2)}\}\} ]_0 = \frac{2}{15} \theta^2 
\frac{(n+3)(n+4)(n+5)}{n+1} \phi^{(2)}_{ab}
\end{align}
\paragraph{Details spin $s$ fields.}
In order to compute the action of $\cI$ on the spin s fields one needs the 
following results, obtained using \eqref{F-x-xi-id-spin-s}:
\begin{align}
 \cQ(\phi^{(s)}) &= \theta^s n \cA^{(3)} + \theta^s s \frac{n+2}{n+1} \cA^{(0)} 
\\
 \cN(\cA^{(0)}) &= 0 &
\{x^a, \cA^{(0)}_a\} &= -\frac{n+1}{\theta^{s-1}} \phi^{(s)} \\
 \cN(\cA^{(2)}) &= \frac{1}{\theta^s } \phi^{(s)} &
\{x^a, \cA^{(2)}_a\} &= 0  \ .
\end{align}
\subsection{Young tableaux and tensor fields}
\label{app:Young}
We discuss some properties of Young diagrams and projectors. For a useful 
resource in this context we refer to \cite[App.\ E]{Didenko:2014dwa}.
\paragraph{Spin 1.}
Let $\phi^{(1)}_{a_1 \ldots a_nd;e}$ have symmetry corresponding to the diagram 
${\tiny \Yvcentermath1 \young(daaa,e) }$.
Then its total symmetrization in $(a_1 \ldots a_nde)$ vanishes.
Since the symmetry in $(a_1 \ldots a_nd)$ is manifest, this means that 
\begin{align}
0 = P_S^{(n+2)} \phi^{(1)}_{a_1 \ldots a_nd;e} &= \frac 
1{n+2}\Big(\phi^{(1)}_{a_1 \ldots a_nd;e} +  \phi^{(1)}_{a_1 \ldots a_ne;d}
  +\sum_n \phi^{(1)}_{a_1 \ldots a_nde;a_i}\Big) \nn\\
  &=  \frac 1{n+2}\Big(\phi^{(1)}_{a_1 \ldots a_nd;e} +  \phi^{(1)}_{a_1 \ldots 
a_ne;d}
  + n \phi^{(1)}_{a_2 \ldots a_nde;a_1}\Big) \; . \nn
\end{align}
Now contracting this with $x^{a_1} \ldots x^{a_n}\xi^d x^e$ yields
\begin{align}
 0 &= \frac 1{n+2}\Big(\phi^{(1)}_{a_1 \ldots a_nd;e} \xi^d x^e+  
\phi^{(2)}_{a_1 \ldots a_ne;d}\xi^d x^e
  + n \phi^{(1)}_{a_2 \ldots a_nde;a_1}\xi^d x^e\Big) x^{a_1} \ldots x^{a_n}   
\nn\\
   &= \frac 1{n+2}\Big((n+1)\phi^{(1)}_{a_1 \ldots a_nd;e} \xi^d x^e+  
\phi^{(1)}_{a_1 \ldots a_ne;d}\xi^d x^e \Big) x^{a_1} \ldots x^{a_n}   \; .
\end{align}
Hence,
\begin{align}
 (n+1)\phi^{(1)}_{a_1 \ldots a_nd;e}x^{a_1} \ldots x^{a_n}  \xi^d x^e+  
\phi^{(1)}_{a_1 \ldots a_ne;d} x^{a_1} \ldots x^{a_n} x^e \xi^d = 0
 \label{F-x-xi-id}
\end{align}
Similarly, we can deduce by contracting with $x^{a_1} \ldots x^{a_n}$ that
\begin{align}
   \phi^{(1)}_{de a_2 \ldots a_n;a_1}x^{a_1} \ldots x^{a_n} = 
  -\frac{1}{n} \left(
  \phi^{(1)}_{a_1 \ldots a_n d;e} +  \phi^{(1)}_{a_1 \ldots a_n e;d}
  \right) x^{a_1} \ldots x^{a_n} \; .
\end{align}
\paragraph{Spin 2.}
Similarly, let $\phi^{(2)}_{a_1 \ldots a_nbd;ce}$ have symmetries corresponding 
to the diagram  ${\tiny \Yvcentermath1 \young(bdaaa,ce)}$.
Then its total symmetrization in $(a_1 \ldots a_nbdc)$ vanishes\footnote{Because 
that would correspond to a Young diagram with a 
row of length $n+3$ i.e.\ an irrep $(n+3,0)$, which upon tensoring with 
$(1,0)$ cannot give $(n+2,4)$}.
Since the symmetry in $(a_1\ldots a_nbd)$ is manifest, this means that 
\begin{align}
0 = P_S^{(n+3)} \phi^{(2)}_{a_1 \ldots a_nbd;ce} &= \frac 
1{n+3}\Big(\phi^{(2)}_{a_1 \ldots a_nbd;ce} 
+ \phi^{(2)}_{a_1 \ldots a_ncd;be} +  \phi^{(2)}_{a_1 \ldots a_nbc;de}
  +\sum_n \phi^{(2)}_{ca_1 \ldots a_nbd;a_ie}\Big) \nn\\
  &= \frac 1{n+3}\Big(\phi^{(2)}_{a_1 \ldots a_nbd;ce} + \phi^{(2)}_{a_1 \ldots 
a_ncd;be} +  \phi^{(2)}_{a_1 \ldots a_nbc;de}
  +n \phi^{(2)}_{ca_2 \ldots a_nbd;a_1e}\Big)   \, . \nn
\end{align}
Now contracting this with $x^{a_1} \ldots x^{a_n}\xi^b x^c\theta^{de}$ gives 
\begin{align}
 0 
  &= \Big(\phi^{(2)}_{a_1 \ldots a_ncd;be}\xi^b x^c\theta^{de}
  + (n+1) \phi^{(2)}_{ca_2 \ldots a_nbd;a_1e}\xi^b x^c\theta^{de}\Big) x^{a_1} 
\ldots 
x^{a_n} \; ,
  \label{F-x-xi-id-spin-2}
\end{align}
which is the spin 2 version of \eqref{F-x-xi-id}.
Similarly, contracting  this with $x^{a_1} \ldots x^{a_n}x^b x^d k^{ce}$ with 
some symmetric tensor $k^{ce}$ gives 
\begin{align}
  0    
  &=\Big(\phi^{(2)}_{a_1 \ldots a_nbd;ce}x^b x^d k^{ce} 
  + (n+2)\phi^{(2)}_{ca_2 \ldots a_nbd;a_1e}x^b x^d k^{ce}\Big) x^{a_1} \ldots 
x^{a_n} \,.
  \label{graviton-anomalous-relation}
\end{align}
Finally, contracting with  $\theta^{b a_1} \theta^{de}$ 
gives
\begin{align}
 0 &=  \Big(\phi^{(2)}_{a_1 \ldots a_nbd;ce} \theta^{de}
 + \phi^{(2)}_{a_1 \ldots a_ncd;be} \theta^{de}+  \phi^{(2)}_{a_1 \ldots 
a_nbc;de}\theta^{de}
  +n \phi^{(2)}_{ca_2 \ldots a_nbd;a_1e}\theta^{de}\Big) \theta^{b 
a_1}\ \nn\\
 &= \Big(\phi^{(2)}_{c \ldots a_nad;be} \theta^{b a} \theta^{de} 
  +n \phi^{(2)}_{ca_2 \ldots a_nbd;ae} \theta^{b a} \theta^{de}\Big) \; .
\end{align}
Thus,
\begin{align}
 0 &= \phi^{(2)}_{a_1 \ldots a_nad;be}x^{a_1} \ldots x^{a_n} \theta^{b a}\ 
\theta^{de} 
  +n \phi^{(2)}_{a_1 \ldots a_nbd;ae}x^{a_1} \ldots x^{a_n} \theta^{b a} 
\theta^{de}  \nn\\
  0 &= \phi^{(2)}_{\ldots ad;be}(x) \theta^{b a} \theta^{de} 
  +n \phi^{(2)}_{\ldots bd;ae}(x)\theta^{b a} \theta^{de} 
\end{align}
This also implies 
\begin{align}
  \phi^{(2)}_{\ldots ad;be}(x) \theta^{b a} \theta^{de} 
  = \phi^{(2)}_{\ldots ad;be}(x)\theta^{ab}\theta^{ed} 
  = - n \phi^{(2)}_{\ldots bd;ae}(x)\theta^{b a}\theta^{de} 
   = - n \phi^{(2)}_{\ldots bd;ae}(x)\theta^{ab }\theta^{ed} \; .
   \label{Ftheta-theta}
\end{align}
Similarly, one deduces the following identities:
\begin{align}
 \phi^{(2)}_{a_1\ldots a_n  d  e; a b} x^{a_1} \cdots x^{a_n}  x^{b} x^d &= - 
\frac{1}{n+2} \phi^{(2)}_{a_1\ldots a_n d b;ae} x^{a_1} \cdots x^{a_n}  x^{b} 
x^d  \; ,\\
  \phi^{(2)}_{a_1 \ldots a_n a b;cd} x^{a_1}\cdots x^{a_n} x^c x^d 
  &= \frac{-1}{n+1} 
\left(\phi^{(2)}_{a_1 \ldots a_n a c  d;cb}
+ \phi^{(2)}_{a_1 \ldots a_n b b;ca} \right)
x^{a_1}\cdots x^{a_n} x^c x^d \nn \\
&=\frac{2}{(n+1)(n+2)} \phi^{(2)}_{a_1 \ldots a_n c ad ;ab} x^{a_1}\cdots 
x^{a_n} x^c x^d  \; ,
\label{marcus-id} \\
 \{x^d ,\{x_d , \phi^{(2)}_{a_1 \ldots  a_{n+2} ;ab} x^{a_1} \cdots 
x^{a_{n+2}} \} \} &= - \theta (n+2)(n+5) \phi^{(2)}_{a_1 \ldots  a_{n+2} ;ab} 
x^{a_1} \cdots 
x^{a_{n+2}} \; .
\end{align}

\paragraph{Spin s.}
We can derive a spin $s$ identity. Consider the symmetrization $P_S^{n+s-1}$ 
of
\begin{align}
{\footnotesize \Yvcentermath1 \young(12{\ldots}s12{\ldots}n,12{\ldots}s) }
\end{align}
in $a_1\ldots a_n b_1 \ldots b_s; c_1 \ldots c_{s-1}$, which vanishes. Thus, we 
obtain
\begin{align}
 0=& \ (s-1) \phi_{a_1 \ldots a_n b_1 \ldots b_s;c_1 \ldots c_{s-1} d} 
 \label{eq:identity_spin_s}\\
 &+ \phi_{a_1 \ldots a_n c_1 b_2 \ldots b_s;b_1 c_2 \ldots c_{s-1} d}
 + \phi_{a_1 \ldots a_n b_1 c_1 b_3 \ldots b_s;b_2 c_2 \ldots c_{s-1} d}
 +\ldots
 + \phi_{a_1 \ldots a_n b_1 \ldots c_1;b_s c_2 \ldots c_{s-1} d} \nn \\
&+ \phi_{a_1 \ldots a_n c_2 b_2 \ldots b_s;c_1 b_1 c_3 \ldots c_{s-1} d}
 + \phi_{a_1 \ldots a_n b_1 c_2 b_3 \ldots b_s;c_1 b_2 c_3 \ldots c_{s-1} d}
 +\ldots
 + \phi_{a_1 \ldots a_n b_1 \ldots c_2;c_1 b_s c_3 \ldots c_{s-1} d} \nn\\
&+\ldots \nn\\
 &+ \phi_{a_1 \ldots a_n c_{s-1} b_2 \ldots b_s; c_1 \ldots c_{s-2} b_1 d}
 + \phi_{a_1 \ldots a_n b_1 c_{s-1} b_3 \ldots b_s; c_1 \ldots c_{s-2} b_{2} d}
 +\ldots
 + \phi_{a_1 \ldots a_n b_1 \ldots c_{s-1}; c_1 \ldots c_{s-2} b_{s} d} \nn \\
&+\sum_{i=1}^n \phi_{c_1 a_1 \ldots a_n b_1 \ldots b_s; a_i c_2 \ldots c_{s-1} 
d} \nn \\
&+\sum_{i=1}^n \phi_{c_2 a_1 \ldots a_n b_1 \ldots b_s; c_1 a_i c_3 \ldots 
c_{s-1} d} \nn \\
&+\ldots \nn \\
&+\sum_{i=1}^n \phi_{c_{s-1} a_1 \ldots a_n b_1 \ldots b_s; c_1  \ldots 
c_{s-2} a_i  d} \nn
\end{align}
Contracting \eqref{eq:identity_spin_s} with $ \xi^{b_1} x^{a_1} \cdots x^{a_n} 
x^{c_1} \cM^{b_2 c_2 } \cdots \cM^{b_s c_s }$ yields
\begin{align}
 0
=& \ (n+s-1 -(s-2)) \phi_{a_1 \ldots a_n b_1 \ldots b_s;c_1 \ldots c_{s-1} d} 
  \xi^{b_1} x^{a_1} \cdots x^{a_n} 
x^{c_1} \cM^{b_2 c_2 } \cdots \cM^{b_s c_s }
 \nn\\
 &+ \phi_{a_1 \ldots a_n c_1 b_2 \ldots b_s;b_1 c_2 \ldots c_{s-1} d}
  \xi^{b_1} x^{a_1} \cdots x^{a_n} 
x^{c_1} \cM^{b_2 c_2 } \cdots \cM^{b_s c_s } \nn \\
=& \ (n+1) \phi_{a_1 \ldots a_n b_1 \ldots b_s;c_1 \ldots c_{s-1} d} 
  \xi^{b_1} x^{a_1} \cdots x^{a_n} 
x^{c_1} \cM^{b_2 c_2 } \cdots \cM^{b_s c_s }
 \nn\\
 &+ \phi_{a_1 \ldots a_n c_1 b_2 \ldots b_s;b_1 c_2 \ldots c_{s-1} d}
  \xi^{b_1} x^{a_1} \cdots x^{a_n} 
x^{c_1} \cM^{b_2 c_2 } \cdots \cM^{b_s c_s } \; ,
\end{align}
i.e.\ it yields the generalization of \eqref{F-x-xi-id} and
\eqref{F-x-xi-id-spin-2}:
\begin{align}
0= \left( (n+1) \phi_{a_1 \ldots a_n b_1 \ldots b_s;c_1 \ldots c_{s-1} d} 
+ \phi_{a_1 \ldots a_n c_1 b_2 \ldots b_s;b_1 c_2 \ldots c_{s-1} d}
   \right)
 \xi^{b_1} 
x^{c_1} x^{a_1} \cdots x^{a_n}  \cM^{b_2 c_2 } \cdots \cM^{b_s c_s }
 \label{F-x-xi-id-spin-s}
\end{align}
\section{Inner product matrix}
\label{app:inner-prod}
First, \eqref{A0-x-phiab} together with \eqref{spin2-deriv-contract-id} gives
\begin{align*}
 \int \cA^{(0)}[\phi^{(2)}]\cA^{(0)}[\phi^{(2)}] &= \frac{1}{\theta^2} 
\frac{1}{(n+2)^2} \int \{x^b , \phi_{ba}^{(2)}\} \{x^c , \phi_{ca}^{(2)}\}
= \frac{(n+3)(n+4)}{3 \theta (n+2)^2} \int \phi_{ab}^{(2)}\phi_{ab}^{(2)} .
\end{align*}
Similarly, \eqref{Q-F-2} allows to compute
\begin{align}
 \int \cA^{(3)}[\phi^{(2)}]\cA^{(3)}[\phi^{(2)}] 
  &= \frac 1{n^2}\int \big(\frac{1}{\theta^2}\cQ(\phi^{(2)})-2 \frac{n+2}{n+1}\cA^{(0)}\big)
   \big( \frac{1}{\theta^2} \cQ(\phi^{(2)})-2  \frac{n+2}{n+1}\cA^{(0)}\big) \nn\\
  &= \frac 1{n^2}\int \phi^{(2)} \Big(-\frac{1}{\theta^4}\Box 
-4\frac{1}{\theta^3}(n+2)\Big)\phi^{(2)} +  \frac{4}3 \frac{(n+3)(n+4) }{(n+1)^2\theta} \phi^{(2)}_{ab}\phi^{(2)}_{ab}  \nn\\
  &= \frac 2{15} \frac{(n+3)(n+4)(n^2+8n+21)}{n(n+1)^2(n+2)}\frac{1}{\theta} 
\int  \phi^{(2)}_{ab}\phi^{(2)}_{ab}  \; ,
\end{align}
where we used 
\begin{align}
 \int \cQ(\phi^{(2)})\cA^{(0)} &= \int \{x^a,\phi^{(2)}\} A_a^{(0)} = - \int 
\phi^{(2)} \{x^a, A_a^{(0)}\} 
  = \frac{1}{\theta} (n+1) \int \phi^{(2)} \phi^{(2)}
\end{align}
with \eqref{A0-2-gaugecond}, and 
\begin{align}
 \int \cQ(\phi^{(2)}) \cQ(\phi^{(2)}) &=\int \{x^a,\phi^{(2)}\}\{x^a,\phi^{(2)}\}
  = -\int \phi^{(2)} \Box \phi^{(2)}  \nn\\
   &=  \theta ((n+2)(n+5)-6)\int \phi^{(2)} \phi^{(2)} \; .
\end{align}
Finally, separating $\cA^{(2)}$ into normal and tangential parts, we have
\begin{align}
 \int \cA^{(2)}[\phi]\cA^{(2)}[\phi] &=   \frac{1}{R^2}\int x^a A^{(2)}_a[\phi] x^b A^{(2)}_b[\phi] 
 + \frac{1}{\theta R^2}\int \cJ\cA^{(2)} [\phi] \cJ\cA^{(2)} [\phi]\nn\\
  &=  \frac{1}{R^2}\frac{1}{\theta^4}\int \phi \phi 
 + \frac{1}{\theta R^2}\int \cA^{(3)} [\phi] \cA^{(3)} [\phi] \; .
\end{align}
Furthermore,
\begin{align}
 n\int A^{(3)}_b[\phi] A^{(0)}_b[\phi] &=  \int 
\big(\frac{1}{\theta^2} \cQ(\phi^{(2)})-2 \frac{n+2}{n+1}\cA^{(0)}\big)^b 
A^{(0)}_b\nn\\
  &=  (n+1) \frac{1}{\theta^3} \int \phi^{(2)} \phi^{(2)} -2 
\frac{n+2}{n+1} \int\cA^{(0)}_b A^{(0)}_b  \nn\\
 \int A^{(3)}_b[\phi] A^{(0)}_b[\phi] &= \frac{2}{15} \frac 
{(n+4)(n+3)}{(n+1)(n+2)} \frac{1}{\theta} \int \phi _{ab}\phi _{ab} 
\end{align}
Moreover, \eqref{N-A2} gives
\begin{align}
 \int A^{(R)}_b[\phi] A^{(2)}_b[\phi] &= \frac{1}{\theta^4} \int  \phi^{(2)}  
\phi^{(2)} 
\end{align}
and 
\begin{align}
 \int A^{(1)}_b[\phi] A^{(2)}_b[\phi] &=  \int \cJ A^{(0)}_b[\phi] A^{(2)}_b[\phi] = -  \int  A^{(0)}_b[\phi] \cJ A^{(2)}_b[\phi] \nn\\
  &= - \int A^{(0)}_b[\phi]  A^{(3)}_b[\phi] \; .
\end{align}
In addition,
\begin{align}
 \int A^{(R)}_b[\phi] A^{(R)}_b[\phi] &= \frac{1}{\theta^4} R^2 \int  
\phi^{(2)}  \phi^{(2)} 
\end{align}
and 
\begin{align}
 \int A^{(1)}_b[\phi] A^{(1)}_b[\phi]  
&= \theta R^2 \int P_T^{bc} A^{(0)}_b A^{(0)}_c 
= \theta R^2\int A^{(0)}_b[\phi] A^{(0)}_b[\phi] \; .
\end{align}
\section{Metric fluctuations}
\label{app:metric-fluct}
The full metric fluctuations $H_{ab}$ \eqref{gravitons-3} for the modes 
$\cA^{(i)}$ are given by
\begin{align}
 H_{ab}[\cA^{(0)}] &=
 (n+1)(  \theta^{ca} \theta^{b' b} + \theta^{cb} \theta^{b' a} )
 \phi^{(2)}_{a_1 \ldots a_n b'd ce}   x^{a_1} \cdots x^{a_n} \cM^{d e}  \\
&\qquad
 +  \phi^{(2)}_{a_1 \ldots a_nb'b; ce} \theta^{ca}  x^{a_1} \cdots x^{a_n} 
x^{b'}  x^e 
 -  \phi^{(2)}_{a_1 \ldots a_n b'd; cb} \theta^{ca}  x^{a_1} \cdots x^{a_n} 
x^{b'} x^d \nn\\
&\qquad
 +  \phi^{(2)}_{a_1 \ldots a_n b'a ;ce} \theta^{cb}  x^{a_1} \cdots x^{a_n} 
x^{b'}  x^e 
 -  \phi^{(2)}_{a_1 \ldots a_n b'd;ca} \theta^{cb}  x^{a_1} \cdots x^{a_n} 
x^{b'} x^d \nn \,,\\
%
%
 H_{ab}[\cA^{(1)}] &= 
 (n+1) \theta R^2 \left(  \phi^{(2)}_{a_1 \ldots a_nb'd;ae}    
\theta^{b' b} 
  +  \phi^{(2)}_{a_1 \ldots a_nb'  d;be}   \theta^{b' a} \right) 
  x^{a_1} \cdots x^{a_n} \cM^{de} \\
 &\qquad
 +\theta R^2 \bigg( 
 \phi^{(2)}_{a_1 \ldots a_n b' b;ad} 
 + \phi^{(2)}_{a_1 \ldots a_nb'a;bd}
-2 \phi^{(2)}_{a_1 \ldots a_nb'd;ab }
   \bigg) 
x^{a_1} \cdots x^{a_n}  x^{b'} x^d \nn \\
&\qquad - 
\theta \left( \cA_a^{(1)} x_b +  \cA_b^{(1)} x_a \right) \,,  \nn \\
%
%
 H_{ab}[\cA^{(2)}] &= 
 2(n-1)  \theta^{a_1 a} \theta^{a_2 b} \phi^{(2)}_{a_1 a_2 \ldots a_n b' d;c'e}
  x^{a_3} \cdots x^{a_n}  \cM^{b' c'} \cM^{de}    \\
&\qquad -2\frac{n+2}{n+1} 
\left(  \theta^{ca} \phi^{(2)}_{c a_2 \ldots a_n b'e;c'b}
+ \theta^{cb} \phi^{(2)}_{c a_2 \ldots a_nb'e;c'a} \right)
  x^{a_2} \cdots x^{a_n} \cM^{b' c'}   x^e \,, \nn \\
%
%
 H_{ab}[\cA^{(3)}]&= 
 (n-1) \theta R^2 \left(  \theta^{a_2 b}  \phi^{(2)}_{a a_2 \ldots a_n b' d';c' 
e}
 \theta^{a_2 a}  \phi^{(2)}_{b a_2 \ldots a_n b'd' ;c'e} 
\right) x^{a_3} \cdots x^{a_n} \cM^{b' c'} \cM^{d' e}
 \\
&\qquad 
- 2 \frac{n+2}{n+1} \theta R^2 
\left( \phi^{(2)}_{a a_2 \ldots a_n b' d'; c'b} +  \phi^{(2)}_{b a_2 \ldots a_n 
b'd';c'a} 
\right)
 x^{a_2} \cdots x^{a_n} \cM^{b' c'}  x^{d'}
\nn \\
&\qquad
+(n-1) \theta (  x^a \cA_b^{(3)} +  x^b \cA_a^{(3)} ) 
+2 \frac{n+2}{n+1}\theta (  x^a \cA_b^{(0)} +  x^b \cA_a^{(0)} ) \,,  \nn \\
%
%
 H_{ab}[\cA^{(R)}] &= 2 \theta R^2   P_T^{ab}  \phi^{(2)} \,.
\end{align}

\section{Eigenvalues of \texorpdfstring{$\cI$}{I} modes}
\label{app:eigenvalues_modes}
According to \cite[Eq.\ (2.42)]{Steinacker:2016vgf}, the eigenvalues of the Poisson Laplacian are given by 
\begin{align}
 \Box (\tilde{n}-m,2m)= \theta \big(- \tilde{n}(\tilde{n}+3)+m(m+1)\big)
\end{align}
and the vector mode Laplacian \eqref{vector-Laplacian} is
$ \cD^2 = -\Box - 2\cI$ .
The properties of the modes $\cB_I$ are summarized in 
Tab.\ \ref{tab:eigenvalues_I-modes}.
\begin{table}[h]
\scriptsize
 \begin{tabular}{c|c|c|c|c|c}
  mode & rep & identification & $\cI$ eigenvalue & $\Box_x$ eigenvalue & 
$\cD^2$ 
eigenvalue \\ \hline
$B^{(s)}$ & $(\tilde{n}-m,2m) $ & $\tilde{n}=n+s$ & $-2$  & 
$-\tilde{n}(\tilde{n}+3)+m(m+1)$ & \\
 &$(n,2s)$ & $ m=s$ & $-2$ &  $-n^2 -2s-n(2s+3) $  & $n(n+3)+2s(n+1)+4 $\\ 
\hline
$C^{(s)}$ & $(\tilde{n}-m-1,2m)$ & $\tilde{n}=n+s+1$ & $-\tilde{n}-3$ & 
$-\tilde{n}(\tilde{n}+3)+m(m+1)$  &\\
&$(n,2s)$ & $ m=s$ & $-n-s-4 $ & $-n^2-n(5 + 2s) -4(s+1) $ & $(n+3) (n+2 s+4) $ 
\\ 
\hline
$D^{(s)}$ & $(\tilde{n}-m+1,2m)$ & $\tilde{n}=n+s-1$ & $\tilde{n}$ & 
$-\tilde{n}(\tilde{n}+3)+m(m+1)$ & \\
&$(n,2s)$ & $m=s$ & $n+s-1 $ & $-n^2 -n(1 +2s) +2 $ & $(n-1)(n+2s) $ \\ \hline
$E^{(s)}$ & $(\tilde{n}-m-1,2m+2) $ & $\tilde{n}=n+s$ & $m-1$ & 
$-\tilde{n}(\tilde{n}+3)+m(m+1)$ & \\
&$(n,2s)$ & $ m=s-1 $ & $s-2 $ & $-n^2-n(2s+3)-4s $ & $n(n+3)+2s(n+1)+4 $ 
\\ 
\hline
$F^{(s)}$ & $(\tilde{n}-m+1,2m-2)$ & $\tilde{n}=n+s$ & $-m-2$ & 
$-\tilde{n}(\tilde{n}+3)+m(m+1)$ & \\
&$(n,2s)$ & $m=s+1$ & $-s-3 $ & $-n^2-n(2s+3)+2$ & $n(n+3) +2s(n+1) +4 $
 \end{tabular}
\caption{The eigenvalues of the $\cI$-modes $B,C,D,E,F$ for any spin level s. 
(All $\cI$, $\Box_x$, and $\cD^2$ eigenvalues are  modulo $\theta$.) Note that 
the notation for the representations follows that of \cite{Steinacker:2016vgf}.}
\label{tab:eigenvalues_I-modes}
\end{table}

\section{Green's functions}
\subsection{Semi-classical}
\label{App:propagators}
We consider the solutions to the equations \eqref{eq:eom_modeB_flat} and 
\eqref{eq:eom_modeCD_flat} with a Gaussian source as inhomogeneity.
\paragraph{Preliminaries}
Define Fourier transforms and their inverse as (in 4d)
\begin{align}
 \tilde{f}(\vec{k}) \coloneqq \frac{1}{4 \pi^2} \int \mathrm{d}^4 x \ 
f(\vec{x}) e^{i \vec{k} \cdot  \vec{x}} \; , \qquad 
 f(\vec{x}) \coloneqq \frac{1}{4 \pi^2} \int \mathrm{d}^4 k \ 
\tilde{f}(\vec{k}) e^{-i \vec{k} \cdot  \vec{x}} \; .
\end{align}
We can then solve inhomogeneous differential equation with 
differential operator $\cD$ as usual
\begin{align}
\mathcal{D} \psi(\vec{x}) &= f(\vec{x})
\quad  \Leftrightarrow \quad 
\widetilde{\mathcal{D}} \widetilde{\psi}(\vec{k}) = 
\widetilde{f}(\vec{x}) 
\quad  \Leftrightarrow \quad 
 \psi(\vec{x}) =  \frac{1}{4 \pi^2} \int \mathrm{d}^4 k \   
\widetilde{\mathcal{D}}^{-1}
\widetilde{f}(\vec{k}) e^{-i \vec{k} \cdot  \vec{x}} \; ,
\end{align}
where $\widetilde{\cD}$ is the algebraic operator representing $\cD$.
We consider a Gaussian source $f$ with 
\begin{align}
 f(\vec{x}) = \frac{1}{4\pi^2 a^4} e^{-\frac{\vec{x}^2}{2 a^2}} 
 \qquad \Rightarrow \qquad
 \widetilde{f}(\vec{k}) = \frac{1}{4\pi^2 } e^{-\frac{a^2 \vec{k}^2}{2 
}} 
\end{align}
such that $\int \mathrm{d}^4 x f(\vec{x}) =1$.
\paragraph{Computation}
We compute the integral in 4d spherical coordinates: (i) radial coordinate $r$ 
and (ii) three angles $\psi_1,\psi_2 \in [0,\pi]$, and $\psi_3 \in [0,2\pi]$. 
Then the volume element becomes $\mathrm{d}^4 x = r^3 \mathrm{d}r \ 
(\sin{\psi_1})^2 \mathrm{d} \psi_1 \ \sin{\psi_2} \mathrm{d} \psi_2 
\ \mathrm{d} \psi_3$, and we can choose a direction such that $\vec{x}\cdot 
\vec{k} = k r \cos{\psi_1}$.
 \begin{align}
  \psi(\vec{x}) &=  \frac{1}{4 \pi^2} \int \mathrm{d}^4 k \   
\widetilde{\mathcal{D}}^{-1}
\widetilde{f}(\vec{k}) e^{-i \vec{k} \cdot  \vec{x}} \nn \\
&=  \frac{1}{4 \pi^2} \int k^3 \mathrm{d}k \ 
(\sin{\psi_1})^2 \mathrm{d} \psi_1 \ \sin{\psi_2} \mathrm{d} \psi_2 
\ \mathrm{d} \psi_3 \   
\widetilde{\mathcal{D}}^{-1}
\widetilde{f}(k) e^{-i  k r \cos{\psi_1}} \nn \\
&=  \frac{2}{(2 \pi)^3} \int k^3 \mathrm{d}k \
\widetilde{\mathcal{D}}^{-1}
e^{-\frac{a^2 \vec{k}^2}{2 }} \
 \underbrace{\int_{-1}^1 \mathrm{d} \xi \sqrt{1-\xi^2}    
 e^{-i  k r \xi}}_{=\pi \frac{J_1(kr)}{kr}} \nn \\
&=  \frac{1}{(2 \pi)^2} \int_{0}^\infty  \mathrm{d}k \ k^3
\widetilde{\mathcal{D}}^{-1}
  \frac{J_1(kr)}{kr} e^{-\frac{a^2 \vec{k}^2}{2 }} \; ,
 \end{align}
where $J_n(z)$ denotes the Bessel functions of first kind.
We consider two examples
\begin{itemize}
 \item $\mathcal{D}$ corresponds to the 4d Laplacian, then 
$\widetilde{\mathcal{D}}^{-1} \sim \frac{1}{k^2}$. Then we obtain
\begin{align}
 \psi(\vec{x})=\frac{1}{(2 \pi)^2} \int_{0}^\infty  \mathrm{d}k \   
\frac{J_1(kr)}{r} e^{-\frac{a^2 \vec{k}^2}{2 }}  
=\frac{1-e^{-\frac{r^2}{2 a^2}}}{4 \pi ^2 r^2}
\end{align}
\item $\mathcal{D}$ corresponds to some (first order) operator such that 
$\widetilde{\mathcal{D}}^{-1} \sim \frac{1}{k}$, then
\begin{align}
\psi(\vec{x}) &=    \frac{1}{(2 \pi)^2} \int_{0}^\infty  \mathrm{d}k \ k
  \frac{J_1(kr)}{r} e^{-\frac{a^2 \vec{k}^2}{2 }} \nn \\
&= \frac{1}{8 \sqrt{2} \pi ^{3/2} a^3}e^{-\frac{r^2}{4 a^2}} 
\left(I_0\left(\frac{r^2}{4 a^2}\right)-I_1\left(\frac{r^2}{4 a^2}\right)\right)
\; ,
\end{align}
where $I_n(z)$ denotes the modified Bessel functions of first kind.

We can use the known asymptotic behavior, see for instance 
\cite{Abramowitz:Stegun},
and obtain the two regimes for $\psi$ as follows:
\begin{align}
\psi(\vec{x}) &=
\begin{cases}
 e^{-\frac{r^2}{4 a^2}} \ \frac{ \left(2 -\frac{r^2}{4a^2}\right)}{16 
\sqrt{2} \pi^{3/2} a^3} &, 0 < r \ll 2a  \; ,\\
 \frac{1}{4 \pi ^2 r^3} 
 +\frac{3 a^2}{8 \pi ^2 r^5} 
 +\frac{95 a^4}{64 \pi ^2 r^7} 
 +\ldots
&, \frac{r}{2a}\to \infty \; .
 \end{cases}
\end{align}
\end{itemize}
\subsection{Induced gravity}
\label{App:induced_gravity}
The scale $\bar{R}$ arising in induced gravity, see Section 
\ref{subsec:induced_gravity}, acts as regulator and one can readily compute the 
Green's function by considering Delta-distribution sources.
\paragraph{Mode B.}
Considering the equations of motion \eqref{eq:eom_modeB_induced} for the 
splitting \eqref{eq:split_modeB}, we derive the Green's functions for 
\eqref{eq:eom_modeB_split}.
\begin{itemize}
 \item For the PDE for $h^{(B,\mathrm{loc})}_{\mu \nu}$ we obtain
 \begin{align}
  G^{(B,\mathrm{loc})}(r,0)= \frac{K_{1}\left(\frac{r}{\bar{R}}\right)}{4 \pi^2 
r \bar{R}}
 \end{align}
where $K_{n}(z)$ denotes the modified Bessel function of second kind.
From the known asymptotic behavior \cite{Abramowitz:Stegun} we arrive at
\begin{align}
\label{eq:Green_local_part}
 G^{(B,\mathrm{loc})}(r,0)=
 \begin{cases}
  \frac{1}{4\pi^2 r^2} &, 0< \frac{r}{\bar{R}} \ll \sqrt{2} \\
 \frac{1}{4 \pi^2 r^2} e^{-\frac{r}{\bar{R}}}  \sqrt{\frac{\pi}{2} 
\frac{r}{\bar{R}}}
  \left( 1 
  + \frac{3 \bar{R}}{8 r} 
  -\frac{15 \bar{R}^2}{128 r^2}
  + \ldots\right)
&, \frac{r}{\bar{R}} \to \infty
 \end{cases}
\end{align}
  \item For the PDE for $h^{(B,\mathrm{grav})}_{\mu \nu}$ we compute
  \begin{align}
  G^{(B,\mathrm{grav})}(r,0)= \frac{1 -\frac{r}{\bar{R}} 
K_{1}\left(\frac{r}{\bar{R}}\right)}{4 
\pi^2 \frac{r^2}{\bar{R}^2}} \; .
 \end{align}
 Again, the known asymptotics \cite{Abramowitz:Stegun} reveals
 \begin{align}
 \label{eq:Green_grav_part}
 G^{(B,\mathrm{grav})}(r,0)=
  \begin{cases}
   \frac{1-2\gamma_E +2\ln(\frac{2\bar{R}}{r})}{16\pi} 
  +\frac{5-4\gamma_E +4\ln(\frac{2\bar{R}}{r})}{(16\pi)^2} 
\frac{r^2}{\bar{R}^2} + \ldots
  &, 0< \frac{r}{\bar{R}} 
\ll \sqrt{2} \\
\frac{1}{4 \pi^2 \frac{r^2}{\bar{R}^2}}
\left( 1 - e^{-\frac{r}{\bar{R}}} \sqrt{\frac{\pi}{2} \frac{\bar{R}}{r}}
\left( \frac{r}{\bar{R}} 
+ \frac{3}{8} 
- \frac{15}{128} 
\frac{\bar{R}}{r}
+ \ldots \right)
\right) &, \frac{r}{\bar{R}} \to \infty \; ,
  \end{cases}
 \end{align}
 where $\gamma_E$ denotes the Euler--Mascheroni constant.
\end{itemize}

\paragraph{Mode C and D.}
Considering the equations of motion \eqref{eq:eom_modeCD_induced} for the 
splitting \eqref{eq:split_modeCD}, we derive the Green's functions for 
\eqref{eq:eom_modeCD_split}.
\begin{itemize}
 \item The behavior of the $h^{(I,\mathrm{loc})}_{\mu \nu}$ part is identical 
to the 
previous one of $h^{(B,\mathrm{loc})}_{\mu \nu}$. Therefore,
  \begin{align}
  G^{(I,\mathrm{loc})}(r,0) =G^{(B,\mathrm{loc})}(r,0) \; .
 \end{align}
  \item All left to check is the Green's function for 
$h^{(I,\mathrm{nonloc})}_{\mu \nu}$. We find
\begin{align}
 G^{(I,\mathrm{nonloc})}(r,0) = \frac{1}{8\pi r} \left( L_{-1}\left( 
\frac{r}{\bar{R}}\right) - 
I_{1}\left( \frac{r}{\bar{R}}\right)  \right) \; .
\end{align}
where $L_\alpha(z)$ denotes the modified Struve function.
Employing the tabulated expansion of \cite{Abramowitz:Stegun},
one can deduce the asymptotic behavior as follows:
\begin{align}
\label{eq:Green_nonloc_part}
 G^{(I,\mathrm{nonloc})}(r,0) =
 \begin{cases}
  \frac{1}{8 \pi  r} \left(\frac{2}{\pi } -\frac{r}{2 \bar{R}}
  +\frac{2 r^2}{3 \pi  \bar{R}^2}
  +\frac{2 r^4}{45 \pi  \bar{R}^4}
  +\ldots
   \right)
  &, 0<\frac{r}{\bar{R}}\ll1 \; ,\\
\frac{ \bar{R}^2}{4 \pi^2 r^3}
+\frac{3 \bar{R}^4}{4 \pi^2 r^5}
+\frac{45 \bar{R}^6}{4 \pi^2 r^7}
+\ldots
&, \frac{r}{\bar{R}} \to \infty \; .
 \end{cases}
\end{align}
\end{itemize}

\stoptocwriting
\resumetocwriting

\bibliographystyle{JHEP}
\bibliography{papers}

\end{document}